\documentclass[11pt]{article}

\usepackage[top=0.6in, bottom=0.6in, left=0.6in, right=0.6in]{geometry}
\usepackage{lineno}
\usepackage{amsmath}
\usepackage{amsfonts}
\usepackage{amssymb}
\usepackage{latexsym}
\usepackage{appendix}
\usepackage{verbatim}
\usepackage{bbold}
\usepackage{bm}
\usepackage{mathrsfs}  
\usepackage{cite}
\numberwithin{equation}{section}
\allowdisplaybreaks
\usepackage{epsf}
\usepackage{color}
\usepackage{graphicx}
\usepackage{dsfont}
\usepackage{subfigure}
\usepackage[export]{adjustbox}
\usepackage{hyperref}
\definecolor{darkred}{rgb}{0.8,0.1,0.1}
\definecolor{purple}{RGB}{82, 92, 157}
\hypersetup{urlcolor=purple,
colorlinks=true, linkcolor=purple, citecolor=purple, linktoc=page}

\usepackage[font={small,it}, labelfont=bf]{caption}
\usepackage{subcaption}    
\usepackage{setspace}

\graphicspath{{Figures/}}

\newcommand{\be}{\begin{equation}}
\newcommand{\ee}{\end{equation}}

\renewcommand{\thefootnote}{\fnsymbol{footnote}}
\renewcommand{\thanks}[1]{\footnote{#1}}
\newcommand{\starttext}{
\setcounter{section}{0}
\renewcommand{\thefootnote}{\arabic{footnote}}}

\newcommand{\bea}{\begin{eqnarray}}
\newcommand{\eea}{\end{eqnarray}}
\newcommand{\bma}{\begin{matrix}}
\newcommand{\ema}{\cr\end{matrix}}

\newcommand*\wc{{\mkern 1mu\cdot\mkern 1mu}}

\def\be#1\ee{\begin{equation}#1\end{equation}}
\def\ba#1\ea{\begin{align}#1\end{align}}
\def\[#1\]{\begin{align}#1\end{align}}

\def\be#1\ee{\begin{equation}#1\end{equation}}
\def\ba#1\ea{\begin{align}#1\end{align}}
\def\[#1\]{\begin{align}#1\end{align}}

\def\det{{\rm det \,}}

\def\sm{\smallskip}

\begin{document}
\starttext

\begin{flushright}
March 2026
\end{flushright}

\bigskip

\begin{center}

{\Large \bf 

Efficient computation of the N-th rank QED polarization tensor:\\ Universal worldline structure of form factors
}

\vskip 0.4in

{\large X. Feal$^{1,4,*}$, A. Tarasov$^{2,3,\dag}$, and R. Venugopalan$^{3,4,5,\ddag}$,}

\vskip 0.4in
 
 { \normalsize \emph{${}^{1}$  Department of Chemistry and Physics, State University of New York at Old Westbury, \\ Old Westbury, NY
11568, United States}}

\vskip 0.05in

 {\normalsize\emph{${}^{2}$Department of Physics, North Carolina State University, \\ Raleigh, NC 27695, USA},}
 
\vskip 0.05in

 {\normalsize\emph{
${}^{3}$Center for Frontiers in Nuclear Science (CFNS) at Stony Brook University,\\
Stony Brook, NY 11794, USA}}

 \vskip 0.05in

 {${}^{4}$\normalsize\emph{Physics Department, Bldg. 510A, 
Brookhaven National Laboratory, \\
 Upton, NY 11973, USA}}
 
 \vskip 0.05in

 {${}^{5}$\normalsize\emph{Higgs Centre for Theoretical Physics, The University of Edinburgh, \\
 Edinburgh, EH9 3FD, UK}}
 
\hskip 0.5in

\begin{abstract}
We derived in \cite{Feal:2022iyn,Feal:2022ufw} a compact expression for the $N$-th rank QED polarization tensor $\Pi_{\mu_1\cdots \mu_N}(k_1,\cdots,k_N)$ in a $(0+1)$-dimensional worldline framework.  This fully off-shell object, a function of $N$ external photon four-momenta, is a key ingredient in high-order computations of cusp anomalous dimensions and lepton anomalous magnetic moments. 
 We demonstrate here that $\Pi_{\mu_1\cdots \mu_N}$ can further be expressed simply in terms of a small number of independent ``head" form factors (each representing $(N-1)!/2$ Feynman diagrams) which have a universal structure in terms of sums over fermion Green functions and (propertime derivative of) their boson worldline superpartners. This worldline representation bypasses explicit Wick contractions and avoids tensor reductions to scalar loop integrals \`a la Passarino and Veltman, order by order in perturbation theory. We give explicit expressions for the $4$-th and $6$-th rank head form factors and provide a computer script generalizing these results to arbitrary $N$ external photons. The multiplicity of heads, and their growth with $N$, can be understood in terms of orbits of the permutation group. We employ the Burnside-Cauchy-Frobenius lemma to show that it scales as 
$e^{N-1}/\sqrt{N}$ terms as opposed to the $e^{N-1} N!/\sqrt{N}$ terms in conventional perturbation theory. We reexpress worldline parameter integrals that define the $4$-th rank heads as Feynman parameter integrals  to reproduce the seminal results by  Karplus and Neuman for the on-shell light-by-light amplitude and extend these to the fully off-shell case in massless QED employing a tailored integration-by-parts procedure. In a follow-up paper, we will discuss the direct computation of worldline integrals,  potentially providing a further $N!$ advantage relative to Feynman diagram computations at high orders in perturbation theory. 

\end{abstract}

\hskip 1.5in

\begin{flushright}
{${}^*$\tt xgarciafe@bnl.gov} \\
{${}^\dag$ \tt ataraso@ncsu.edu}\\
{${}^\ddag$ \tt rajuv@bnl.gov}
\end{flushright}

\end{center}

\newpage

\setcounter{tocdepth}{2} 

\noindent\rule{\textwidth}{0.4pt}
{\setstretch{1.2}
\tableofcontents
}
\noindent\rule{\textwidth}{0.4pt}

\baselineskip=16pt

\setcounter{equation}{0}
\setcounter{footnote}{0}

\section{Introduction}
\label{sec:intro}

In \cite{Feal:2022iyn,Feal:2022ufw}, we 
developed a worldline reformulation of Quantum Electrodynamics (QED) as a Lorentz-covariant theory of  superpairs of $(0+1)$-dimensional worldlines interacting nonlocally through the Lorentz force. In the worldline approach to Quantum Field Theory (QFT), the gauge and matter fields are integrated out explicitly, with the result being reexpressed as 0+1-dimensional quantum mechanical path integrals of point-like trajectories of boson and Grassmann variables in propertime. The internal degrees of freedom (spin and helicity in the QED case) are naturally described by the commutation relations of these degrees of freedom. An attractive feature of the worldline formalism is that the coupling of the bosonic and Grassmann variables to external gauge fields (respectively, particle rotation and spin precession therein) can be exponentiated in path ordered form to all orders in the gauge coupling. For a comprehensive review of the worldline formalism, see \cite{Schubert:2001he}. 

We discussed further an all-order $S$-matrix worldline framework which provides a  powerful mechanism to discuss nonperturbative features of gauge interactions that elude a transparent treatment in conventional field theory. In particular, we showed how soft theorems and their Abelian exponentiation follow naturally in this formulation. These are encoded as long distance interactions between the charged 0+1-dimensional worldline currents;  we used this insight to demonstrate for the first time an all-order proof in QED of the cancellation of virtual infrared (IR) divergences in the Faddeev and Kulish (FK) $S$-matrix. This $S$-matrix, in contrast to the standard Dyson $S$-matrix,  provides a manifestly IR-safe formulation of QED at the amplitude level \cite{Kulish:1970ut,Zwanziger:1973if}. For a modern discussion, along the same lines, see also \cite{Hannesdottir:2019opa}.

In particular, in \cite{Feal:2022ufw}, we extended the all-order proof of the IR safety of the FK $S$-matrix to include transitions accompanied by the emission and/or absorption of arbitrary numbers of real photons. We showed how Low's soft theorem applies in the worldline framework~\cite{Low:1958sn}, and recovered Weinberg's theorem~\cite{Weinberg:1965nx} for the exponentiation of soft infrared (IR) divergences. The difference between the Dyson and FK formalisms can be traced to the ordering of limits of how the initial and final times, $x_{i,f}^0 \rightarrow \mp \infty$, and momentum $k\rightarrow 0$ are taken. In the Dyson framework, the former limit is taken first, setting the contribution from asymptotic worldline currents to zero, while in the FK framework, the latter limit is taken first. In this case, the IR divergences cancel between the asymptotic currents and the Dyson current. We argued that such a description can be accommodated within a Wilsonian RG formulation of the worldline approach; this reasoning, for the general case, is implicit in Weinberg's paper, albeit at the cross-section level. The existence of such asymptotic currents/charges (with the appropriate ordering of limits) has been argued to arise as a consequence of these being the Noether 
charges corresponding to  generators of large gauge transformations that do not annihilate the ``Dyson" QED vacuum~\cite{Kapec:2017tkm,Strominger:2017zoo}. The corresponding in-out asymptotic states are coherent states of soft photons~\cite{Kulish:1970ut}. Analogous arguments hold for the gravitational $S$-matrix~\cite{Strominger:2013jfa,Choi:2017ylo}.

Another feature of the worldline formalism we discussed in  \cite{Feal:2022ufw} was the computational advantage (in additional to the conceptual clarity provided) of this all-order $S$-matrix worldline formalism. It provides a powerful alternative approach for high-order perturbative calculations in QFT without using Feynman diagrams~\cite{Strassler:1992zr}, a program first laid out using string-inspired methods in \cite{Kosower:1987ic}. We showed explicitly that {\it multi-loop QED vacuum-vacuum amplitudes with $l$-fermion and $n$-photon loops can be expressed as the convolution of $l$ one-loop gauge invariant $N$th rank polarization tensors}, where $N$ denotes the number of photons that are attached to a particular fermion sub-graph in the amplitude. These $N$th-rank polarization tensors can be expressed as the worldline expectation value of a product on $N$ charged particle currents. A well-known limiting case~\cite{Strassler:1992zr} is that one recovers the Bern-Kosower expression~\cite{Bern:1991aq} for the one-loop polarization tensor with an arbitrary number of external photon legs. The Grassmann integrals in the $N$th-rank polarization tensor can be performed explicitly, allowing us to obtain in \cite{Feal:2022ufw} a novel compact expression containing $N$-unordered propertime integrals and products of very simple boson and fermion worldline propagators and their proper time derivatives. This expression simultaneously represented $(N-1)!/2$ Feynman diagrams (for $N>2$) in perturbation theory that correspond to the combinatorics of the ordering of the propertime integrals. 

In this work, we will explore the concrete implications of applying the worldline formalism to compute high orders in perturbation theory, connect these to well-known results using Feynman perturbation theory, and provide several novel and promising worldline results as part of this exploration. Specifically, in this paper, as an interesting test of the computational potential of the worldline formalism relative to techniques developed for conventional perturbation theory, we will focus 
on computing explicitly the $4$th and $6$th rank polarization tensors, that are, respectively, the key building blocks of the 
four-loop, five-loop and six-loop cusp anomalous dimension in QED.  In a follow-up paper~\cite{PaperII}, we will develop a novel approach to directly compute the worldline integrals corresponding to these high rank polarization tensors. This is  significant because, as we will discuss later in this paper, the usual computation of these integrals required ordering of the $N!$ diagrams involved, thereby losing some of the computational advantage relative to conventional perturbation theory. 
 
An important application of our work is the computation of high order cusp anomalous dimensions. At energies much higher than the electron mass, the cusp anomalous dimension governs the resummation of IR singularities appearing in scattering amplitudes to all orders in perturbation theory. Further, a straightforward if nontrivial generalization of the cusp worldline result corresponds to the computation of the anomalous magnetic moment of the electron and the muon; these provide the ultimate benchmark for precision computations employing Feynman diagrams in QFT~\cite{Aoyama:2019ryr,Aoyama:2020ynm,Aliberti:2025beg}.  Not least, the cusp anomalous dimension in QCD plays a central role in collider physics phenomenology \cite{Korchemskaya:1994qp,Korchemskaya:1996je,Dixon:2008gr,Becher:2009cu,Caron-Huot:2017zfo,Henn:2016jdu,Caron-Huot:2017fxr,Caola:2021rqz,Falcioni:2021buo}. 

To provide context for the importance of such studies, we note that the study of the universal behavior of soft theorems and their all-order exponentiation has endured as an important area of research ever since the aforementioned pioneering work of Weinberg~\cite{Weinberg:1965nx}. He demonstrated how the exponentiation of soft IR divergences in charged-particle transitions (to one-loop order) is entirely determined by the appearance of the cusp anomalous dimension in the renormalization group evolution (RGE) equations for both amplitudes and for cross sections. These depend only on the momenta of the charged particles in the initial and final states and thus can be systematically factorized into a soft factor\footnote{In theories with massless gauge bosons such as QCD, collinear IR divergences further depend on the spin and representation of the emitting parton, and are analogously factorized from the hard $S$-matrix into a jet function.}. In Abelian theories such as QED, where gauge bosons are charge neutral, the cusp anomalous dimension reduces to a sum of pairwise contributions --- or \textit{dipole} terms --- of anomalous dimensions between all pairs of charges in the initial and final states. Each of these terms encodes the IR divergent part of the gauge interaction for the pair over infinitely long distances, with the real part describing the radiative modes of the virtual soft-photon exchanges, and the imaginary part corresponding to the classical Li\'enard--Wiechert/Coulomb potentials. In non-Abelian theories, because gauge bosons themselves carry charge, the cusp anomalous dimension receives additional contributions from simultaneous long-distance exchanges among more than two colored particles in the initial or final states. These are sometimes referred to as \textit{quadrupole} terms, and they first appear at three and four loops~\cite{Henn:2023pqn}.

Before we discuss the worldline strategy for computing cusp anomalous dimensions, we will first  summarize the state-of-the-art in Feynman perturbation theory. The full angle dependence of the cusp anomalous dimension is presently known to four loops ($\text{N}^3\text{LO}$) in massless QED~\cite{Bruser:2020bsh}. In QCD, the angle-dependent cusp anomalous dimension was computed analytically to three loops ($\text{N}^2\text{LO}$)~\cite{Grozin:2014hna,Grozin:2015kna}, 
while more recently the light-like cusp anomalous dimension was obtained to four loops in massless QCD~\cite{Henn:2019swt,vonManteuffel:2020vjv,Agarwal:2021zft}. The high-order computation of cusp anomalous dimensions is a challenging task. In massless theories, for instance, the two-loop (NLO) non-Abelian cusp anomalous dimension was known for a while from the first available next-to-leading computations of the structure functions in deep inelastic scattering (DIS)~\cite{Floratos:1977au,Gonzalez-Arroyo:1979guc,Korchemsky:1987wg}. More than two decades elapsed before first results at the next order were obtained from the three-loop computation of the DGLAP splitting functions~\cite{Moch:2004pa,Vogt:2004mw}. It took yet another decade for first results at four and five loops ($\text{N}^3\text{LO}$ and $\text{N}^4\text{LO}$) to appear, either numerically~\cite{Moch:2018wjh}, analytically~\cite{vonManteuffel:2020vjv,Herzog:2018kwj,Agarwal:2021zft}, or analytically from an extrapolation of the $\mathcal{N}=4$ supersymmetric Yang--Mills calculation of cusp anomalous dimensions~\cite{Henn:2019swt}. The state-of-the-art approaches to these high-order computations rely on a combination of different methods—such as Integration by Parts (IBP) identities and difference or differential equation methods\cite{Chetyrkin:1980pr,Chetyrkin:1981qh,Laporta:2000dsw,Henn:2013pwa}, Mellin--Barnes integral representations \cite{Barnes:1908,Bailey:1935,Boos:1990rg,Davydychev:1990jt,Tausk:1999vh,Dubovyk:2022}, techniques from Heavy Quark Effective Field Theory (HQET) \cite{Peskin:1979va} or the Wilson-loop operator product expansions \cite{Wilson:1969zs}, among others.

Dedicated computer-algebra developments \cite{Strubbe:1974vj,vanOldenborgh:1989wn,Vermaseren:2000nd,Kuipers:2012rf} are then essential for the symbolic manipulation of the large number of terms obtained in these computations, to efficiently solve the systems of difference/differential equations arising from IBP identities, including algorithms to reduce them to a smaller basis of (scalar) master integrals \cite{Laporta:2000dsw,Lee:2012cn,Anastasiou:2004vj,Maierhofer:2017gsa,Smirnov:2019qkx,Henn:2013pwa}, or for finally implementing (numerically stable) analytic solutions for these one-loop scalar $N$-point functions  \cite{tHooft:1978jhc,vanOldenborgh:1989wn,Hahn:1998yk,Ellis:2007qk} in terms of polylogarithm functions and/or multiple nested expansions of transcendental functions \cite{Remiddi:1999ew,Moch:2001zr,Moch:2005uc}. For instance, the family of integrals corresponding to a specific four-loop contribution to the QED cusp anomalous dimension involving the fourth-rank light-by-light QED polarization tensor (whose external photon legs are coupled to the external charged particle) consists of roughly 500 master integrals \cite{Bruser:2020bsh}. For the QCD case, alternative techniques reduce the computation to that of 294 master integrals~\cite{Agarwal:2021zft}. Both results required extensive numerical analysis employing high performance computing. 

Despite these state-of-the-art calculations becoming extremely sophisticated, the feasibility of extending such techniques to higher loops remains unclear, as the number of Feynman diagrams is known to explode factorially. This characteristic feature of conventional perturbation theory poses a formidable challenge, with the existing theoretical predictions perhaps approaching the limits of what can be achieved analytically. The computation of scattering amplitudes becomes quite involved also in subnuclear physics, where accurate QCD predictions require summing a large number of Feynman diagrams in nontrivial background gauge fields when dealing with quarks and gluons. In particular, large logarithmic edge of phase space contributions require one to go beyond fixed order perturbation theory and resum systematically such contributions, to all loop orders, order-by-order in perturbation theory. Another major challenge in multiloop calculations is the appearance of elliptic obstructions, which already arise at two loops with massive propagators and nontrivial thresholds, for example in massive Bhabha scattering and Higgs- or jet-production amplitudes with internal masses. In such cases, no $\epsilon$-factorized dlog form exists for the canonical differential equation procedure, and the expansion of the Feynman diagrams in terms of multiple polylogarithms breaks down \cite{Bloch:2013tra,Duhr:2014woa,Duhr:2025lbz,Vanhove:2026fth}.

Spurred in part by these challenges, there has recently been a surge of developments and exploratory work aimed at deepening our understanding and reworking our formulations of gauge theories and the structure of scattering amplitudes, invoking significant developments in mathematical physics. The study of the structure of spacetime as a twistor theory \cite{Penrose:1972ia}, and its connection to string theory \cite{Nair:1988bq,Witten:2003nn}, has motivated over the last two decades the creation (exploiting its Grassmannian nature) of a calculus for scattering amplitudes in the twistor geometric approach \cite{Arkani-Hamed:2009hub}. Recent developments along these lines, and sharing the same goals, include studies of the connections of Feynman integrals to projective geometry  \cite{Artico:2023jrc}, and to cohomology and intersection theory \cite{Mastrolia:2018uzb,Frellesvig:2019kgj,Frellesvig:2019kgj}. 

\begin{figure}
\includegraphics[width=1.0\textwidth]{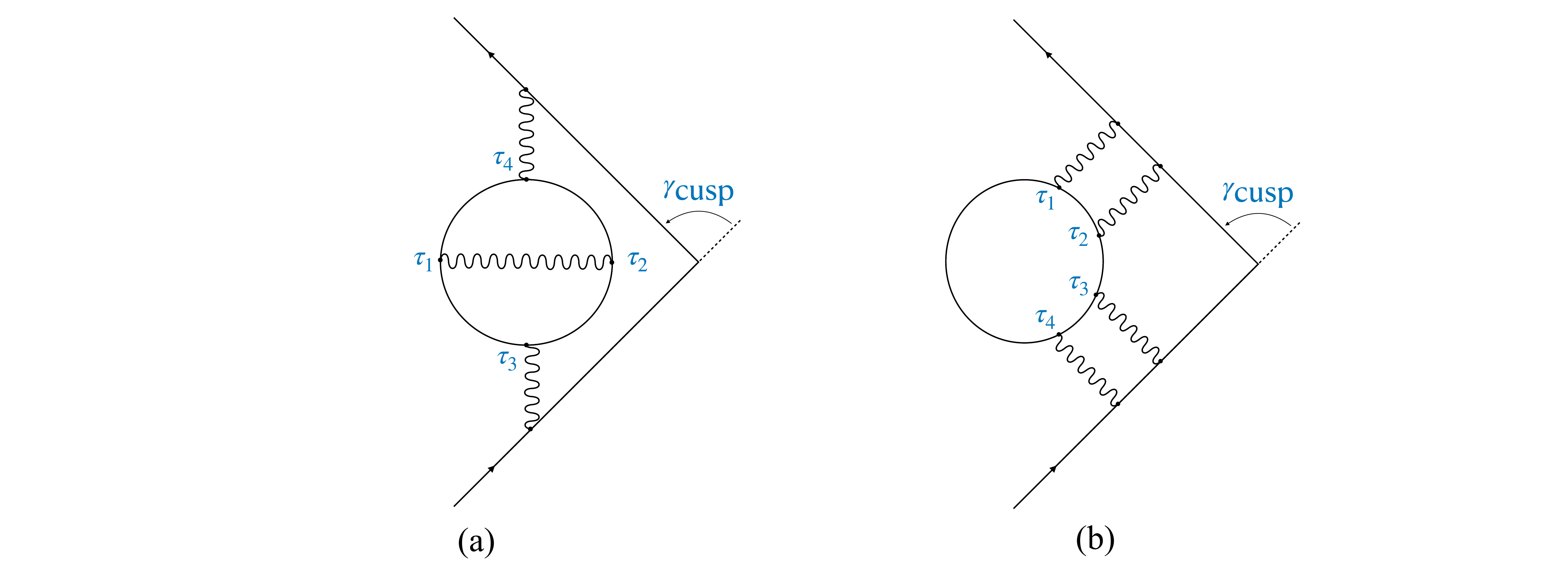}
\caption{Light-by-light  worldline diagrams with $N=4$ photons entering the three loop (a),  and four-loop (b), $N=4$ calculation of the anomalous dimension in the RGE of a cusped Wilson line  $\Gamma_\text{cusp}$. Notice that each photon vertex is located at a worldline parameter $\tau_i$ that will be integrated over $\tau_i=[0,1]$. These worldline diagrams encode all Feynman diagram topologies contributing to  $\Gamma_\text{cusp}$.}
\label{fig:figure_2}
\end{figure}

Our focus in this paper, and in \cite{PaperII}, is to provide an additional viewpoint, using methods that long predate \cite{Feynman:1950ir,Feynman:1951gn,Schwinger:1951nm} the modern era of high-order computations, to discuss possible solutions to the long standing issue of the factorial growth of Feynman diagrams in perturbation theory. As a starting point, we will focus on a computation of the fully off-shell light-by-light amplitude in QED in the worldline formalism. We are motivated by the fact that the four-photon amplitude serves as a fundamental building block for higher-loop amplitudes: it contributes, in particular, to the three-loop term of the cusp anomalous dimension 
through a self-energy correction to the leading-order vacuum polarization, and at four loops, through light-by-light scattering among virtual photons. See Fig. \ref{fig:figure_2}. Our framework generalizes to the $N$-th rank polarization tensor, which is a powerful ingredient in the strategy to compute cusps to even high orders in the worldline formalism. 

As noted earlier, the off-shell light-by-light polarization tensor also provides one of the most nontrivial contributions to the four-loop cusp anomalous dimension in QCD, when one replaces the external photon legs by gluon legs. Within the worldline formalism itself, it  provides the first nontrivial example of a gauge invariant set (or family) of several Feynman diagrams, corresponding in perturbation theory to different topologies representing permutations of photon vertices and their mirror diagrams. 

The first calculation of the light-by-light amplitude employing Feynman diagrams dates back to the seminal work of Karplus and Neuman, who provided the explicit structure\footnote{In some of the literature on the subject, the Karplus-Neuman results are understood as applying only to the on-shell light-by-light scattering amplitude. This is however a misunderstanding. The derivation of the rank-4 QED vacuum polarization tensor in the original papers, as also noted in \cite{Costantini:1971cj}, is completely general and remains valid when all four external photons are off shell. The four-photon amplitude is obtained in those works simply by taking limits of the general result for the polarization tensor for the specialized cases of interest.} of the fourth-rank QED vacuum polarization tensor~\cite{Karplus:1950zza}, together with the corresponding scattering amplitude and cross section for nonlinear light-by-light interactions, where the photons are on-shell~\cite{Karplus:1950zz}. Several additional interesting features were subsequently added to the Karplus-Neuman computation by Dolen~\cite{Dolen:1965gfx}. A detailed calculation of the four-point amplitude was carried out with two off-shell photon legs in \cite{Costantini:1971cj}; see also \cite{Ahmadiniaz:2020jgo,Ahmadiniaz:2023vrk,Ahmadiniaz:2020wlm}. 
We note that in the standard perturbative approach, the full off-shell light-by-light diagram contribution to the cusp anomalous dimension was only computed recently  in massless QED~\cite{Bruser:2020bsh}. An analytical result at zero angle can be extracted indirectly from the full massless calculation of the (mass-independent) electron anomalous magnetic moment\footnote{The mass-independent four-loop QED contribution to the electron $g-2$ is known analytically, 
a staggering calculation of 891 Feynman master integrals by Laporta~\cite{Laporta:2017okg}, completing a twenty-year effort since the completion  of the three-loop calculation in~\cite{Laporta:2009olb}. The muon's QED contributions to $g-2$ are known numerically to five loops~\cite{Aoyama:2012wk}.} $g-2$.

We will compute here for the first time the off-shell light-by-light amplitude in the 
worldline formalism and shall provide the tensor structure of the fourth-rank tensor for off-shell photons. There is currently no calculation of the light-by-light scattering amplitude  in this formalism  for the fully off-shell case - see \cite{Schubert:2024heu} for a nice recent review. (We will review as well related efforts within the worldline framework~\cite{Ahmadiniaz:2020jgo,Ahmadiniaz:2023vrk}.)  We will show that key features of the Karplus-Neuman formalism, namely, the extensive use of Ward identities, can be fully ported to the worldline formalism. In particular, the $N$th rank polarization tensor can be classified into head, shoulder and tail form factors, with the latter two determined by the head contributions. We will further explain how one systematizes this procedure, and subsequently, apply it to the computation of the sixth-rank vacuum polarization tensor in the worldline formalism. 

It is important to analyze this full tensor structure of the off-shell $N$-photon amplitudes (and more generally, their $N$-gluon  extensions) for processes where the usual simplifications obtained by tensorial reduction and/or on-shell symmetrization methods in the helicity basis of external states are not available. This is essential for instance in many applications in QCD that are sensitive to gauge invariant products of field-strength tensors. A key example is the extraction of nonperturbative operators in deeply inelastic scattering (DIS) measurements at a future Electron-Ion Collider (EIC) that provide fundamental insight into the dynamics of chiral symmetry breaking and confinement~\cite{Aschenauer:2017jsk}.
Towards this end, a novel and powerful result we will derive is that the head form factors can be expressed in a universal form that does not require extensive use of (worldline) integration-by-part identities. 

The paper is organized as follows. In Section~\ref{sec:nthrankpoltensors}, we will review our derivation of the master formula for the QED $N$-th rank polarization tensor (with $N$ external photon legs) within the worldline formalism. We will then demonstrate in detail how to construct the tensor structure for $N = 2$ and $N = 4$ by exploiting the symmetries of the problem to any given order in perturbation theory. These  allow us to simplify the computation to the evaluation of only six gauge invariant head form factors. We provide the exact expressions for these form factors to lowest order in the worldline formalism. We then discuss the extension of this procedure to obtain the structure for $N=6$. We give the explicit form of the $40$ irreducible form factors and a simple script that automatizes this task to generate all $N$th-rank tensors beyond $N=6$. 

In Section~\ref{sec:cauchy-frobenius-burnside}, we formally relate this problem to counting equivalence classes, or orbits, under symmetries of the permutation group. Using the Burnside-Cauchy–Frobenius lemma\footnote{This lemma was popularized by Burnside who attributed its derivation to Frobenius, though it is now widely credited to Cauchy.}, we directly obtain the minimal set of form factors that fully determine the $N$-photon amplitude at arbitrary order in perturbation theory. We also derive the large-$N$ asymptotics, demonstrating how the proposed method leads to an $N!$ factorial reduction in the number of tensor form factors.

In Section~\ref{sec:feynman_diagram_correspondence}, we match our results to the detailed construction of the on-shell four-photon amplitude originally obtained by Karplus and Neuman using Feynman diagrams in QED. In order to recover the Karplus-Neuman results, we see that we need to introduce ordering of the worldline diagrams. An unfortunate consequence is that this procedure removes an $N!$ advantage the worldline approach has relative to Feynman diagram computations. How to restore this advantage has been a question of considerable interest for sometime~\cite{Edwards:2021elz,Edwards:2022dbd,Ahmadiniaz:2020jgo,Ahmadiniaz:2023vrk,Schubert:2024heu}. We will discuss this issue at length in \cite{PaperII} where we develop a novel approach to the problem. 

Here however we make the explicit connection to the Feynman diagram computation and recover the expressions for the head form factors given by Karplus and Neuman. We introduce an integration-by-parts (IBP) procedure suited for this purpose that can be used to efficiently recover their on-shell results systematically. This approach, which differs from the IBP procedures employed in the literature, is very powerful, and we demonstrate how it can be used to extend the Karplus and Neuman results to the completely off-shell amplitude.

Finally, in Section~\ref{sec:conclusions}, we summarize our results. Applications include the aforementioned  high order worldline computations of the cusp anomalous dimension, as well as refinements of existing practical applications in high energy nuclear physics\footnote{For related recent treatments employing Feynman perturbation theory, see \cite{Bhattacharya:2022xxw,Bhattacharya:2023wvy,Bhattacharya:2024geo} (and references therein) for the chiral anomaly in polarized DIS, and \cite{Bhattacharya:2022xxw,Bhattacharya:2023wvy,Coriano:2024qbr} for DVCS.} of the worldline formalism, to chiral kinetic theory~\cite{Mueller:2017lzw,Mueller:2017arw,Mueller:2019gjj}, small $x$ physics in deeply inelastic scattering (DIS)~\cite{Mueller:2019qqj,Tarasov:2019rfp}, the role of the chiral anomaly in polarized DIS~\cite{Tarasov:2020cwl,Tarasov:2021yll,Tarasov:2025mvn}, and to deeply virtual Compton scattering (DVCS). These will be developed in follow-up work~\cite{Feal-Tarasov-Venugopalan}. 

Several key results are fleshed out in Appendices \ref{app:rank4_nosymmetryreduction} through \ref{app:oneloopscalarintegrals}. Appendix~\ref{app:rank4_nosymmetryreduction} provides an explicit derivation of all the gauge invariant form factors for the $4$-th rank vacuum polarization tensor in the worldline formalism. Appendix~\ref{app:transversedecomposition} presents the full transverse decomposition of the rank-4 tensor form factors using Ward identities and current conservation. Appendix~\ref{app:notational_invariance} lists the equivalence classes of head form factors defined by the symmetry-group action encoding the permutation symmetries of the tensors. Appendix~\ref{app:fullresult} provides the full result for the \(N=4\) photon amplitude, or rank-4 tensor in general kinematics, after reduction to the minimal set of six independent form factors. Appendix~\ref{app:N6list} presents the complete list of worldline form factors that fully determine the sixth-rank polarization tensor. Finally, Appendix~\ref{app:oneloopscalarintegrals} reviews, and includes for completeness, a calculation of the basis of one-loop scalar four-point functions in massive and massless QED using conventional methods, which are required to obtain the corresponding \(N=4\) form factors in closed form. 

\sm

\section[The N-th rank vacuum polarization tensor in the worldline formalism.]{The $N$-th rank vacuum polarization tensor in the worldline formalism.}\label{sec:nthrankpoltensors}
In this section, we will perform a detailed computation of the fourth-rank vacuum polarization tensor in QED in the worldline formalism and discuss further its generalization to $N=6$ and beyond\footnote{For a completely bottom-up treatment that the present work builds on, we refer the reader to our earlier papers \cite{Feal:2022iyn} and \cite{Feal:2022ufw}.}. Our goal is to illustrate how this framework can be employed systematically for analytic high-order evaluations of phenomenologically relevant processes. The approach and results provided here will serve as key building blocks for constructing high-order analytical computations with applications of phenomenological interest. As discussed in the introduction, outstanding applications for which highly precise results exist include cusp anomalous dimensions and the electron and muon anomalous magnetic moments.
\subsection[Worldline master formula for the one-loop N-th photon amplitude.]{Worldline master formula for the one-loop $N$-th photon amplitude.}
\label{subsec:worldlinemaster}
Our starting point is the closed form expression for the $N^\text{th}$-rank QED vacuum polarization tensor we derived in \cite{Feal:2022ufw}. It can be expressed as the normalized worldline expectation value of a product of $N$-charged electromagnetic worldline currents,
\begin{align}
&\Pi_{\mu_1\ldots \mu_N}(k_1,\ldots,k_N) = - \big\langle i\tilde{J}_{\mu_1}(k_1)\cdots i\tilde{J}_{\mu_N}(k_N)\big\rangle \label{eq:pi_mu1_muN}
\end{align}
Each $i$-th current insertion is given in terms of a worldline superpair $\{x_\mu(\tau),\psi_\mu(\tau)\}$ with worldline parameter $\tau \in [0,1]$ as
\ba 
\tilde{J}_{\mu}(k)=g\int^1_0 d\tau \Big[\dot{x}_\mu(\tau) +i\varepsilon_0 \psi_\mu(\tau)\psi_\nu(\tau)k_\nu\Big]e^{-ik\wc x(\tau)}\,.
\label{eq:j_current_def}
\ea 
Here the commuting (bosonic) worldline $x_\mu(\tau)$ describes the trajectory of a pointlike virtual particle with charge $g$ propagating in $d$-dimensional loop in Euclidean spacetime. Its anti-commuting (fermionic) superpartner $\psi_\mu(\tau)$ encodes the local spin preccession along the path. Besides, $k_i$ is the 4-momentum of the photon emitted or absorbed at the corresponding current insertion along the path and the parameter $\varepsilon_0$ denotes the propertime of the fermion. The normalized worldline expectation value in Eq.~\eqref{eq:pi_mu1_muN} can then be written as a first-quantized but fully Lorentz invariant worldline path integral as\footnote{We assume conventional UV renormalization and therefore omit the overall factor associated with the free energy contribution of the QED vacuum, which serves as a Schwinger propertime counterterm in the worldline formalism.} 
\ba 
&\left\langle \mathcal{O}\left[x(\tau),\psi(\tau)\right]\right\rangle = \frac{1}{2}\int_0^\infty \frac{d\varepsilon_0}{\varepsilon_0} e^{-\varepsilon_0m^2}\nonumber\\
&\times\int_\text{P} \mathcal{D}^4 x \int_\text{AP} \mathcal{D}^4\psi\exp\bigg\{-\frac{1}{4\varepsilon_0}\int^1_0 d\tau \dot{x}^2_\mu(\tau)-\frac{1}{4}\int^1_0 d\tau \psi_\mu(\tau)\dot{\psi}_\mu(\tau)\bigg\} \mathcal{O}[x(\tau),\psi(\tau)]\,,
\ea 
with periodic (P) boundary conditions for $x_\mu(\tau)$ and anti-periodic (AP) boundary conditions for $\psi_\mu(\tau)$, reflecting the fact that the virtual particle describes a closed loop in both Euclidean spacetime and spin coordinates. A final integration is performed over all possible propertimes $\varepsilon_0$, weighted by $\exp(-\varepsilon_0 m^2)$, where $m$ is the virtual fermion mass, effectively suppressing spacetime regions with wavelengths much larger than its Compton wavelength. It is an exact expression of the one-loop $N$-photon amplitude  as a path integral over all possible closed worldline contours (representing quantum fluctuations) of a point-like fermion interacting $N$ times with the gauge field.

The path integrals can be evaluated for general $N$ in $d$-dimensional Euclidean spacetime, as explained in \cite{Feal:2022iyn}, by introducing a pair of auxiliary Grassmann variables and reexponentiating each current insertion, recovering an earlier result due to Bern and Kosower \cite{Bern:1991aq}. In \cite{Feal:2022ufw}, we showed further that the resulting Grassmann integrals can also be performed for arbitrary $N$. This step is essential because it yields a universal compact expression for QED vacuum polarization tensors of any rank. Else, one is faced with the increasingly complex procedure, at each order in perturbation theory, of performing explicit Wick contractions of $(0+1)$-dimensional worldline superpair insertions $\{x_\mu(\tau),\psi_\mu(\tau)\}$ in terms of free bosonic and fermionic worldline Green functions defined in a circle with Dirichlet boundary conditions. The final result we obtained for the $N$th rank polarization tensor is~\cite{Feal:2022ufw} 
\ba 
&\Pi_{\mu_1\ldots \mu_N}(k_1,\ldots,k_N) = -(2\pi)^d\delta^d(k_1+\cdots+k_N) \frac{2g^N\mu^{2N-Nd/2}}{(4\pi)^{d/2}} \nonumber\\
&\times \int_0^1d\tau_1\ldots\int^1_0 d\tau_N \int^\infty_0\frac{d\varepsilon_0}{\varepsilon_0^{1+d/2}}\exp\bigg\{-\varepsilon_0m ^2+\frac{\varepsilon_0}{2}\sum_{i,j=1}^N k_i\cdot k_j G^B_{ij}\bigg\} I_{\mu_1\cdots \mu_N}(k_1,\ldots,k_N)\,,
\label{eq:pi_mu1_muN_solution}
\ea 
where the polynomial in the integrand is defined to be  
\begin{align}
I_{\mu_1\cdots \mu_N}(k_1,\ldots,k_N)& = \sum_{2N_a+N_b+2N_c=N}\frac{(-1)^{N_c}}{N_a!N_b!N_c!N_c!} \epsilon_{i_1j_1\ldots i_{N_a}j_{N_a}m_1\ldots m_{N_b}p_1q_1\ldots p_{N_c}q_{N_c}}\epsilon_{i_1j_1\ldots i_{N_a}j_{N_a}l_1\ldots l_{N_b}r_1s_1\ldots r_{N_c}s_{N_c}}\nonumber\\
&\times \bigg\{\prod_{\alpha=1}^{N_a} A_{i_\alpha j_\alpha}\bigg\}\bigg\{\prod_{\alpha=1}^{N_b} B_{m_\alpha l_\alpha}\bigg\}\bigg\{\prod_{\alpha=1}^{N_c} C_{p_\alpha q_\alpha}\bigg\}\bigg\{\prod_{\alpha=1}^{N_c} D_{r_\alpha s_\alpha}\bigg\}\,.\label{eq:I_mu1_muN}
\end{align}
\begin{figure}
\includegraphics[width=1.0\textwidth]{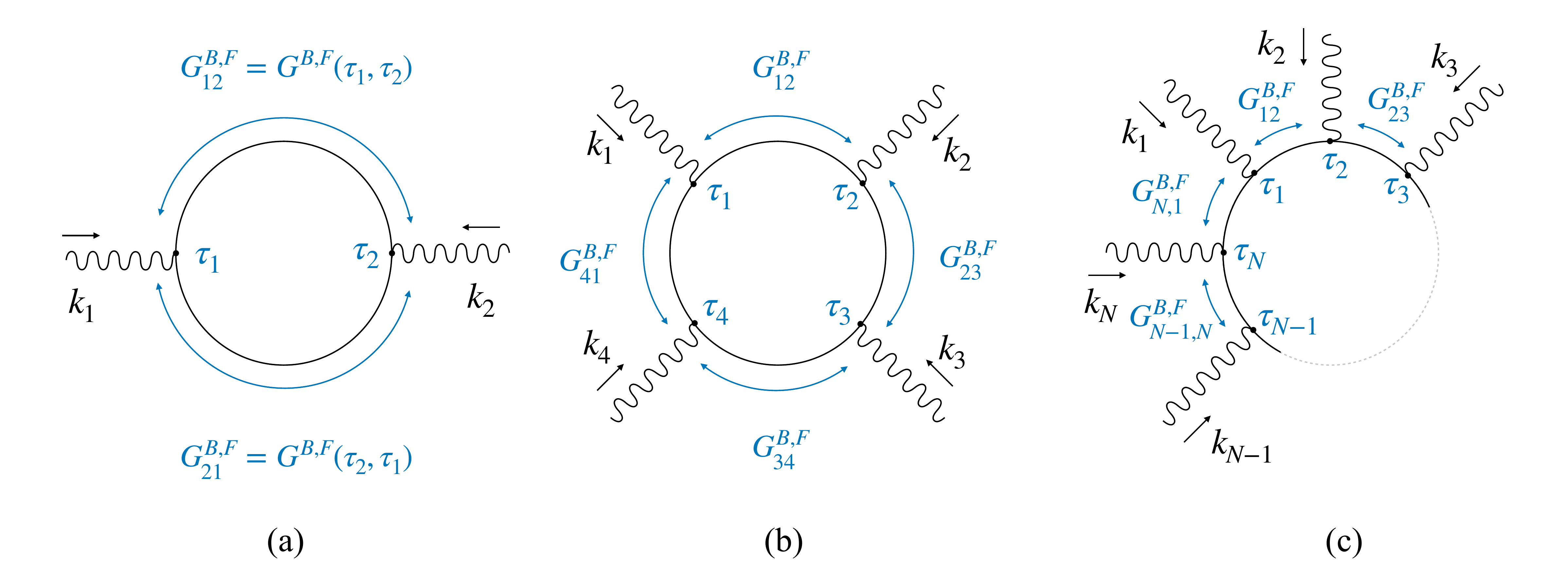}
\caption{Worldline diagrams for the QED vacuum polarization tensors at lowest order for (a) $N=2$, (b) $N=4$, and (c) general $N$ photons. Diagrams with odd $N$ insertions vanish by Furry's theorem. Each photon vertex, located at the worldline parameter $\tau_i$, will be integrated over $\tau_i=[0,1]$, with both numerator and denominator expressions depending on all boson and fermion worldline Green functions $G^{B,F}(\tau_i,\tau_j)$ in the problem. In this manner, the worldline diagrams encode all $(N-1)!$ possible ordering permutations of Feynman diagrams in conventional perturbation theory.}
\label{fig:figure_1}
\end{figure}
Here $\epsilon_{i_1i_2\ldots i_n}$ is the conventional Levi-Civita symbol, with the sum over repeated indices left implicit.

The coefficient functions $A_{ij}$, $B_{ij}$, $C_{ij}$ and $D_{ij}$ are defined in terms of worldline Green functions as
\begin{align}
\label{eq:ABCD_def}
A_{ij}=\frac{\varepsilon_0}{2} \delta_{\mu_i\mu_j}\ddot{G}^B_{ij}\,,\,\,\, B_{ij}= \varepsilon_0 \sum_{\ell=1}^N (k_{\ell})_{\mu_i} \Big[\delta_{ij} \dot{G}^B_{i\ell}+\delta_{\ell j}G^F_{\ell i}\Big]\,,\,\,\, C_{ij}=\frac{1}{2}\delta_{\mu_i\mu_j}G^F_{ij}\,,\,\,\, D_{ij}= \frac{\varepsilon_0^2}{2} k_i\cdot k_j G^F_{ij}\,.
\end{align}
The periodic and antiperiodic worldline Green functions are given by\footnote{In what follows, we will ignore the constant ($1/6$) zero-mode term in the boson Green function. It drops from the calculation because $\sum_i k_i=0$ and because the polynomial $I_{\mu_1\ldots\mu_N}$ involves only derivatives of $G_B$.}
\ba
\label{eq:g_b_ij_g_f_ij}
G^B_{ij}=|\tau_i-\tau_j|(1-|\tau_i-\tau_j|)-\frac{1}{6}\,,\,\,\,\, G^F_{ij}=\text{sign}(\tau_i-\tau_j)\,.
\ea  
The numerators (corresponding $I_{\mu_1\cdots \mu_N}$) are theory specific and depend on the details of the interaction vertices in QED (couplings, spin structures, derivatives, etc.). The denominator (expressed here as the Schwinger proper time integral over the variable $\varepsilon_0$) is universal in that it follows directly from the free part of the QED Lagrangian, containing field propagators already integrated over fermion loop momenta. 

We will employ shorthand notations for compact definitions of derivatives and double derivatives of the worldline Green functions:
\ba 
\dot{G}^B_{ij}\equiv dG^B_{ij}/d\tau_i\,,\,\,\,\,\,\ddot{G}^B_{ij}\equiv d^2G^B_{ij}/d\tau_id\tau_j\,.
\ea
These satisfy
\ba 
\label{eq:g_b_ij_g_f_ij_derivatives}
\dot{G}^B_{ij}= \text{sign} (\tau_i-\tau_j) -2(\tau_i-\tau_j)=-\dot{G}^B_{ji}\,, \,\,\,\,\ddot{G}^B_{ij}=2-2\delta(\tau_i-\tau_j) = \ddot{G}^B_{ji}\,,\,\,\,\,\, G^F_{ij}=-G^F_{ji}\,.
\ea 
Eq.~\ref{eq:pi_mu1_muN_solution} is an exact result in QED and encodes the (gauge invariant) set of $(N-1)!/2$ Feynman diagrams\footnote{The $N!$ counts the total number of permutations, the factor $1/N$ represents the $N$ configurations related by cyclic rotations $\tau_i\to\tau_{i+1}$, and the factor $1/2$ accounts for the two identical configurations related by reversal of the loop orientation $\tau_i\to 1-\tau_i$. Its generalization to multi-loop diagrams is straightforward \cite{Feal:2022iyn}.} (for $N>2$) required to construct the $N$-th rank vacuum polarization tensor, with each Feynman diagram corresponding to one particular permutation of the photon lines.  Figure~\eqref{fig:figure_1} provides a schematic representation of the polarization tensor and illustrates the role of the boson and fermion worldline Green functions for the $N = 2$, $N = 4$, and general $N$ cases. We recall that odd-$N$ tensors vanish by virtue of Furry's theorem; a detailed discussion can be found in Appendix A of \cite{Feal:2022ufw}. The main purpose of the rest of this section will be to show that in Eq.~\eqref{eq:pi_mu1_muN} one only needs to evaluate the $(N_a,N_b,N_c)=(0,N,0)$ terms in the tensor, as the rest of terms in  Eq.~\eqref{eq:I_mu1_muN} follow from current conservation. 
\subsection[N=2: QED second-rank vacuum polarization tensor.]{$N=2$: QED second-rank vacuum polarization tensor.}
\label{subsec:N2}
As a simple example illustrating the usage of Eq.~\eqref{eq:pi_mu1_muN}, we will consider the second-rank QED polarization tensor with $N = 2$ photons in worldline diagram~(a) of Figure~\eqref{fig:figure_1}.
A direct evaluation of Eq.~\eqref{eq:I_mu1_muN} leads to three contributions $(N_a, N_b, N_c)$ such that the total number of $A$, $B$, and $C$ factors satisfies $2N_a + N_b + 2N_c = 2$, namely $(1, 0, 0)$, $(0, 2, 0)$, and $(0, 0, 1)$. Thus, from 
Eq.~\eqref{eq:I_mu1_muN} one gets\footnote{To avoid clutter, we leave implicit here the indices and arguments in the r.h.s. of the equation. These are made explicit when substituting the definitions in Eqs.~\eqref{eq:ABCD_def}.}\begin{align}
I_{\mu_1\mu_2}(k_1,k_2) = \epsilon_{i_1j_1}\epsilon_{i_1j_1} A_{i_1 j_1} + \frac{1}{2} \epsilon_{m_1m_2}\epsilon_{l_1 l_2} B_{m_1 l_1}B_{m_2 l_2} - \epsilon_{p_1q_1}\epsilon_{r_1s_1} C_{p_1 q_1} D_{r_1 s_1}\,,
\end{align}
which yields
\ba 
I_{\mu_1\mu_2}(k_1,k_2)= A_{12}+A_{21}+(B_{11}B_{22}-B_{12}B_{21})-(C_{12}-C_{21})(D_{12}-D_{21})\,,\label{eq:I_mu1_mu2}
\ea 
Substituting Eqs.~\eqref{eq:ABCD_def} then leads to
\ba 
I_{\mu_1\mu_2}(k_1,k_2)= \varepsilon_0 \delta_{\mu_1\mu_2}\ddot{G}^B_{12}+
\varepsilon_0^2(k_2)_{\mu_1}(k_1)_{\mu_2}\dot{G}^B_{12}\dot{G}^B_{21}-\varepsilon_0^2\Big[\delta_{\mu_1\mu_2} k_1\wc k_2-(k_2)_{\mu_1}(k_1)_{\mu_2}\Big]\Big[G^F_{12}\Big]^2\,.
\ea 
It is possible now to obtain a Lorentz-invariant structure of this tensor which is transverse to the external photon momenta; this is achieved  by integrating by parts the double derivatives of the bosonic worldline Green function. Indeed, noting $\dot{G}^B_{12} = -\dot{G}^B_{21}$ from Eq.~\eqref{eq:g_b_ij_g_f_ij_derivatives} and using $k_2 = -k_1$ one gets
\ba 
I_{\mu_1\mu_2}(k_1,k_2)=  \varepsilon_0^2 \Big\{\Big[{G}^F_{12}\Big]^2-\Big[\dot{G}^B_{12}\Big]^2\Big\} \Big\{k_1^2\delta_{\mu_1\mu_2}-(k_1)_{\mu_1}(k_1)_{\mu_2}\Big\}\,.\label{eq:I_mu1_mu2_intbyparts}
\ea 
Upon replacement of Eqs.~\eqref{eq:g_b_ij_g_f_ij} and \eqref{eq:g_b_ij_g_f_ij_derivatives}, this leads to the standard Feynman diagram result for the $N=2$ tensor--see \cite{Feal:2022iyn}. 

Our goal here is to show that one can obtain the result of the previous integration by parts procedure by only considering  so-called \textit{head} form factors, following the convention of Karplus and Neuman \cite{Karplus:1950zza}.  For $N=2$, these are the $B$ terms in Eq.~\eqref{eq:I_mu1_mu2}, which were defined in Eq.~\eqref{eq:ABCD_def}. The procedure we will propose here is desirable at large orders in perturbation theory, as our formula in Eq.~\eqref{eq:pi_mu1_muN_solution} is universal for general $N$. It allows for the construction of the $N$-photon or $N$-gluon one-loop amplitude (the latter with the appropriate color decomposition) without requiring any set of rules for the integration-by-parts procedure, which becomes increasingly complex at large $N$. Indeed, following Karplus and Neuman, one starts by noting that the tensor Eq.~\eqref{eq:pi_mu1_muN} is orthogonal to all photon momenta. In the worldline formalism, this is naturally enforced due to current conservation of worldline currents in Eq.~\eqref{eq:j_current_def}. 

Specifically, upon contracting the $N$th rank tensor in Eq.~\eqref{eq:pi_mu1_muN} with any external photon momentum, one can factor out overall terms of the form
\ba 
k_\mu \tilde{J}_\mu(k) = g\int^1_0 d\tau \Big[k_\mu \dot{x}_\mu(\tau) +i\varepsilon_0k_\mu \psi_\mu(\tau) k_\nu \psi_\nu(\tau)\Big] e^{-ik\wc x(\tau)}=0\,.
\ea 
The first contribution in the bracket cancels because it can be expressed as a total derivative with respect to the worldline parameter $\tau$, and the commuting worldline $x_\mu(\tau)$ traverses a closed loop trajectory with periodic boundary conditions $x_\mu(1)=x_\mu(0)$. The second  contribution cancels because $k_\mu k_\nu$ is symmetric whereas $\psi_\mu(\tau)\psi_\nu(\tau)$ antisymmetric. The Quantum Field Theory (QFT) counterpart of the previous Ward identity is of course well known. It follows from the integration by parts of matrix elements or expectation values of time ordered products of fermion currents. The previous observation leads to the two expected Ward identities 
\ba 
(k_1)_{\mu_1}\Pi_{\mu_1\mu_2}(k_1,k_2)=0\,,\,\,\,\,\,\, (k_2)_{\mu_2}\Pi_{\mu_1\mu_2}(k_1,k_2)=0\,.
\ea 
Now, either from Eqs.~\eqref{eq:pi_mu1_muN} and \eqref{eq:I_mu1_muN} --- or from a direct tensor analysis --- the $2^\text{nd}$-rank tensor has the general form
\ba 
\Pi_{\mu_1\mu_2}(k_1,k_2)&=(2\pi)^d\delta(k_1+k_2)\Big\{F(k_1,k_2) \delta_{\mu_1\mu_2}+\sum_{\substack{i_1,i_2=1,2}}(k_{i_1})_{\mu_1}(k_{i_2})_{\mu_2}H_{i_1i_2}(k_1,k_2)\Big\}\nonumber\\
&=(2\pi)^d\delta(k_1+k_2)\Big\{F(k_1,k_2) \delta_{\mu_1\mu_2}+(k_2)_{\mu_1}(k_1)_{\mu_2}H_{21}(k_1,k_2)\Big\}\label{eq:pi_mu1_mu2_generalform}
\ea 
The so-called \textit{tail} form factor $F(k_1,k_2)$ (in the Karplus--Neuman convention) can be constructed explicitly from the $N_a=1$ and $N_c=1$ terms in Eq.~\eqref{eq:I_mu1_muN} with $N=2$; these are the contributions containing the $A$, $C$ and $D$ functions in Eq.~\eqref{eq:I_mu1_mu2}. The \textit{head} form factor $H_{21}(k_1,k_2)$ arises solely from the $N_b=2$ terms, which are the contributions containing the $B$ functions  in Eq.~\eqref{eq:I_mu1_mu2}. Note that the \textit{head} form factors corresponding to $i_1=1$ and $i_2=2$ in Eq.~\eqref{eq:pi_mu1_mu2_generalform} vanish identically by the definition of $B$ in Eqs.~\eqref{eq:ABCD_def}. Instead of constructing the form factors explicitly, we project the tensor in the direction $(k_1)_{\mu_1}$ and observe that from $(k_1)_{\mu_1}\Pi_{\mu_1\mu_2}(k_1,k_2)=0$, one immediately gets $F(k_1,k_2)=-(k_1\wc k_2)H_{21}(k_1,k_2)$. Hence
\ba 
\Pi_{\mu_1\mu_2}(k_1,k_2)=(2\pi)^d\delta(k_1+k_2) \Big\{(k_1)^2\delta_{\mu_1\mu_2}-(k_1)_{\mu_1}(k_1)_{\mu_2}\Big\} H_{21}(k_1,k_2)\,,
\label{eq:ts2-gc}
\ea 
where we used $k_1=-k_2$. In this way, we avoid having to evaluate the  \textit{tail} form factor. 
From the $N_b=2$ terms in Eq.~\eqref{eq:I_mu1_muN}, one obtains for the \textit{head} form factor 
\ba 
H_{21}(k_1,k_2)= -2\frac{(g\mu^{2-d/2})^2}{(4\pi)^{d/2}} \Gamma(2-d/2) \int^1_0 d\tau_1d\tau_2 \Big[\big(G^F_{12}\big)^2-\big(\dot{G}^B_{12}\big)^2\Big]\Big[m^2+k_1^2G^B_{12}\Big]^{d/2-2}\,.
\ea
As promised, it is now easy to see that the procedure outlined above (for the decomposition of a tensor manifestly 
transverse to the external photon momenta) is equivalent to a direct evaluation of Eq.~\eqref{eq:pi_mu1_muN_solution} for $N=2$ by integration by parts in $\tau_1$ or $\tau_2$, as shown explicitly in Eq.~\eqref{eq:I_mu1_mu2_intbyparts}. Of course, in both cases one should recover the standard form for the $2^\text{nd}$-rank tensor. In particular, upon substitution of Eqs.~\eqref{eq:g_b_ij_g_f_ij} and \eqref{eq:g_b_ij_g_f_ij_derivatives}, 
\ba 
H_{21}(k_1,k_2)= -8\frac{(g\mu^{2-d/2})^2}{(4\pi)^{d/2}} \Gamma(2-d/2) \int^1_0 d\tau_1d\tau_2 \frac{|\tau_1-\tau_2|(1-|\tau_1-\tau_2|)}{\big[m^2+k_1^2|\tau_1-\tau_2|(1-|\tau_1-\tau_2|)\big]^{2-d/2}}
\ea 
However, as we will see in the next section, the procedure described here, relying only on the knowledge of the head form factors, admits a closed universal expression for arbitrary $N$, and therefore offers a much more efficient route to computing the polarization tensor for large $N$.

Another important aspect of the proposed procedure is that it allows one  to efficiently reconstruct\footnote{See also \cite{Karplus:1950zz,Karplus:1950zza} and \cite{Costantini:1971cj}.} gauge invariant  operator structures  
when the off-shell tensor is used as part of a larger amplitude, such as Feynman diagrams in the presence of background or dynamical gauge fields. Indeed, in such problems, the polarization tensor in Eq.~\eqref{eq:pi_mu1_muN} is 
projected onto arbitrary products of gauge fields, 
\begin{align}
&\Pi_{\mu_1\ldots \mu_N}(k_1,\ldots,k_N)A_{\mu_1}(k_1)\dots A_{\mu_N}(k_N) = - \big\langle i\tilde{J}_{\mu_1}(k_1)\cdots i\tilde{J}_{\mu_N}(k_N)\big\rangle A_{\mu_1}(k_1)\dots A_{\mu_N}(k_N)\,. \label{eq:pi_mu1_muN-A}
\end{align}
Typically, for on-shell amplitudes and cross-sections, it is necessary to present the final result of computing the correlator of the worldline currents in Eq. (\ref{eq:pi_mu1_muN-A}) in terms of combinations of gauge invariant operators of the gauge fields, for example products of field strengths $F_{\mu\nu}$. 
Obtaining such gauge invariant structures directly from off-shell amplitude computations is nontrivial, since gauge invariance is not manifest at the level of single Feynman diagrams. The result of the procedure discussed here, however, yields gauge invariant results which appear naturally by construction of the tensor as a product of conserved charged electromagnetic worldline currents in Eq.~\eqref{eq:pi_mu1_muN}.

For the simple example of interest here, we get directly from Eq. (\ref{eq:ts2-gc}),
\ba 
&\Pi_{\mu_1\mu_2}(k_1,k_2)A_{\mu_1}(k_1)A_{\mu_2}(k_2)= (2\pi)^d\delta(k_1+k_2) \Big\{ \partial_\mu A_{\nu}(k_1) \partial_\mu A_{\nu}(k_2) -\partial_\mu A_{\nu}(k_1) \partial_\nu A_{\mu}(k_2) \Big\} H_{21}(k_1,k_2)
\nonumber\\
&= (2\pi)^d\delta(k_1+k_2) \frac{1}{2}F_{\mu\nu}(k_1)F_{\mu\nu}(k_2) H_{21}(k_1,k_2)\,.\label{eq:ts2-gc-bf}
\ea
We will show that the $N$-th rank polarization tensors can in general be written in terms of gauge invariant form factors, which will allow us to project out gauge invariant combinations of background fields, providing an elegant solution to this long-standing problem.

\subsection[N=4: QED fourth-rank vacuum polarization tensor.]{$N=4$: QED fourth-rank vacuum polarization tensor.}
\label{subsec:N4}
The   $4$-th rank ($N=4$) polarization tensor is a key quantity in  evaluating the light-by-light scattering amplitude and cross-section. We will adopt the compact notation of \cite{Karplus:1950zz} where the $i$-th photon momentum symbol is replaced by its label $i$ when there is no ambiguity. In other words, we will replace the momentum symbols $k_1,k_2,k_3$ and $k_4$ of photons by $1$, $2$, $3$, and $4$, respectively. We will also replace the $\mu_i$-th Lorentz index by $i$; subindices $\mu_1\mu_2\mu_3\mu_4$ become $1234$. With this compact notation, the $4$-th rank tensor can be rewritten as 
\ba 
\Pi_{\mu_1\mu_2\mu_3\mu_4}(k_1,k_2,k_3,k_4)\rightarrow \Pi_{1234}(1234)\,,
\ea 
where we also got rid of the commas separating the momentum labels in the arguments of the function. The direct substitution of Eqs.~\eqref{eq:ABCD_def} into Eq.~\eqref{eq:I_mu1_muN} leads to the following Lorentz invariant form for the $4^\text{th}$-rank QED polarization tensor in terms of the allowed tensorial structures\footnote{As in the $N=2$ case addressed in the previous section, the various restrictions on the summations over $i_1,i_2,i_3,i_4$ in Eq.~\eqref{eq:Pi_mu1_mu2_mu3_mu4_result} are obtained directly in our approach because the contributions $i_1=1$, $i_2=2$, $i_3=3$ and $i_4=4$ vanish by construction of Eqs.~\eqref{eq:ABCD_def}. In contrast, in the conventional perturbative construction, to maintain such a symmetrical form for the tensor, one must sum over Feynman diagrams, impose momentum conservation, and choose three independent momentum variables for each term in the tensor. Further, the latter must be done such that it becomes independent of the momentum index they carry.}:
\ba 
\label{eq:Pi_mu1_mu2_mu3_mu4_result}
&\Pi_{1234}(1234) = (2\pi)^d\delta^d(k_1+k_2+k_3+k_4) \\
&\times \bigg\{ \delta_{12}\delta_{34} F_1(1234)+\delta_{13}\delta_{24}F_2(1234)+ \delta_{14}\delta_{23}F_3(1234)\nonumber\\
&+\sum_{\substack{i_3=1,2,4\\i_4=1,2,3}}\delta_{12} ({i_3})_{3} ({i_4})_{4} G_{1,i_3i_4}(1234)+\sum_{\substack{i_2=1,3,4\\i_4=1,2,3}}\delta_{13} ({i_2})_{2} ({i_4})_{4} G_{2,i_2i_4}(1234)+ \sum_{\substack{i_2=1,3,4\\i_3=1,2,4}}\delta_{14}({i_2})_{2}({i_3})_{3}G_{3,i_2i_3}(1234)\nonumber\\
&+\sum_{\substack{i_1=2,3,4\\i_4=1,2,3}}\delta_{23}({i_1})_{1}({i_4})_{4}G_{4,i_1i_4}(1234)+\sum_{\substack{i_1=2,3,4\\i_3=1,2,4}}\delta_{24}({i_1})_{1}({i_3})_{3}G_{5,i_1i_3}(1234)+\sum_{\substack{i_1=2,3,4\\i_2=1,3,4}}\delta_{34}({i_1})_{1}({i_2})_{2}G_{6,i_1i_2}(1234) \nonumber\\
&+\sum_{\substack{i_1=2,3,4\\i_2=1,3,4}} \sum_{\substack{i_3=1,2,4\\i_4=1,2,3}} ({i_1})_{1}({i_2})_{2}({i_3})_{3}({i_4})_{4} H_{i_1i_2i_3i_4}(1234) \bigg\}\,.\nonumber
\ea 
Here, as shown explicitly in Appendix \ref{app:rank4_nosymmetryreduction}, the $F$, $G$, and $H$ ``form factors" are linear combinations of the $A$, $B$, $C$ and $D$ expressions in terms of worldline Green functions that we introduced in Eq.~\eqref{eq:ABCD_def}.
In the nomenclature of Karplus and Neuman~\cite{Karplus:1950zza,Karplus:1950zz}, the three $F(1234)$ Lorentz invariant form factors in Eq.~\eqref{eq:Pi_mu1_mu2_mu3_mu4_result} are \textit{tails}, the $6*3^2$($=54$) $G_{ij}(1234)$ terms are \textit{shoulders}, and the $3^4$ ($=81$) $H_{ijkl}(1234)$ terms, the \textit{heads}. Pseudovector (antisymmetric tensor) contributions lead to terms that can either be expressed as combinations of the above tensor structures, or yield pseudoscalar quantities when contracted with the photon polarization vectors. Imposing current conservation and permutation symmetry further constrains the decomposition of the tensor into its independent tensor structures. In particular, because of current conservation, it suffices to know only the \textit{heads} (the 81  $H_{ijkl}$ invariants for $N=4$ photons) to fully determine any polarization tensor. One can obtain expressions for the \textit{tails} and \textit{shoulders} in terms of the \textit{heads} and thereby greatly simplify the calculation of the tensor. 

This point is extremely relevant to the general worldline formulation of the QED $N$-th rank polarization tensors because it is straightforward, from Eqs.~\eqref{eq:pi_mu1_muN} and Eq.~\eqref{eq:I_mu1_muN}, to extract an identity for the \textit{head} form factors which has a universal structure in terms of worldline Green functions for arbitrary $N$.
To obtain this identity, we simply note that the \textit{head} factors accompany the tensor forms $(i_1)_1(i_2)_2\cdots (i_N)_N$ in Eq.~\eqref{eq:Pi_mu1_mu2_mu3_mu4_result}, and therefore can only arise from the $(N_a, N_b, N_c) = (0, N, 0)$ term for arbitrary $N$ in the sum of Eq.~\eqref{eq:I_mu1_muN}, namely
\ba 
I^{(0,N,0)}_{\mu_1\cdots \mu_N}(12\cdots N)& =  \frac{1}{N!} \epsilon_{m_1\ldots m_{N}}\epsilon_{l_1\ldots l_{N}}B_{m_1 l_1}B_{m_2l_2}\cdots B_{m_Nl_N}=\det B\,,
\ea
so that they are given in closed form by
\ba 
&H_{i_1i_2\ldots i_N}(12\cdots N) \nonumber\\
&= -2\frac{\big(g\mu^{2-d/2}\big)^N}{(4\pi)^{d/2}}\Gamma(N-d/2)\int_0^1 d\tau_1d\tau_2 \cdots d\tau_N \big[D(12\cdots N;\tau_1,\ldots,\tau_N)\big]^{N-d/2} P_{i_1i_2\cdots i_N}(\tau_1,\tau_2,\ldots,\tau_N)\,.\label{eq:H_i1_iN}
\ea 
Here we carried out the trivial integration over propertime parameter $\varepsilon_0$ to obtain the standard quadratic form of the worldline  denominator as a function of the available kinematic invariants (constructed out of the external photon momenta), which takes the well-known form 
\ba 
D(12\cdots N;\tau_1,\ldots,\tau_N) =m^2-\frac{1}{2}\sum_{i,j=1}^N G_B(\tau_i,\tau_j)k_i\cdot k_j\,.
\label{eq:d_n_def}
\ea 
The numerator is a polynomial constructed solely from worldline Green functions. It admits the following universal representation for a general number of photons $N$,
\ba 
&P_{i_1i_2\ldots i_N}(\tau_1,\tau_2,\ldots,\tau_N) \nonumber\\
&=\sum_{j_1,j_2,\ldots j_N=1}^N \epsilon_{j_1j_2\cdots j_N} \Big[\delta_{1j_1}\dot{G}^B_{1i_1}+\delta_{j_1i_1}G^F_{i_11}\Big]\Big[\delta_{2j_2}\dot{G}^B_{2i_2}+\delta_{j_2i_2}G^F_{i_22}\Big]\cdots \Big[\delta_{Nj_N}\dot{G}^B_{Ni_N}+\delta_{j_Ni_N}G^F_{i_NN}\Big]\label{eq:P_i1_iN_worldline}\,.
\ea 
This expression is a Levi-Civita weighted sum over bosonic and fermionic Green functions (the dot represents the propertime derivative) automatically implementing the $(\mathcal{N}=1)$ SUSY algebra of bosonic and fermionic worldline contractions in the tensor. Because $\epsilon_{j_1\ldots j_N}$ is nonzero only when $(j_1,\ldots,j_N)$ is a permutation of $(1,\ldots,N)$, we can also write this expression in terms of permutations $\sigma\in S_N$, where $S_N$ denotes the permutation group. Letting $j_k=\sigma(k)$ and noting $\epsilon_{j_1\ldots j_N}=\text{sgn}(\sigma)$ in particular,
\ba 
P_{i_1i_2\ldots i_N}(\tau_1,\tau_2,\ldots,\tau_N)=\sum_{\sigma\in S_N} \text{sgn}(\sigma)\prod_{k=1}^N\Big\{\delta_{k,\sigma(k)}\dot{G}^B_{k,i_k}+\delta_{\sigma(k),i_k}G^F_{i_k,k}\Big\}\,.\label{eq:P_i1_iN_worldline_alternativeform}
\ea 

Eqs.~\eqref{eq:H_i1_iN} and~\eqref{eq:P_i1_iN_worldline}-\eqref{eq:P_i1_iN_worldline_alternativeform} constitute  the central result of this paper. They allow one to explicitly construct and identify, 
using the symmetries of the problem, the unique \textit{head} form factors completely specifying the $N$-th rank vacuum polarization tensor in QED. As we will demonstrate shortly, the novel procedure outlined here bypasses the factorial growth of form factors $\sim \exp(N\log N)\sim  N!$, as well as the $N!$ Feynman diagrams in each form factor, at arbitrary $N$-th order in perturbation theory.

Returning to  $N=4$, and following the previously outlined procedure for the $N=2$ case, we write the four Ward identities for the tensor,
\ba 
 (1)_1 \Pi_{1234}(1234)= 0\,,\label{eq:ward_1}\\
 (2)_2 \Pi_{1234}(1234)=0\,,\label{eq:ward_2}\\
 (3)_3 \Pi_{1234}(1234)=0\,,\label{eq:ward_3}\\
 (4)_4 \Pi_{1234}(1234)=0\,.\label{eq:ward_4}
\ea
These relations will enable the three tails $F$ and the 54 shoulders $G_{ij}$ to be expressed in terms of the 81 heads $H_{ijkl}$. We will only outline here this tensor decomposition and reduction procedure for the first of the required terms; we will elaborate on it further and present the final results in Appendix \ref{app:transversedecomposition}. 

Starting from the  Ward identity in Eq. \eqref{eq:ward_1}, and substituting then Eq.\eqref{eq:Pi_mu1_mu2_mu3_mu4_result} into Eq.~\eqref{eq:ward_1}, the reduced tensor involves only the components $\mu_2$, $\mu_3$, and $\mu_4$. For this tensor to be zero for any value of $k_i$, $i=1,2,3,4$, the factor multiplying each tensor structure has to vanish. In particular, collecting the factors multiplying the structure containing one metric tensor,  plugging Eq.~\eqref{eq:Pi_mu1_mu2_mu3_mu4_result} into Eq.~\eqref{eq:ward_1}, one obtains the following identities:
\ba
(1)_2\delta_{34}\times\bigg[F_1(1234)+\sum_{i_1\neq 1} (1\wc i_1)G_{6,i_11}(1234)\bigg]=0\,,\label{eq:F_1_ward1}\\ (3)_2\delta_{34}\times\bigg[\sum_{i_1\neq 1}(1\wc i_1)G_{6,i_13}(1234)\bigg]=0\,,\label{eq:F_1_ward1b}\\
(4)_2\delta_{34}\times\bigg[\sum_{i_1\neq 1} (1\wc i_1)G_{6,i_14}(1234)\bigg]=0\,.\label{eq:F_1_ward1c}\\
(1)_3\delta_{24}\times \bigg[F_2(1234)+\sum_{i_1\neq 1} (1\wc i_1)G_{5,i_11}(1234)\bigg]=0\,,\label{eq:F_2_ward1}\\
(2)_3\delta_{24}\times\bigg[\sum_{i_1\neq 1} (1\wc i_1) G_{5,i_12}(1234)\bigg]=0\,,\label{eq:F_2_ward1b}\\ (4)_3\delta_{24}\times\bigg[\sum_{i\neq 1} (1\wc i_1) G_{5,i_14}(1234)\bigg]=0\,.\label{eq:F_2_ward1c}\\
(1)_4\delta_{23}\times\bigg[F_3(1234)+\sum_{i\neq 1} (1\wc i_1) G_{4,i_11}(1234)\bigg]=0\,,\label{eq:F_3_ward1}\\ (2)_4\delta_{23}\times\bigg[\sum_{i\neq 1} (1\wc i_1)G_{4,i_12}(1234)\bigg]=0\,,\label{eq:F_3_ward1b}\\ (3)_4\delta_{23}\times\bigg[ \sum_{i\neq 1} (1\wc i_1)G_{4,i_13}(1234)\bigg]=0\,.\label{eq:F_3_ward1c}
\ea 
Eqs.~\eqref{eq:F_1_ward1}, \eqref{eq:F_2_ward1}, and \eqref{eq:F_3_ward1} express the tails \(F_1(1234)\), \(F_2(1234)\), and \(F_3(1234)\) as linear combinations of the shoulders \(G_{6,ij}(1234)\), \(G_{5,ij}(1234)\), and \(G_{4,ij}(1234)\). We note that additional but equivalent expressions are obtained from the use of any of the other three Ward identities in Eqs.~\eqref{eq:ward_2}-\eqref{eq:ward_4}. Further identities satisfied among the shoulders are also obtained in this way from Eqs.\eqref{eq:F_1_ward1b}-\eqref{eq:F_1_ward1c}, Eqs.~\eqref{eq:F_2_ward1b}-\eqref{eq:F_2_ward1c} and Eqs.~\eqref{eq:F_3_ward1b}-\eqref{eq:F_3_ward1c}. These can be used to simplify the resulting expressions when needed. 

Likewise, collecting terms without metric tensors yields the identities:
\ba 
(1)_2(1)_3(1)_4\times\bigg[G_{1,11}(1234)+G_{2,11}(1234)+G_{3,11}(1234)+\sum_{i_1\neq 1} (1\wc i_1) H_{i_1111}(1234)\bigg]&=0\label{eq:G_ward1_1}\,,\\
(1)_2(i_3)_3(1)_4\times\bigg[G_{1,i_31}(1234)+G_{3,1i_3}(1234)+\sum_{i_1\neq 1} (1\wc i_1) H_{i_11i_31}(1234)\bigg]&=0\,,\,\,\, i_3=2,4\,,\label{eq:G_ward1_2}\\
(1)_2(1)_3(i_4)_4\times\bigg[G_{1,1i_4}(1234)+G_{2,1i_4}(1234)+\sum_{i_1\neq 1} (1\wc i_1)H_{i_111i_4}(1234)\bigg]&=0\,,\,\,\, i_4=2,3\,,\label{eq:G_ward1_3}\\
(1)_2(i_3)_3(i_4)_4\times\bigg[G_{1,i_3i_4}(1234)+\sum_{i_1\neq 1}(1\wc i_1) H_{i_11i_3i_4}(1234)\bigg]&=0\,,\,\,\, i_3=2,4\,,\,\,\, i_4=2,3\,.\label{eq:G_ward1_4}
\ea 
These relations now allow us to express some of the shoulders in terms of the heads, and therefore some of the tails in terms of the heads. By collecting the rest of terms in the tensor coming from the contraction with \(k_1\) one finds additional relations generated by Eq.~\eqref{eq:ward_1}. We streamline this transverse decomposition and collect all such identities using Eqs.~\eqref{eq:ward_1}---\eqref{eq:ward_4} to solve for all tails and shoulders in terms of the heads in Appendix \ref{app:transversedecomposition}. This procedure allows us, as a first step, to reduce the calculation of $138$ form factors down to the evaluation of the $3^4=81$ heads. 

As noted by Karplus and Neuman \cite{Karplus:1950zz}, because the worldline currents in Eq.\eqref{eq:pi_mu1_muN} commute (essentially because photons are Bose particles), the tensor in Eq. \eqref{eq:Pi_mu1_mu2_mu3_mu4_result} is invariant under the 24 simultaneous permutations of momenta $k_i, k_j$ and indices $\mu_i, \mu_j$. This invariance provides identities amongst \textit{head} form factors reducing the calculation further, from 81 down to only 6. In conventional QFT, the same symmetry is realized only by including all  three $(4-1)!/2$ Feynman diagrams, already encoded in the master formula of Eq.~\eqref{eq:Pi_mu1_mu2_mu3_mu4_result}. In other words, a given Feynman diagram (a time ordered product of fermion currents after Wick expansion of the action) never satisfies this Bose symmetry on its own.

Note also that these symmetries are general. They remain valid for the $4$-th rank tensor at arbitrary loop order in perturbation theory.
In order to find relations of these type, let us consider for instance the form factor $H_{4443}(4321)$.  By exchanging the third momentum $k_2$ with the fourth momentum $k_1$, we obtain the identity\footnote{Note that the specific labeling of the arguments as $(4321)=(k_4,k_3,k_2,k_1)$ is irrelevant; they could equally well be denoted $(a_1 a_2 a_3 a_4)$. The subscripts $4443$ here refer to the fourth and third momentum arguments, namely $k_1$ and $k_2$, in this case. $H_{4443}(4321)$ multiplies the tensor structure $(1)_1(1)_2(1)_3(2)_4=(k_1)_{\mu_1}(k_1)_{\mu_2}(k_1)_{\mu_3}(k_2)_{\mu_4}$.}
\ba 
H_{4443}(4321)=H_{3343}(4312)\label{eq:h4443_h3343}
\ea 
This identity can be extracted directly from the worldline master formula in Eq.~\eqref{eq:H_i1_iN} as follows. The polynomial numerators of the two heads $H_{4443}(k_4,k_3,k_2,k_1)$ and $H_{3343}(k_4,k_3,k_1,k_2)$ are given by Eq.~\eqref{eq:P_i1_iN_worldline} and take the explicit form
\ba
P_{4443}(\tau_1,\tau_2,\tau_3,\tau_4)
=
\dot G^B_{14}\,\dot G^B_{24}\,
\Big[(G^F_{34})^2-(\dot G^B_{34})^2\Big],
\nonumber\\
P_{3343}(\tau_1,\tau_2,\tau_3,\tau_4)
=
\dot G^B_{13}\,\dot G^B_{23}\,
\Big[(G^F_{34})^2-(\dot G^B_{34})^2\Big] \, .
\ea
After exchanging the third and fourth momentum arguments in the first head $H_{4443}(k_4,k_3,k_2,k_1)$, its denominator given by Eq.~\eqref{eq:d_n_def} changes, since the momentum invariants accompanying the $G^B(\tau_i,\tau_j)$ functions when $\tau_{i,j}=\tau_{3,4}$ are now changed. One may then relabel $\tau_3 \leftrightarrow \tau_4$ to restore the denominator to its original form. Under the exchange $\tau_3 \leftrightarrow \tau_4$, the factors $(G^F_{34})^2$ and $(\dot G^B_{34})^2$ in the numerator $P_{4443}$ remain invariant, while the product $\dot G^B_{14}\dot G^B_{24}$ transforms into $\dot G^B_{13}\dot G^B_{23}$, thereby reproducing the polynomial numerator $P_{3343}$, leading thus to the relation in Eq.~\eqref{eq:h4443_h3343}.

This permutation symmetry satisfied by the tensor head form factors can be formally defined and generalized for arbitrary $N$ as follows. Let $H_{i_1\ldots i_N}(a_1\ldots a_N)$ be a head in the $N$-th rank tensor, where we recall, by construction, each index $i_k$ refers to the (momentum) argument in the $k$-th position, as seen in Eq.~\eqref{eq:Pi_mu1_mu2_mu3_mu4_result}. Let also $\sigma=(pq)$ be a two element transposition exchanging $p$ and $q$ in the permutation group $S_N$. Then $\sigma(p)=q$, $\sigma(q)=p$ and $\sigma(k)=k$ for $k\neq p,q$ (all other elements unchanged). Current conservation implies
\ba 
\label{eq:transposition_def}
H_{i_1i_2\ldots i_N}(a_1a_2\ldots a_N)= H_{\sigma(i_{\sigma^{-1}(i_1)})\sigma(i_{\sigma^{-1}(i_2)})\ldots\sigma(i_{\sigma^{-1}(i_N)})}(a_{\sigma(i_1)}a_{\sigma(i_2)}\ldots a_{\sigma(i_N)})\,.
\ea 
Here $\sigma(k)$ specifies which original argument moves into position $k$ after the exchange. Similarly, $\sigma^{-1}(k)$ identifies where the $k$-th position came from, so $i_{\sigma^{-1}(k)}$ is the index that originally sat in that position. Finally, $\sigma(i_{\sigma^{-1}(k)})$ mirrors the same transposition in the indices. To apply this to the example we just considered, the transposition of the third and fourth momentum arguments $\sigma=(1)(2)(34)$ corresponds to the map 
\[
\sigma(1)=1,\qquad \sigma(2)=2,\qquad \sigma(3)=4,\qquad \sigma(4)=3 .
\]
Accordingly, its inverse satisfies
$
\sigma^{-1}(1)=1, \sigma^{-1}(2)=2, \sigma^{-1}(3)=4, \sigma^{-1}(4)=3.
$ Substituting this permutation into Eq.~\eqref{eq:transposition_def}, then the group element action on a head element defines the equivalence, in this case, to be 
\[
H_{i_1 i_2 i_3 i_4}(a_1 a_2 a_3 a_4)
=
H_{\sigma(i_1)\sigma(i_2)\sigma(i_4)\sigma(i_3)}(a_1 a_2 a_4 a_3).
\]
For the example under consideration, one immediately finds the previous identification
$H_{4443}(4321) = H_{3343}(4312)$.

The usefulness of the general expression in Eq.~\eqref{eq:transposition_def}, formulated in terms of permutation group symmetries for arbitrary $N$, will become much clearer in the next section when we address the computation of higher rank tensors.  We will apply there the Burnside-Cauchy–Frobenius lemma to count the number of equivalence classes and  determine the number of independent form factors that fully specify the $N$-photon amplitude at arbitrary loop order in perturbation theory.

For $N=4$, starting with any arbitrary head, there are $4! = 24$ permutations of the arguments in $S_4$ that can be realized always as a composition of operations of the type in Eq.~\eqref{eq:transposition_def}. Repeated application of Eq.~\eqref{eq:transposition_def} then generates the list of 24 head form factors that are related by virtue of the permutation symmetry of the polarization tensor, and hence belong to the same equivalence class defined by this symmetry. For $N=4$, this reduces the original list of 81 head form factors needed to define the tensor down to the computation of only 6 representatives of each orbit under Eq.~\eqref{eq:transposition_def}. As a particular illustration, starting with $H_{2111}(1234)$ and applying Eq.~\eqref{eq:transposition_def} sequentially, one generates the set of relations\footnote{Since the manipulation of these identities may be unfamiliar, let's further pick at random amongst the identities in Eq.~\eqref{eq:class_1_representative} and demonstrate explicitly the equality $H_{3313}(2314)=H_{2422}(4132)$. We first observe
\begin{align}
\sigma(1) = 4\,\,; \,\, \sigma(2)=3\,\,; \,\,\sigma(3) =2 \,\,;\,\, \sigma(4)=1\,,
\end{align}
indicating that the term in the first position within the parenthesis of the head on the l.h.s., is moved to the 4th position in the parenthesis in the head on the r.h.s., and so on. Conversely, the inverse permutations (the positions where the expressions in the parenthesis on the r.h.s. came from) are 
\begin{align}
\sigma^{-1}(1) =4\,\,;\,\,\sigma^{-1}(2) =3\,\,;\,\,\sigma^{-1}(3)=2\,\,;\,\,\sigma^{-1}(4)=1\,.
\end{align}
The r.h.s of the formula in Eq.~\eqref{eq:transposition_def} then gives 
\begin{align}
H_{\sigma(i_{\sigma^{-1}(i_1)}),\sigma(i_{\sigma^{-1}(i_2)}),\sigma(i_{\sigma^{-1}(i_3)}),\sigma(i_{\sigma^{-1}(i_4)})}(4132)=H_{\sigma(i_4),\sigma(i_3),\sigma(i_2),\sigma(i_1)}(4132)= H_{\sigma(3),\sigma(1),\sigma(3),\sigma(3)}(4132)=H_{2422}(4132)\,,
\end{align}
for $i_1=i_2=i_4=3$ and $i_3=1$, which proves the identity.}
\ba 
&H_{2111}(1234)= H_{2111}(1243) = H_{3111}(1324) = H_{4111}(1342) = H_{3111}(1423)=H_{4111}(1432)\nonumber\\
=&H_{2122}(2134)=H_{2122}(2143)=H_{3313}(2314)=H_{4441}(2341)=H_{3313}(2413)=H_{4441}(2431)\nonumber\\
=&H_{2322}(3124)=H_{2422}(3142)=H_{3323}(3214)=H_{4442}(3241)=H_{4443}(3421)=H_{3343}(3412)\nonumber\\
=&H_{2322}(4123)=H_{2422}(4132)=H_{3323}(4213)=H_{4442}(4231)=H_{4443}(4321)=H_{3343}(4312)\,.\label{eq:class_1_representative}
\ea 
We next select a head form factor not present in the above set, and continue the procedure. The complete list of identities among the head form factors for $N=4$ is given in Appendix \ref{app:notational_invariance}. The explicit expression for the $4$-th rank polarization tensor is given in Appendix \ref{app:fullresult}. This follows after 
 substitution in Eq.~\eqref{eq:Pi_mu1_mu2_mu3_mu4_result} of all tail and shoulder form factors in terms of heads (by means of the Ward identities in Eqs.~\eqref{eq:sol_g1_11} to ~\eqref{eq:sol_f3}), and after substitution of the resulting heads by one of the six representatives of each class using Eqs.~\eqref{eq:class_1_representative}, and Eqs.~\eqref{eq:class_2_representative}--\eqref{eq:class_6_representative}. 

Repeated heads appear within each equivalence class generated in this way; they share the same indices but differ by permutations of the photon-momentum arguments. For example, the set in  Eq.~\eqref{eq:class_1_representative} contains only twelve independent heads, while the set in Eq.~\eqref{eq:class_6_representative} contains only three independent heads. For this reason, counting equivalence classes (independent heads) is less trivial than it may seem and must proceed either by explicit bookkeeping (as done in here and Appendix \ref{app:notational_invariance} following Karplus and Neuman \cite{Karplus:1950zza}) or by using the Burnside-Cauchy-Frobenius lemma, as we will discuss in detail in Section \ref{sec:cauchy-frobenius-burnside}. We emphasize that the central result of this work, the worldline master formula for the tensor form factors in Eq.~\eqref{eq:H_i1_iN}, automatically satisfies the aforementioned permutation group symmetries.

Our treatment here of the computation of the $4^\text{th}$-rank polarization tensor in QED is valid to any arbitrary loop order in perturbation theory and completely general. We shall focus on giving the explicit expressions to lowest order $\mathcal{O}(g^4)$ in perturbation theory using the worldline master expression. Specializing Eq.~\eqref{eq:P_i1_iN_worldline} to $N=4$ one immediately finds
\ba 
&H_{ijkl}(1234)=-2\frac{g^4\mu^{8-2d}}{(4\pi)^{d/2}} \Gamma(4-d/2)\int^1_0 d\tau_1d\tau_2d\tau_3d\tau_4 \big[D(1234;\tau_1,\tau_2,\tau_3,\tau_4)\big]^{d/2-4} P_{ijkl}(\tau_1,\tau_2,\tau_3,\tau_4)\,.\label{eq:H_def_worldline}
\ea
where the polynomial numerator is defined in terms of periodic and anti-peridic worldline Green functions as 
\ba 
P_{ijkl}(\tau_1,\ldots,\tau_4)=\sum_{m,n,r,s=1}^4\epsilon_{mnrs}\Big\{\delta_{1m}\dot{G}^B_{1i}+\delta_{mi}G^F_{i1}\Big\}\Big\{\delta_{2n}\dot{G}^B_{2j}+\delta_{nj}G^F_{j2}\Big\}\Big\{\delta_{3r}\dot{G}^B_{3k}+\delta_{rk}G^F_{k3}\Big\}\Big\{\delta_{4s}\dot{G}^B_{4l}+\delta_{sl}G^F_{l4}\Big\}\,.\label{eq:4_numerator}
\ea 
and the denominator
\ba 
D(1234;\tau_1,\ldots,\tau_4)=m^2 -k_1\wc k_2 G^B_{12}-k_1\wc k_3 G^B_{13}-k_1\wc k_4 G^B_{14} -k_2\wc k_3 G^B_{23}-k_2\wc k_4 G^B_{24}-k_3\wc k_4 G^B_{34}\,.\label{eq:4_denominator}
\ea 
It is straightforward to check at this point that Eq.~\eqref{eq:H_def_worldline} preserves the permutation symmetry due to the commutation of any two electromagnetic currents in the tensor, automatically satisfying the large set of identities we advanced in Eq.~\eqref{eq:class_1_representative}, and Eqs.~\eqref{eq:class_2_representative} to \eqref{eq:class_6_representative}  of Appendix \ref{app:notational_invariance}, from purely symmetry considerations. As we will see in a moment, while this is true for Eq.~\eqref{eq:H_def_worldline}, it is not true for the $(4-1)!/2=3$ Feynman diagrams this expression encodes: each Feynman diagram does not preserve permutation symmetry, but only their gauge invariant sum. 
Because the expression we have for the \textit{heads} is universal and valid for any $N$, the result for the vacuum polarization tensor to $\mathcal{O}(g^N)$ encoding $(N-1)!/2$ Feynman diagrams is straightforward in this approach.

It only remains to evaluate explicitly the functional form of the $6$ independent heads, $H_{2111}(1234)$, $H_{2121}(1234)$, $H_{2123}(1234)$, $H_{2311}(1234)$, $H_{2143}(1234)$ and $H_{2341}(1234)$. The other $75$ heads  are obtained from these by permutations of momentum arguments. These enter the expression only through the polynomial denonominator in Eq.~\eqref{eq:4_denominator}. The full result for the tensor is listed in Appendix \ref{app:fullresult}. The six unique polynomial numerators are easy to obtain, and given by
\ba 
P_{2111}(\tau_1,\tau_2,\tau_3,\tau_4)&=\dot{G}^B_{13}\dot{G}^B_{14}\Big[(G_{12}^{F})^2-(\dot{G}_{12}^{B})^2\Big]\,,\nonumber\\
P_{2121}(\tau_1,\tau_2,\tau_3,\tau_4)&=\dot{G}^B_{14}\dot{G}^B_{23}\Big[(G_{12}^{F})^2-(\dot{G}_{12}^{B})^2\Big]\,,\nonumber\\
P_{2123}(\tau_1,\tau_2,\tau_3,\tau_4)&=\dot{G}^B_{23}\dot{G}^B_{34}\Big[(G_{12}^{F})^2-(\dot{G}_{12}^{B})^2\Big]\,,\nonumber\\
P_{2311}(\tau_1,\tau_2,\tau_3,\tau_4)&=\dot{G}^B_{14}\Big[\dot{G}^B_{12}\dot{G}^B_{13}\dot{G}^B_{23}-G^F_{12}G^F_{13}G^F_{23}\Big]\,,\nonumber\\
P_{2143}(\tau_1,\tau_2,\tau_3,\tau_4)&= \Big[(G_{12}^{F})^2-(\dot{G}_{12}^{B})^2\Big]\Big[(G_{34}^{F})^2-(\dot{G}_{34}^{B})^2\Big]\,,\nonumber\\
P_{2341}(\tau_1,\tau_2,\tau_3,\tau_4)&=G^F_{12}G^F_{14}G^F_{23}G^F_{34}-\dot{G}^B_{12}\dot{G}^B_{14}\dot{G}^B_{23}\dot{G}^B_{34}\,.\label{eq:6numerators}
\ea 
Recall from Eqs.~\eqref{eq:g_b_ij_g_f_ij} and \eqref{eq:g_b_ij_g_f_ij_derivatives}, that the worldline Green functions are defined as
\ba 
G^B_{ij}\equiv |\tau_i-\tau_j|(1-|\tau_i-\tau_j|)\,,\,\,\,\dot{G}^B_{ij}=\text{sign}(\tau_i-\tau_j)-2(\tau_i-\tau_j)\,,\,\,\,G^F_{ij}=\text{sign}(\tau_i-\tau_j)\,.\nonumber
\ea 

The structure of these $6$ master heads contains a numerator that always expands into a small sum of monomials built from free bosonic and fermionic worldline Green functions 
(Eqs.~\eqref{eq:6numerators}) and, in the denominator, a quadratic form in the available invariants and bosonic Green functions 
(Eq.~\eqref{eq:4_denominator}). Because the integrand is invariant under global shifts 
$\tau_i' = \tau_{i}+ \delta$ due to the functional form of the worldline Green functions, one may therefore fix this invariance by choosing, for instance, $\delta = 1 - \tau_1$, 
which sets one of the parameters to $\tau_1' = 1$ and allows one then to perform the corresponding integral trivially. As a result, for general $N$ the worldline integrals depend on only $N-1$ independent $\tau$-parameters, reflecting the cyclic symmetry of the diagram.

Further, as promised, the manifest $\mathcal{N}=1$ supersymmetry of the $6$ \textit{head} form factors entering the tensor allows for a simple check of the results: Replace each $\dot{G}^B_{ij}$ function by $G^F_{ij}$ in the expresions above, and check that the amplitude yields zero. This $\mathcal{N}=1$ supersymmetry will hold for general $N$ and can be traced back to the fully symmetric form of the polynomials in Eq.~\eqref{eq:P_i1_iN_worldline}.

Eqs.~\eqref{eq:6numerators} can alternatively be obtained by a symmetric partial integration of the tail form-factor contributions, as explained in the previous section, followed by a final application of the corresponding $\mathcal{N}=1$ supersymmetry substitution rules for the spinor amplitude; see \cite{Bern:1991aq,Schubert:1997ph} for the on-shell case. The resulting expressions in terms of worldline Green functions can also be classified and interpreted in terms of cyclic permutations together with ``tail'' contributions\footnote{This ``tail'' terminology shall not be confused with the tail form factors we define in this work following the usage introduced by Karplus and Neuman \cite{Karplus:1950zza}.}. 
However, it remains unclear how the latter integration-by-parts approach (see also \cite{Ahmadiniaz:2020jgo,Ahmadiniaz:2023vrk}) can be generalized to the full reconstruction of the off-shell amplitude and tensor, and/or used to determine for instance the minimal number of head form factors required to fully specify this physical object for arbitrary $N$.

The novel approach we have outlined here has a straightforward logic and is extremely powerful, especially at high orders, and large $N$, as it allows one to easily identify the number of independent head form factors that are required to completely specify the $N$-photon amplitude, and further, to construct them through the universal form in Eqs.~\eqref{eq:H_i1_iN} and \eqref{eq:P_i1_iN_worldline}-\eqref{eq:P_i1_iN_worldline_alternativeform} valid for arbitrary $N$. These expressions automatically encode the result of this $\mathcal{N}=1$ integration-by-parts procedure and generalize it unambigously at arbitrary loop order. 

As noted previously, the  structure of the six irreducible head form factors from Eq.~\eqref{eq:H_def_worldline} is deeply tied to gauge invariant  field-strength operator structures in QED, that appear when the rank-4 tensor is projected onto arbitrary (on-shell or off-shell) gauge-field configurations, as in the context of a larger amplitude,
\ba 
\Pi_{1234}(1234)A_1(1)A_2(2)A_3(3)A_4(4)\,,\quad A_i(j)\equiv A_{\mu_i}(k_j)e^{ik_j\wc x}\,.
\ea 
For instance, projecting the part of the rank-4 tensor accompanying the head term $H_{2143}(1234)$ in the equivalence class of Eq.~\eqref{eq:class_5_poltensorterms} and substituting $(j)_i A_\ell(j) = -i\partial_i A_\ell(j)$ leads to\footnote{After projection, each term looks like 
\ba
(\partial_{\mu} A_i(\cdot))(\partial_{\nu} A_j(\cdot))(\partial_{\rho} A_l(\cdot))(\partial_{\sigma} A_m(\cdot))\,.
\ea 
Because of momentum conservation total derivatives vanish
\ba 
\partial_\alpha\big[A_i(1)A_j(2)A_l(3)A_m(4)\big]
=
i(k_1+k_2+k_3+k_4)_\alpha\,A_i(1)A_j(2)A_l(3)A_m(4)=0\,,
\ea 
and this allows each derivative to be freely reassigned between fields to reconstruct it in terms of antisymmetric and gauge invariant field-strength combinations.}
\ba 
H_{2143}(1234)\Big[ (2)_{1}(3)_{2}(4)_{3}(1)_{4}-\cdots\Big]A_1(1)A_2(2)A_3(3)A_4(4)= H_{2143}(1234)\Big[(\partial_1A_2)(\partial_2A_3)(\partial_3A_4)(\partial_4A_1)+\cdots \Big]
\ea 
where $\cdots$ in the brackets on the r.h.s indicate the remaining Lorentz tensors appearing in the corresponding term of Eq.~\eqref{eq:class_5_poltensorterms}.
The result can be cast in terms of gauge invariant combinations of field-strength tensors yielding\footnote{
Likewise, one can demonstrate that the projection of the $H_{2143}(1324)$ terms of Eq.~\eqref{eq:class_5_poltensorterms} produce
\ba 
H_{2143}(1324)\Big[(3)_{1}(4)_{2}(1)_{3}(2)_{4}-\cdots \Big]A_1(1)A_2(2)A_3(3)A_4(4)=\frac{1}{4}H_{2143}(1324)F_{13}(1)F_{31}(3)F_{24}(2)F_{42}(4)\,.
\ea 
and those corresponding to $H_{2143}(4321)$ of Eq.~\eqref{eq:class_5_poltensorterms}
\ba 
H_{2143}(4321)\Big[ &(4)_{1}(3)_{2}(2)_{3}(1)_{4}-\cdots\Big] A_1(1)A_2(2)A_3(3)A_4(4)=\frac{1}{4}H_{2143}(4321)F_{14}(1)F_{41}(4)F_{23}(2)F_{32}(3)\,.
\ea }
\ba 
H_{2143}(1234)\Big[ (2)_{1}(3)_{2}(4)_{3}(1)_{4}-\cdots\Big]A_1(1)A_2(2)A_3(3)A_4(4)=\frac{1}{4}H_{2143}(1234) F_{12}(1)F_{21}(2)F_{34}(3)F_{43}(4)\,.
\ea 
These features become particularly relevant to the extraction of non-perturbative physics in the broader context of perturbative QCD calculations, which take the form of gauge invariant products of field-strength operators. A nice example of this is seen in the analogous computation of the one-loop quark box diagram, respectively in polarized DIS and in DVCS, where the non-perturbative operators carry information on the chiral anomaly (polarized DIS) and trace anomaly (unpolarized DVCS), both of which are central to understanding chiral symmetry breaking and confinement \cite{Bhattacharya:2022xxw,Bhattacharya:2023wvy,Tarasov:2020cwl,Tarasov:2021yll,Tarasov:2025mvn,Feal-Tarasov-Venugopalan}.

The fact that six is the minimal number of polynomial numerators in the worldline formalism—and hence the minimal number of head form factors required to fully determine the rank-4 QED vacuum-polarization tensor admits a deeper explanation in terms of the symmetries of the permutation group\footnote{Our construction determines the number of independent Bose-symmetric gauge invariant  structures compatible with $S_4$ and the Ward identities for products of $N=4$ field-strength tensors. These structures form a basis, although not necessarily an orthogonal one. An orthogonal basis could be obtained by applying the appropriate Young symmetrizers to project the tensors onto irreducible representations of $S_4$. In that case, the same amplitude could be expressed equivalently in terms of six orthogonal Bose-symmetric and gauge invariant  structures.}  that we will give in Section \ref{sec:cauchy-frobenius-burnside}, and generalize for arbitrary $N$. We will demonstrate there a drastic $N!$ reduction of the number of head form factors in perturbation theory by using this procedure.
As it is well known \cite{Schubert:2001he} each of these head form factors (with unordered $\tau$-parameters) encode themselves $(N - 1)!/2$ number of Feynman diagrams. This will be demonstrated explicitly in Section \ref{sec:feynman_diagram_correspondence}, where we recover exactly the results of Karplus and Neuman. 

\subsection[N=6 and beyond: QED sixth- and higher-rank vacuum polarization tensors.]{$N=6$ and beyond: higher rank QED vacuum polarization tensors.}

When the number of photons $N$ becomes large, the above procedure can—and must—be made systematic. The transverse decomposition of the polarization tensor, dictated by current conservation, follows from general principles, as we illustrated explicitly for the $N=4$ case. In this section, we therefore focus on the efficient use of the worldline master formula in Eq.~\eqref{eq:Pi_mu1_mu2_mu3_mu4_result} to compute higher-rank tensors, or equivalently, $N$-photon amplitudes.

In particular, using Eq.~\eqref{eq:Pi_mu1_mu2_mu3_mu4_result}, we will provide, for the first time, the explicit worldline representation of the 40 independent form factors—out of the original $(N-1)^N\rightarrow 5^6 = 15625$ terms—that completely determine the rank-6 QED vacuum-polarization tensor. We will also provide the $N=8$ and $N=10$ results in the form of an algorithm that automatically implements Eq.~\eqref{eq:Pi_mu1_mu2_mu3_mu4_result} for general \(N\) and identifies the minimal set of head form factors required to fully specify the tensor. The script is provided as an ancillary file with this manuscript. It generates all head form factors for arbitrary $N$ and then explicitly counts equivalence classes, or orbits, by sequentially applying permutations $\sigma \in S_N$ to each element, generalizing the procedure outlined in the previous section.

For $N=6$, the solution follows from Eq.~\eqref{eq:P_i1_iN_worldline} to be, 
\ba 
&H_{i_1i_2i_3i_4i_5i_6}(123456) \nonumber\\
&= -2\frac{\big(g\mu^{2-d/2}\big)^N}{(4\pi)^{d/2}}\Gamma(6-d/2)\int_0^1 d\tau_1d\tau_2 \cdots d\tau_N \big[D(123456;\tau_1,\ldots,\tau_6)\big]^{6-d/2} P_{i_1i_2i_3i_4i_5i_6}(\tau_1,\ldots,\tau_6)\,,
\ea 
The denominator is universal, given by 
\ba 
D(123456;\tau_1,\ldots,\tau_6) = m^2-\frac{1}{2}\sum_{i,j=1}^6 G^B_{ij} k_i \wc k_j\,.
\ea 
Therefore, the main task is to determine the polynomial numerators for this case from the general expression in Eq.~\eqref{eq:P_i1_iN_worldline_alternativeform} comprising Levi-Civita weighted sums of fermion and (derivatives of) boson worldline Green functions:
\ba 
&P_{i_1i_2i_3i_4i_5i_6}(\tau_1,\ldots,\tau_6)=\sum_{\sigma\in S_6} \text{sgn}(\sigma)\prod_{k=1}^6\Big\{\delta_{k,\sigma(k)}\dot{G}^B_{k,i_k}+\delta_{\sigma(k),i_k}G^F_{i_k,k}\Big\}
\ea 

We will demonstrate in the next section, using the Burnside/Cauchy–Frobenius lemma, the minimal set of head form factors that fully determines the rank-6 tensor contains 40 elements. Each of these is a sum of two monomials built from six bosonic and fermionic Green functions, for example
\ba 
P_{2 1 1 1 1 1}(\tau_1,\ldots,\tau_6) = \dot{G}^B_{1 2}\dot{G}^B_{2 1}\dot{G}^B_{3 1}\dot{G}^B_{4 1}\dot{G}^B_{5 1}\dot{G}^B_{6 1} -\dot{G}^B_{3 1}\dot{G}^B_{4 1}\dot{G}^B_{5 1}\dot{G}^B_{6 1}G^F_{1 2}G^F_{2 1}\,.
\ea 
The complete list of the remaining 39 form factors that fully determine the tensor is provided in Appendix~\ref{app:N6list}. For $N=8$ and $N=10$, the head form factors are likewise linear combinations of monomials involving products of eight and ten boson and fermion worldline Green functions, respectively. This list too can be directly obtained using the computer script accompanying this manuscript. We will also discuss in the next section the general problem of counting the number of equivalence classes at large $N$ and shall determine the asymptotic scaling of the reduction in the number of independent form factors achieved by this method.

\section[Large N asymptotics of orbit counting: Burnside-Cauchy-Frobenius lemma.]
{Large $N$ asymptotics of orbit counting: Burnside-Cauchy-Frobenius lemma.}
\label{sec:cauchy-frobenius-burnside}
In this section, we will employ the Burnside-Cauchy-Frobenius lemma to count the number of orbits (corresponding to each of the independent head form factors) that are required to fully specify the QED vacuum-polarization tensor at arbitrary order in perturbation theory. The Burnside–Cauchy-Frobenius  lemma allows one to count distinct heads modulo an equivalence relation, which in our case is the (Bose) permutation symmetry of the tensor. We will derive in this section a compact expression for this orbit counting for arbitrary $N$, corresponding to the number of distinct head form factors needed to completely determine the $N$-th rank vacuum polarization tensor (or $N$-photon amplitude). 

As we saw in Section~\ref{subsec:N4} --- see also Appendix~\ref{app:notational_invariance}--- heads are classified into equivalence classes according to the pattern of their subindex structure. We therefore focus solely on studying these subindex patterns and disregard for our current discussion the momentum arguments. We formally define the set of all index patterns in the head form factors as
\ba 
X_N=\{(i_1,\ldots,i_N) \in\{1,\ldots,N\}^N \,:\, i_k\neq k \,\,\text{for all}\,\, k\}.
\ea 
Note that there are $|X_N|=(N-1)^N$ such $N$-tuples, which corresponds to the total number of tensor heads before reduction. Let $S_N$ be the group of permutations of $\{1,\ldots,N\}$. Then any two heads with subindices given by $N$-tuples $(i,j)\in X_N$ are said to be equivalent, or related by permutation symmetry, if there exists a permutation operation $\sigma \in S_N$ which sends one set element into another,
\ba 
j_k\equiv (\sigma\cdot i)_k=\sigma(i_{\sigma^{-1}(k)})\,,\qquad k=1,\ldots,N\,.\label{eq:equivalence_def}
\ea 
where in what follows $\sigma\cdot i$ denotes formally the action of a permutation $\sigma$ on an $N$-tuple of subindices $i$. This is what we implemented explicitly in Section \ref{subsec:N4} and in Appendix \ref{app:notational_invariance}.

The Burnside-Cauchy-Frobenius lemma states that the number of equivalence classes  $\#(X_N/S_N)$, or orbits, is given by 
\ba 
\# \text{orbits} = \# (X_N/S_N) = \frac{1}{|S_N|}\sum_{\sigma\in S_N}|\text{Fix}(\sigma)|\label{eq:burnside_lemma}
\ea 
where $|S_N|=N!$ and $\text{Fix}(\sigma)$ is the set of elements that are fixed (left invariant) by $\sigma$:
\ba 
\text{Fix}(\sigma) = \{i\in X_N: \sigma\cdot i = i\}\,.
\ea 
As opposed to Eq.~\eqref{eq:equivalence_def}, which relates different $N$-tuples of indices $j$ and $i$, thereby establishing an equivalence among them, the invariance conditions are the set of $N$ equations,
\ba 
i_k = (\sigma \cdot i)_k =\sigma(i_{\sigma^{-1}(k)})\,,\quad k=1,2,\ldots,N\,,
\label{eq:fixed_point}
\ea 
that define the $N$-tuples that remain fixed under a permutation $\sigma$. Therefore, rather than explicitly generating all heads and counting equivalence classes, as done in Appendix~\ref{app:notational_invariance}—a procedure that becomes impractical for large $N$—the lemma proceeds by first classifying permutations according to their cycle structure,  determining how many elements of $X_N$ are fixed (left invariant) by each permutation type, and finally summing over all  permutations $\sigma\in S_N$. 

A cycle of length $m$ ($m<N$), or an $m$-cycle, is a permutation of $m$ elements:
\ba 
\sigma(i_1)=i_2\,, \sigma(i_2)=i_3\,,\ldots \,,\sigma(i_m)=i_1\,,
\ea 
In cycle notation $(i_1i_2\ldots i_m)$ denotes the permutation $i_1\to i_2$, $i_2\to i_3$ ... $i_m\to i_1$. For clarity, we illustrate the usage of this lemma for orbit counting first in the case $N=4$, thereby  recovering our previous result. The set of head index patterns is given in this case by
\ba 
X_4 = \left\{(i_1,i_2,i_3,i_4) \in \{1,2,3,4\}^4 \quad | \quad i_1\neq 1 , i_2\neq 2, i_3\neq 3, i_4\neq 4\right\}\,,
\ea 
so that $|X_4|=(4-1)^4=81$, corresponding to the total number of head form factors before reduction. As discussed above, the goal is to determine how many inequivalent $4$-tuples exist under permutation symmetry. 

Towards this end, one starts
classifying each $\sigma\in S_4$ by its cycle structure, in order to sum first over cycle types. For $N=4$ the possible cycle structures are:

\textit{1) Trivial permutation.}
This corresponds to the identity, $\sigma=(1)(2)(3)(4)=e$, i.e. $\sigma(k)=k$ for all $k$. All elements of $X_4$ (indices of the head form factors) are invariant under this action, hence
\ba 
\left|\text{Fix}(e)\right|=|X_4|=81.
\ea 
and the contribution to the orbit count is therefore $1\cdot 81=81$. 

\textit{2) Single 2-cycle.}
There are six such permutations. Consider, for instance, $\sigma=(12)(3)(4)$, with  $\sigma(1)=2$, $\sigma(2)=1$, $\sigma(3)=3$ and $\sigma(4)=4$. The invariance condition $i_k=\sigma(i_{\sigma^{-1}(k)})$ gives $i_1=\sigma(i_2)$, $i_2=\sigma(i_1)$,  $i_3=\sigma(i_3)$ and $i_4=\sigma(i_4)$. Because $\sigma(3)=3$ and $\sigma(4)=4$, together with $i_3\neq 3$ and $i_4\neq 4$, one must have $i_3=4$ and $i_4=3$. The remaining conditions $i_1=\sigma(i_2)$ and $i_2=\sigma(i_1)$ yield $(i_1,i_2)=(2,1),\,(3,3)\,, \text{or} \, (4,4)$, excluding $(1,2)$ because of the constraints $i_1\neq 1$ and $i_2\neq 2$, and only one 2-cycle is permitted. Thus
\ba 
|\text{Fix}(\text{2-cycle})|=3
\ea 
With six such permutations, the total contribution is $6\cdot 3=18$.

\textit{3) Single 3-cycle.}
These are permutations consisting of one $3$-cycle and one fixed point; there are three such permutations. Consider $\sigma=(123)(4)$, with $\sigma(1)=2$, $\sigma(2)=3$, $\sigma(3)=1$, $\sigma(4)=4$. The invariance conditions imply $i_4=\sigma(i_4)$ which forces $i_4=4$. Since this violates $i_4\neq 4$ there are no invariant elements. 
\ba 
|\text{Fix}(\text{3-cycle})|=0
\ea 
The total contribution is therefore $3\cdot 0=0$. 

\textit{4) Two 2-cycles.} There are three such permutations. Consider $(12)(34)$, with $\sigma(1)=2$, $\sigma(2)=1$, $\sigma(3)=4$ and $\sigma(4)=3$. Invariance requires $(i_1,i_2)=(i,\sigma(i))$ with $i\neq 1$, so $i\in {2,3,4}$. Likewise $(i_3,i_4)=(j,\sigma(j))$ with $j\neq 3$, so $j\in {1,2,4}$. Hence
\ba 
\left|\text{Fix}(\text{two 2-cycles}) \right|= 3\cdot 3 = 9.
\ea 
With three such permutations, the total contribution is $3\cdot 9=27$.

\textit{5) Single 4-cycle}. There are six such permutations. Consider $\sigma=(1234)$, with $\sigma(1)=2$, $\sigma(2)=3$, $\sigma(3)=4$ and $\sigma(4)=1$. Invariance forces $(i_1,i_2,i_3,i_4)=(i,\sigma(i),\sigma^2(i),\sigma^3(i))$. Since $i_1\neq 1$ we have $i\in\{2,3,4\}$ and all the resulting tuples satisfy $i_k\neq k$. Therefore,
\ba 
\left|\text{Fix}(\text{4-cycle})\right|=3\,,
\ea 
There are $4!/4=6$ such 4-cycles, so the total contribution is $3\cdot 6=18$.

Finally, applying the Burnside lemma, Eq.~\eqref{eq:burnside_lemma},
\ba 
\# \, \text{orbits} = \# (X_4/S_4) = \frac{1}{24} \left(81+6\cdot 3+ 3\cdot 0+3\cdot 9 +6\cdot 3\right) = \frac{144}{24}=6\,,
\ea 
in agreement with the direct counting we performed in Appendix~ \ref{app:notational_invariance}.

Remarkably, the previous discussion can be generalized for arbitrary $N$. To this end, we note that, as with $N=4$, any permutation $\sigma\in S_N$ can be written as a product of $n_m$ cycles of length $m$, $m=1,\ldots,N$. The cycle structure is fully specified by the integers $n_1,\ldots,n_m$. It follows that the sum over $m$ and $n_m$, for each particular cycle structure, is subject to the constraint
\ba 
\sum_{m=1}^N n_m m = N\,.
\ea 
For a given cycle structure, the number of distinct permutations in $S_N$ with that structure is
\ba 
\frac{N!}{\prod_{m=1}^N m^{n_m}n_m!}\,.
\label{eq:numberperm_cyclestructure}
\ea 
Here, we start from the total number of orderings $|S_N| = N!$ and divide by two sources of overcounting.
First, for each fixed $m$, the $n_m$ cycles of length $m$ can be permuted among themselves in $n_m!$ indistinguishable ways.
Second, each individual cycle of length $m$ admits exactly $m$ equivalent cyclic rotations that represent the same permutation. Since there are $n_m$ such cycles, this produces a factor $m^{n_m}$.
Dividing by the product of these factors yields Eq.~\eqref{eq:numberperm_cyclestructure}.

Consider an $m$-cycle of $\sigma$ on the set of positions   $(k_1,\ldots,k_m)$. The fixed-point condition in Eq.~\eqref{eq:fixed_point} implies 
\ba
&i_{k_1}=\sigma(i_{k_m})\,,\nonumber\\
&i_{k_2}=\sigma(i_{k_1})\,,\nonumber\\
&\vdots\nonumber\\
&i_{k_m}=\sigma(i_{k_{m-1}})\,.\label{eq:fixedpoint_mcycle}
\ea 
By successive substitution one obtains
\ba 
i_{k_1}=\sigma(i_{k_m})=\sigma^2(i_{k_{m-1}})=\cdots=\sigma^m(i_{k_1}),
\ea 
hence the system \eqref{eq:fixedpoint_mcycle} is equivalent to the single condition
\ba 
\sigma^{m}(i_{k_1})=i_{k_1}\,.\label{eq:m_parity}
\ea 
Then all remaining values $i_{k_j}$ are determined by Eq.~\eqref{eq:fixedpoint_mcycle} and there is no extra freedom once the integer $i_{k_1}$ is chosen. 

We now count the number of integers $i \in \{1,\ldots,N\}$ satisfying $\sigma^m(i)=i$. Every element $i$ belongs to exactly one of the disjoint cycles of $\sigma$. If $i$ lies in a cycle of $\sigma$ of length $d$, then $\sigma^m(i)=i$ holds if and only if $d \mid m$ (where we use the notation for $d$ is a divisor of $m$). Since there are $n_d$ cycles of length $d$,  the number of solutions of $\sigma^m(i)=i$ is
\ba 
S_m=\sum_{d|m}n_d\, d < N\,.
\ea 
Imposing the constraint $i_{k_1}\neq k_1$ removes exactly one solution, so the number of allowed values for $i_{k_1}$ is $S_m-1$. Because there are $n_m$ such cycles of length $m$ for $m=1,\ldots,N$, the number of  fixed points of $\sigma$ is
\ba 
|\mathrm{Fix}(\sigma)|=\prod_{m=1}^N\left(\sum_{d\mid m} d\,n_d-1\right)^{n_m}.\label{eq:numberfixedpointsbymcycle}
\ea 
Using $|S_N|=N!$ and combining Eqs.~\eqref{eq:numberperm_cyclestructure} and \eqref{eq:numberfixedpointsbymcycle}, we obtain
\begin{equation}
\#(X_N/S_N) =
\sum_{\substack{n_1,\ldots,n_N\ge 0\\ \sum_{m=1}^N m n_m = N}}
\left[\prod_{m=1}^N \frac{1}{m^{n_m} n_m!}\right]
\left[\prod_{m=1}^N\left(\sum_{d\mid m} d\,n_d-1\right)^{n_m}\right].
\label{eq:burnside_result_generalN}
\end{equation}
Eq.~\eqref{eq:burnside_result_generalN} yields the orbit count directly,
without explicitly constructing all head form factors and organizing them
into equivalence classes — a task that becomes rapidly impractical as $N$ grows. An implementation of Eq.~\eqref{eq:burnside_result_generalN} is also included in the computer script accompanying this work, and its validity can be verified through direct comparison with the explicit counting of the equivalence classes for any value of $N$.

We can now use Eq.~\eqref{eq:burnside_result_generalN} to treat larger values of $N$.
The resulting number of orbits —  the number of equivalence classes,
or independent \emph{heads} required to fully specify the off-shell rank-$N$
tensor in QED — is summarized in Table~\ref{tab:ratio_orbits_total}.
\begin{table}[ht]
\centering
\begin{tabular}{c|c|c}
$N$ & $\#(X_N/S_N)$ & $\#(X_N/S_N)/(N-1)^N$ \\ \hline
2  & 1     & $1$ \\
4  & 6     & $7.41\times 10^{-2}$ \\
6  & 40    & $2.56\times 10^{-3}$ \\
8  & 291   & $5.05\times 10^{-5}$ \\
10 & 2273  & $6.52\times 10^{-7}$ \\
12 & 18264 & $5.82\times 10^{-9}$
\end{tabular}
\caption{Number of orbits $\#(X_N/S_N)$ of index patterns for head tensors under permutation symmetry as a function of $N$, and their ratios to the total number of tensor heads $(N-1)^N$. Odd $N$ configurations vanish by Furry's theorem.}
\label{tab:ratio_orbits_total}
\end{table}
The Burnside sum in Eq.~\eqref{eq:burnside_result_generalN} shows that $\#(X_N/S_N)$ grows exponentially with rate $e^N$, with slow subexponential corrections coming from the cycle-structure–dependent factors. In contrast, the total number of head form factors before reduction, using the Stirling formula, grows faster than factorially: 
\ba 
\#(X_N/S_N)\sim \frac{1}{\sqrt{N}}e^{N-1}\,, \qquad (N-1)^N \sim \frac{e^{N-1}}{\sqrt{2\pi N}} N!
\ea 
\section{Feynman diagram correspondence.}
\label{sec:feynman_diagram_correspondence}

In this section, we will discuss the computation of the light-by-light scattering amplitude. We will   
demonstrate how worldline master integrals generate the Feynman diagrams of conventional perturbation theory, and connect these results with the previously known expressions due to Karplus and Neuman \cite{Karplus:1950zz} for the four-point amplitude. Their derivation for the amplitude however only considered the case where the four external photons in the polarization tensor are put on-shell, and the squared amplitude admits a familiar first-line simplification in the spinor-helicity basis. Here we will discuss the efficiency of such calculations for the full tensor amplitude, allowing all photon momenta to be off shell. This is relevant for applications where the 
amplitude enters as a subgraph of higher-point diagrams, such as in the computation of cusp anomalous dimensions or lepton anomalous magnetic moments, or in processes
where interference effects may become relevant; examples include  ultraperipheral nuclear collisions (UPCs) at RHIC and the LHC, or in DIS at the EIC. A more extensive treatment of the off-shell case, and its role in determining the three- and four-loop contributions to the cusp anomalous dimension, will appear in \cite{PaperII}.  

Towards recovering the Karplus and Neuman expressions explicitly, and connecting directly with Feynman diagram results, we will split the integration of the corresponding worldline integrals into ordered sectors.\footnote{In \cite{PaperII}, we will discuss a framework whereby one bypasses this step of converting worldline integrals into Feynman parameter integrals, and instead, computes the gauge invariant form factors directly. This opens up a path towards addressing the long-standing issue of the factorial growth of Feynman diagrams in perturbation theory.} For the $N=4$ photon amplitude, there are $(4 - 1)! /2= 3$  integration regions in the worldline master integrals for the head form factors in Eq.~\eqref{eq:H_def_worldline}. These correspond to the orderings
\ba 
{(a)} \,\,\, \tau_1=1>\tau_2>\tau_3>\tau_4 \,,\nonumber\\
{(b)} \,\,\,\tau_1=1>\tau_3>\tau_2>\tau_4\,,\nonumber\\
{(c)}\,\,\,\tau_1=1>\tau_2>\tau_4>\tau_3\,, \label{eq:orderings}
\ea 
plus the three contributions corresponding to the invariance of the integrand under $\tau_i'=1-\tau_i$ transformations, that are equal to $(a)$, $(b)$ and $(c)$, respectively, under reflection symmetry. For the $(a)$ contribution, we change worldline integration variables to  Feynman parameter variables, with the substitutions
\ba 
y_1=\tau_1-\tau_2=1-\tau_2\,,\,\,\,\, y_2=\tau_2-\tau_3\,,\,\,\,\, y_3=\tau_3-\tau_4\,,\,\,\,\, y_4=\tau_4\,,
\ea  
such that $y_1,y_2,y_3,y_4\ge 0$ and $y_1+y_2+y_3+y_4=1$. The contributions from (b) and (c) can be obtained directly by exchanging $k_3\leftrightarrow k_2$ and $k_4\leftrightarrow k_3$ in the arguments in the expression for (a). As a result, we can write
\ba 
H_{ijkl}(1234)= H^{(a)}_{ ijkl}(1234)+H^{(a)}_{ijkl}(1324)+H^{(a)}_{ ijkl}(1243)\,,
\label{eq:three_feyndiag_headdecomposition}
\ea 
where we used the superscript $(a)$ to denote the restriction of the integrations in the head form factors to the $(a)$ region. The r.h.s. therefore corresponds to three Feynman diagrams while the l.h.s. corresponds to the full worldline diagram, at this particular order $\mathcal{O}(g^4)$ in perturbation theory. We find
\ba 
&H_{ijkl}^{(a)} (1234)= -2\frac{g^4\mu^{8-2d}}{(4\pi)^{d/2}} \Gamma(4-d/2)\int_{-\infty}^\infty  dy_1dy_2dy_3dy_4 \theta(y_1)\theta(y_2)\theta(y_3)\theta(y_4)\delta(1-y_1-y_2-y_3-y_4)\nonumber\\
&\times 2 P_{ijkl}(y_1,y_2,y_3,y_4)\big[D(1234;y_1,y_2,y_3,y_4)\big]^{d/2-4}\,,
\ea 
where we carefully accounted for the factor $2$ coming from the use of the symmetry of $H_{ijkl}(1234)$ under $\tau_i'\to 1-\tau_i$ to reduce the number of integrals. The polynomial denominator becomes
\ba 
D(1234;y_1,y_2,y_3,y_4)=m^2+ y_1y_2k_2^2+y_1y_3 (k_2+k_3)^2+y_1y_4 k_1^2 +y_2y_3k_3^2 +y_2y_4 (k_1+k_2)^2 +y_3y_4 k_4^2\,.
\ea 
We employed here four-momentum conservation $k_1+k_2+k_3+k_4=0$, as specified the $\delta$-function in Eq.~\eqref{eq:Pi_mu1_mu2_mu3_mu4_result}. The 6 polynomial numerators in Eq.~\eqref{eq:6numerators} transform to
\ba 
&2P_{2111}(y_1,y_2,y_3,y_4)=8y_1 (y_2+y_3+y_4)(y_1+y_2-y_3-y_4)(y_1+y_2+y_3-y_4)\,,\nonumber\\
&2P_{2121}(y_1,y_2,y_3,y_4)=8y_1(y_2+y_3+y_4)(y_4-y_1-y_2-y_3)(y_1+y_3+y_4-y_2)\,,\nonumber\\
&2P_{2123}(y_1,y_2,y_3,y_4)=8y_1(y_2+y_3+y_4)(y_1+y_3+y_4-y_2)(y_1+y_2+y_4-y_3)\,,\nonumber\\
&2P_{2311}(y_1,y_2,y_3,y_4)=2(y_4-y_1-y_2-y_3)\big[(y_2+y_3+y_4-y_1)(y_3+y_4-y_1-y_2)(y_1+y_3+y_4-y_2)-1\big]\,,\nonumber\\
&2P_{2143}(y_1,y_2,y_3,y_4)=32y_1y_3(y_2+y_3+y_4)(y_1+y_2+y_3)\,,\nonumber\\
&2P_{2341}(y_1,y_2,y_3,y_4)=2\big[1-(y_2+y_3+y_4-y_1)(y_4-y_1-y_2-y_3)(y_1+y_3+y_4-y_2)(y_1+y_2+y_4-y_3)\big]\,.
\label{eq:6numerators_feynman}
\ea 
{The equations in Eq.~\eqref{eq:6numerators_feynman} reproduce exactly the polynomial numerators originally derived by Karplus and Neuman (see Eqs.~(30a) to (35b) of \cite{Karplus:1950zza}) through an evaluation of the $4^\text{th}$-rank tensor using the corresponding Feynman diagrams.} 

A convenient basis for evaluating the integrals outlined above is provided by the one-loop scalar $4$-point function (see Appendix~ \ref{app:oneloopscalarintegrals} for details), 
\ba 
&I^d_{n_1,n_2,n_3,n_4}\left(1234\right)=\frac{\Gamma(4-\frac d2)}{(4\pi)^{d/2}}\int^\infty_{-\infty} d^4 y\, \theta(y_1)\theta(y_2)\theta(y_3)\theta(y_4)\delta(1-y_{1234})y_1^{n_1-1}y_2^{n_2-1}y_3^{n_3-1}y_4^{n_4-1}\nonumber\\
&\times\bigg[m^2+y_1y_2k_2^2+y_1y_3k_{23}^2+y_1y_4k_1^2+y_2y_3k_3^2+y_2y_4k_{12}^2+y_3y_4k_4^2\bigg]^{d/2-4}\,,\label{eq:J_definition}
\ea 
where $k_{12}=k_1+k_2$, $k_{23}=k_2+k_3$, and $y_{1234}=y_1+y_2+y_3+y_4$. We take all photon external momenta $k_i$ to be incoming and work in Euclidean spacetime, so that all Lorentz invariants entering the above expression are non-negative. The analytic continuation to Minkowski spacetime, which restores the branch cut structure of the amplitude, can always be performed at the end of the calculation. The one-loop scalar four-point basis function in Eq.~\eqref{eq:J_definition} with  $n_i=1$, $i=1,2,3,4$, (no powers of Feynman parameters in the polynomial numerator) corresponds to 
\ba 
&I_{1,1,1,1}^d\left(1234\right)= \int\frac{d^d p}{(2\pi)^d}\frac{1}{\left[(p+k_1)^2+m^2\right]\left[(p+k_1+k_2)^2+m^2\right]\left[(p+k_1+k_2+k_3)^2+m^2\right] \left[p^2+m^2\right]} \,.\label{eq:J_definition_momentum}
\ea 
The values of $n_i$, with $i=1,2,3,4$, can be directly inferred from the polynomials appearing in the numerators of Eq.\eqref{eq:6numerators_feynman}, and all head form factors expressed as linear combinations of $I_{n_1,n_2,n_3,n_4}^d(1234)$. Omitting, for clarity, the photon-momentum arguments $(1234)$, one obtains in this basis:
\ba 
H_{2111}^{(a)}(1234)&=16g^4\Big[\,
I^d_{2,1,1,4}
+ I^d_{2,1,2,3}
- I^d_{2,1,3,2}
- I^d_{2,1,4,1}
- I^d_{2,2,1,3}
- 2\,I^d_{2,2,2,2}
- I^d_{2,2,3,1}
- I^d_{2,3,1,2}
+ I^d_{2,3,2,1}
\nonumber\\ 
&+ I^d_{2,4,1,1}- 2\,I^d_{3,1,1,3}
- 2\,I^d_{3,1,2,2}
+ 2\,I^d_{3,2,2,1}
+ 2\,I^d_{3,3,1,1}
+ I^d_{4,1,1,2}
+ I^d_{4,1,2,1}
+ I^d_{4,2,1,1}
\,\Big]\,,\label{eq:H2111_basisexp}\\
H_{2121}^{(a)}(1234)&=-16g^4\Big[
I^d_{2,1,1,4}
+ I^d_{2,1,2,3}
- I^d_{2,1,3,2}
- I^d_{2,1,4,1} 
 - I^d_{2,2,1,3}
- 2\, I^d_{2,2,2,2}- I^d_{2,2,3,1} 
- I^d_{2,3,1,2}
+ I^d_{2,3,2,1}\nonumber\\
&+ I^d_{2,4,1,1} 
 - 2\, I^d_{3,1,2,2}
- 2\, I^d_{3,1,3,1}
- 2\, I^d_{3,2,2,1} 
- I^d_{4,1,1,2}
- I^d_{4,1,2,1}
- I^d_{4,2,1,1}\Big]\,,\label{eq:H2121_basisexp}\\
H_{2123}^{(a)}(1234)&=-16g^4\Big[\;
I^d_{2,1,1,4}
+ I^d_{2,1,2,3}
- I^d_{2,1,3,2}
- I^d_{2,1,4,1} 
+ I^d_{2,2,1,3}
+ 2\,I^d_{2,2,2,2}
+ I^d_{2,2,3,1}
- I^d_{2,3,1,2}
+ I^d_{2,3,2,1}\nonumber\\
&- I^d_{2,4,1,1} 
+ 2\,I^d_{3,1,1,3}
+ 2\,I^d_{3,1,2,2}
+ 2\,I^d_{3,2,1,2} 
+ I^d_{4,1,1,2}
+ I^d_{4,1,2,1}
+ I^d_{4,2,1,1}\Big]\,,\label{eq:H2123_basisexp}\\
H_{2311}^{(a)}(1234)&=4g^4\Big[
- I^d_{1,1,1,2}
+ I^d_{1,1,1,5}
+ I^d_{1,1,2,1}
+ 2\,I^d_{1,1,2,4} 
- 2\,I^d_{1,1,4,2}
- I^d_{1,1,5,1}
+ I^d_{1,2,1,1}
- 2\,I^d_{1,2,1,4}- 4\,I^d_{1,2,2,3} \nonumber\\
&- 2\,I^d_{1,2,3,2}
+ 2\,I^d_{1,3,2,2}
+ 2\,I^d_{1,3,3,1} + 
2\,I^d_{1,4,1,2}
- I^d_{1,5,1,1}
+ I^d_{2,1,1,1}
- 2\,I^d_{2,1,1,4}-4\,I^d_{2,1,2,3}
- 2\,I^d_{2,1,3,2}  \nonumber\\
&+ 4\,I^d_{2,2,1,3}
+ 4\,I^d_{2,2,2,2} 
- 2\,I^d_{2,3,1,2}
+ 2\,I^d_{3,1,2,2}
+ 2\,I^d_{3,1,3,1}- 2\,I^d_{3,2,1,2} 
+2\,I^d_{3,3,1,1}
+ 2\,I^d_{4,1,1,2}
- I^d_{5,1,1,1}\,\Big]\,,\label{eq:H231_basisexp}\\
H^{(a)}_{2143}(1234)&=-64g^4\Big[
 I^d_{2,1,3,2}
+ I^d_{2,1,4,1}
+ I^d_{2,2,2,2}
+ 2\,I^d_{2,2,3,1}
+ I^d_{2,3,2,1} 
+ I^d_{3,1,2,2}
+ I^d_{3,1,3,1}
+ I^d_{3,2,2,1}\,\Big]\,,\label{eq:H2143_basisexp}\\
H_{2341}^{(a)}(1234)&=-4g^4\Big[
 I^d_{1,1,1,1}
- I^d_{1,1,1,5}
+ 2\,I^d_{1,1,3,3}
- I^d_{1,1,5,1} 
+ 2\,I^d_{1,3,1,3}
+ 2\,I^d_{1,3,3,1}
- I^d_{1,5,1,1} 
+ 8\,I^d_{2,2,2,2}\nonumber\\
&+ 2\,I^d_{3,1,1,3}
+ 2\,I^d_{3,1,3,1}
+ 2\,I^d_{3,3,1,1}
- I^d_{5,1,1,1}\,\Big]\,.\label{eq:H2341_basisexp}
\ea
Note that the above decomposition is general and valid for arbitrary external photon kinematics\footnote{For on-shell kinematics, identities like $I^4_{n_1,n_2,n_3,n_4}(1234)
=
I^d_{n_3,n_4,n_1,n_2}(1234)$, and $
I^d_{n_1,n_2,n_3,n_4}(1234)
=
I^d_{n_2,n_1,n_4,n_3}(3214)$ can be further used, if desired, to partially simplify the number of terms entering Eqs.~\eqref{eq:H2111_basisexp} to \eqref{eq:H2341_basisexp}.}. 
This procedure can be made systematic for general $N$ using first Eq.~\eqref{eq:H_def_worldline} to identify the independent head form factors that completely determine the rank-$N$ polarization tensor, and then the polynomial numerators in Eq.~\eqref{eq:P_i1_iN_worldline} to translate them directly to Feynman diagrams; this 
allows for a systematic procedure 
in the above basis of scalar integrals. 

In Appendix~\ref{app:oneloopscalarintegrals}, we present an explicit computation of $I^4_{1,1,1,1}$ for on-shell photon kinematics with  finite fermion mass, where the integrals are free of both UV and IR singularities in $d=4$ dimensions. The result is given by Eq.~\eqref{eq:I4_massiveonshell_result} in terms of logarithms and dilogarithms of the kinematic invariants $k_{12}^2$ and $k_{23}^2$, which become Mandelstam $s$ and $t$ variables, respectively, after Wick rotation to Minkowski time. The analytic structure of the integral is determined by the standard branch cuts associated with Landau singularities, given in this case by the roots in Eqs.~\eqref{eq:y_t_y_s_roots} and \eqref{eq:y_ts_roots}:
\ba 
y_{t}^\pm = \frac{1}{2}\pm \frac{1}{2}\bigg[1+\frac{4m^2}{k_{23}^2}\bigg]^{1/2}\,,\,\,\,\, y_{s}^\pm =\frac{1}{2}\pm \frac{1}{2}\bigg[1+\frac{4m^2}{k_{23}^2}\bigg]^{1/2}\,,\,\,\, y_{ts}^\pm = \frac{1}{2}\pm \frac{1}{2}\bigg[1+4m^2\bigg(\frac{1}{k_{23}^2}+\frac{1}{k_{12}^2}\bigg)\bigg]^{1/2}\,.
\ea 
The computation of the IR divergent massless limit of $I_{1,1,1,1}^4$ is also explicitly performed in Appendix \ref{app:oneloopscalarintegrals}. The result is given in  Eq.~\eqref{eq:I_4_massless_onshell1} in $d=4$ dimensions, or analogously, by Eq.~\eqref{eq:I_4_massless_onshell2} in $d=4-2\epsilon$ dimensions. 

The remaining integrals $I^d_{n_1,n_2,n_3,n_4}$ could, in principle, be evaluated case-by-case, as done in \cite{Karplus:1950zz}, with the results  expressed, in each case, in terms of rational functions of the above kinematic invariants and polylogarithms of maximum weight $2$. 
In this way, one will recover explicitly the Karplus–Neuman expressions for the six head form factors. A direct comparison, however, is extremely cumbersome due to the complexity and length of the resulting expressions\footnote{See in particular, Eqs. A1–A10 in Appendix A of~\cite{Karplus:1950zz}, expressed in terms of three transcendental functions defined in Eqs. B1–B3 of Appendix B; explicit results were also derived in the corresponding dispersive framework in \cite{DeTollis:1964una,Costantini:1971cj}.}. Moreover, it is unclear how this method can be readily extended to the computation of scalar integrals in more complicated cases with general kinematic configurations, such as the fully off-shell case, or to higher-point functions with $N>4$.

It is therefore preferable to adopt an alternative general procedure. A more systematic approach~\cite{Bern:1992em,Bern:1993kr} consists of reducing the problem to the computation of a single scalar basis integral $I^d_{1,1,1,1}$—for which analytic results are known~\cite{tHooft:1978jhc} (see also~\cite{Denner:1991kt}) and implemented in several publicly available software packages~\cite{vanOldenborgh:1989wn,Hahn:1998yk,Ellis:2007qk}—with all other integrals obtained by differentiation with respect to the kinematic invariants. This method, however, requires knowledge of the basis integral $I^d_{1,1,1,1}$ in general $d=4-2\epsilon$ dimensions, {\it through $\mathcal{O}(\epsilon)$}, even in cases where it is free from UV and IR singularities. In the massive case, it also requires extending Eq.~\eqref{eq:J_definition_momentum} to include four distinct (dummy) internal fermion masses during the differentiation procedure. In practice, this approach can be as involved—if not more so—than computing the required integrals directly in $d=4$ on a case-by-case basis, where their evaluation is typically simpler. Moreover, in the fully off-shell case, the basis integral $I^d_{1,1,1,1}$ is known in general $d=4-2\epsilon$ dimensions only in terms of a generalized hypergeometric series~\cite{Davydychev:2017bbl} without fully explicit analytic continuation, or, analytically, in terms of a basis of 32 dilogarithms~\cite{Denner:2010tr}. The resulting expressions are difficult to implement, as they require a case-by-case treatment of different kinematic regions before the differentiation procedure of~\cite{Bern:1992em,Bern:1993kr} can be applied effectively.

We introduce here an alternative systematic method to evaluate the remaining integrals. In what follows, we will extend the light-by-light calculation of Karplus and Neuman to the fully off-shell case and outline a general strategy to compute all contributions entering Eqs.~\eqref{eq:H2111_basisexp}–\eqref{eq:H2341_basisexp}. As mentioned earlier, a novel approach that streamlines this task directly in the worldline formalism—thereby avoiding the proliferation of Feynman diagrams—will be presented in \cite{PaperII}. The procedure developed here is instead suited for mapping and translating our worldline results into standard Feynman diagram techniques  efficiently. We present the strategy for the computation of the massless off-shell case; however, the extension to massive configurations, including the on-shell calculation of Karplus and Neuman discussed above, is in principle straightforward.

The key observation is that the power of the polynomial denominator in all terms of Eqs.~\eqref{eq:H2111_basisexp}–\eqref{eq:H2341_basisexp} is fixed, with $N - d/2 = 2 + \epsilon$ for all contributions. As a consequence, our definition of the basis integrals in Eq.~\eqref{eq:J_definition} differs from the standard one used in integration-by-parts (IBP) procedures that is given in Eq. \eqref{eq:Npoint_scalar_function} of Appendix~\ref{app:oneloopscalarintegrals}. In conventional IBP bases, that we denote with  $\mathcal{I}_{n_1,n_2,n_3,n_4}^d$ in Appendix~\ref{app:oneloopscalarintegrals}, the indices $n_i$ encode both powers of propagator denominators and, in the equivalent representation of Eq.~\eqref{eq:I_N_feynmanpar}, also generate powers of Feynman parameters in the numerator. The choice in Eq.~\eqref{eq:Npoint_scalar_function} is therefore better suited to conventional Passarino–Veltman and IBP reductions in Feynman diagram calculations, where the reduction of tensor structures generate integrals with shifted propagator powers. 

In contrast, in our worldline approach, the rank-N tensor is systematically reduced to a smaller set of independent head form factors, scaling relatively as $\sim 1/N!$ at large $N$, with the power $N-d/2$ of the polynomial denominator fixed for all terms contributing to the rank-$N$ tensor\footnote{This feature avoids the appearance of spurious UV divergences in individual Feynman diagrams, which are only canceled after summing over different diagrams in conventional approaches.}. The tensor structure is instead absorbed directly into the polynomial numerators of worldline Green functions (specified in Eq.~\eqref{eq:P_i1_iN_worldline}), which can, as we discussed, be translated into Feynman parameters after splitting the worldline integrals into ordered sectors. As a result,  complexity in the worldline approach is shifted systematically from denominators to numerators. 

Remarkably, this shift does not introduce additional complications; rather, it enables a more transparent and systematic organization of the computation. In particular, it allows for the identification of a minimal set of independent head form factor numerators, as encoded in the universal expression for the polynomial numerators in the worldline approach in Eq.\eqref{eq:P_i1_iN_worldline}, valid for arbitrary $N$. 

A further advantage of this approach emerges at this stage, when translating these results into the language of Feynman diagrams: it simplifies the implementation of an equivalent IBP reduction in this basis, leading to a finite linear algebraic system of equations relating integrals with shifted indices. To demonstrate this, we first perform the integral over one of the Feynman parameters in Eq.~\eqref{eq:J_definition}, obtaining
\ba 
I^4_{n_1,n_2,n_3,n_4}(1234)=\frac{1}{(4\pi)^2}\int_{\substack{y_1,y_2,y_3\geq 0\\ y_1+y_2+y_3\leq 1}} dy_1 dy_2 dy_3 \frac{y_1^{n_1-1}y_2^{n_2-1}y_3^{n_3-1}(1-y_1-y_2-y_3)^{n_4-1}}{D(y_1,y_2,y_3)^2}\,,\label{eq:J_definition_y4integrated}
\ea 
where in the off-shell massless case
\ba 
&D(y_1,y_2,y_3)\label{eq:Dy1y2y3}\\
&=y_1y_2k_2^2+y_1y_3k_{23}^2+y_1(1-y_1-y_2-y_3)k_1^2+y_2y_3k_3^2+y_2(1-y_1-y_2-y_3)k_{12}^2+y_3(1-y_1-y_2-y_3)k_4^2\,.\nonumber
\ea 
The result for the massless integral $I_{1,1,1,1}^4(1234)$ in off-shell kinematics is well-known (see for instance\cite{Davydychev:1990jt,Bern:1993kr}) and is computed explicitly in Euclidean spacetime in Appendix \ref{app:oneloopscalarintegrals};  the result is given in  Eq.~\eqref{eq:I_4_massless_offshell}. 

Having determined $I_{1,1,1,1}^4$ there, we now consider the evaluation of $I_{2,1,1,1}^4$. The latter differs from the former by a single power of $y_1$ in the numerator. The idea is to express this factor $y_1$ in terms of a suitable polynomial in the kinematic invariants and $\partial_{y_1} D$. Indeed, from Eq.~\eqref{eq:Dy1y2y3}
\ba 
\partial_{y_1} D(y_1,y_2,y_3)=-2k_1^2y_1+a(y_2,y_3)\,, \quad a(y_2,y_3)=y_2(k_2^2-k_1^2-k_{12}^2)+y_3(k_{14}^2-k_1^2-k_4^2)+k_1^2\,,
\ea 
where we used $k_1+k_2+k_3+k_4=0$, so
\ba 
y_1=\frac{1}{2k_1^2}\Big\{a(y_2,y_3)-\partial_{y_1}D\Big\}\,,
\ea 
and
\ba 
\frac{y_1}{D^2}= \frac{1}{2k_1^2}\Big\{\frac{y_2}{D^2}(k_2^2-k_1^2-k_{12}^2)+\frac{y_3}{D^2}(k_{14}^2-k_1^2-k_4^2)+\frac{1}{D^2}k_1^2-\frac{1}{D^2}\partial_{y_1}D\Big\}\,.
\ea 
Integrating both sides of the previous equation for $y_1,y_2,y_3\geq 0$ and $y_1+y_2+y_3\leq 1$ and dividing by $(4\pi)^2$, following our definition in Eq.~\eqref{eq:J_definition_y4integrated}, one obtains
\ba 
I_{2,1,1,1}^4(1234)&=\frac{1}{2k_1^2}(k_2^2-k_1^2-k_{12}^2)I_{1,2,1,1}^4(1234)+\frac{1}{2k_1^2}(k_{14}^2-k_1^2-k_4^2)I_{1,1,2,1}^4(1234)+\frac{1}{2}I_{1,1,1,1}^4(1234)\nonumber\\
&+\frac{1}{2k_1^2}\frac{1}{(4\pi)^2}\int_{\substack{y_1,y_2,y_3\geq 0\\ y_1+y_2+y_3\leq 1}} dy_1dy_2dy_3 \partial_{y_1}\frac{1}{D(y_1,y_2,y_3)}\,.
\ea 
The integral over $y_1$ of the term in the second line readily produces
\ba 
 \int_{\substack{y_1,y_2,y_3\geq 0\\ y_1+y_2+y_3\leq 1}} dy_1dy_2dy_3 \partial_{y_1}\frac{1}{D(y_1,y_2,y_3)} = \int_{\substack{y_2,y_3\geq 0\\ y_2+y_3\leq 1}} dy_2dy_3 \left\{\frac{1}{D(1-y_2-y_3,y_2,y_3)}-\frac{1}{D(0,y_2,y_3)}\right\}\,.
\ea 
From the structure of Eq.~\eqref{eq:Dy1y2y3}, the two terms on the right-hand side can be identified with standard triangle integrals. The first involves propagators built from the momenta $k_3$, $k_2$, and $k_2+k_3$, yielding $I_{1,1,1}^4(3,2,2+3)$, while the second involves $k_3$, $k_3+k_4$, and $k_4$, giving $I_{1,1,1}^4(3,3+4,4)$. These contributions can be interpreted as boundary terms of the four-point (box) configuration, where two photons are effectively merged (or pinched) into a single one, producing a three-point (triangle) configuration. We therefore write
\ba 
I_{2,1,1,1}^4(1,2,3,4)&=\frac{1}{2k_1^2}(k_2^2-k_1^2-k_{12}^2)I_{1,2,1,1}^4(1,2,3,4)+\frac{1}{2k_1^2}(k_{14}^2-k_1^2-k_4^2)I_{1,1,2,1}^4(1,2,3,4)\nonumber\\
&+\frac{1}{2}I_{1,1,1,1}^4(1,2,3,4)+\frac{1}{2k_1^2}\left[I_{1,1,1}^4(3,2,2+3)-I_{1,1,1}^4(3,3+4,4)\right]\,.
\ea 
The above equation expresses $I_{2,1,1,1}^4$ in terms $I_{1,1,1,1}^4$ and  $I_{1,2,1,1}^4$ and $I_{1,1,2,1}^4$ the latter two still being unknown. To derive corresponding equations for these integrals, one proceeds analogously by solving for $y_2$ and $y_3$ in terms of $\partial_{y_2}D$ and $\partial_{y_3}D$, respectively. One obtains,
\ba 
I_{1,2,1,1}^4(1,2,3,4)&=\frac{1}{2k_{12}^2}(k_2^2-k_1^2-k_{12}^2)I_{2,1,1,1}^4(1,2,3,4)+\frac{1}{2k_{12}^2}(k_3^2-k_{34}^2-k_4^2)I_{1,1,2,1}^4(1,2,3,4)\nonumber\\
&+\frac{1}{2}I_{1,1,1,1}^4(1,2,3,4)+\frac{1}{2k_{12}^2}\left[I_{1,1,1}^4(2+3,2,3)-I_{1,1,1}^4(1+4,1,4)\right]\,,
\ea 
and
\ba 
I_{1,1,2,1}^4(1,2,3,4)&=\frac{1}{2k_{4}^2}(k_{14}^2-k_1^2-k_{4}^2)I_{2,1,1,1}^4(1,2,3,4)+\frac{1}{2k_{4}^2}(k_3^2-k_{34}^2-k_4^2)I_{1,2,1,1}^4(1,2,3,4)\nonumber\\
&+\frac{1}{2}I_{1,1,1,1}^4(1,2,3,4)+\frac{1}{2k_{4}^2}\left[I_{1,1,1}^4(2,2+3,3)-I_{1,1,1}^4(2,1,1+2)\right]\,.
\ea 
Note that, since there are only three independent Feynman parameters, with $y_4=1-y_1-y_2-y_3$, the fourth integral $I_{1,1,1,2}^4$ can be expressed directly in terms of the others using Eq.~\eqref{eq:J_definition_y4integrated}:
\ba 
I_{1,1,1,2}^{4}(1,2,3,4)=I_{1,1,1,1}^{4}(1,2,3,4)-I_{1,2,1,1}^{4}(1,2,3,4)-I_{1,1,2,1}^{4}(1,2,3,4)\,.
\ea 
The resulting linear algebraic system of equations can now be compactly written in matrix form as
\ba 
G \begin{pmatrix}I_{2,1,1,1}(1,2,3,4)\\I_{1,2,1,1}(1,2,3,4)\\I_{1,1,2,1}(1,2,3,4)\end{pmatrix} = \frac{1}{2} \begin{pmatrix}J_1\\J_2\\J_3\end{pmatrix}\,,\quad \text{or}\quad  \begin{pmatrix}I_{2,1,1,1}(1,2,3,4)\\I_{1,2,1,1}(1,2,3,4)\\I_{1,1,2,1}(1,2,3,4)\end{pmatrix} = \frac{1}{2} G^{-1}\begin{pmatrix}J_1\\J_2\\J_3\end{pmatrix}\,,\label{eq:gram_solution}
\ea 
where the inhomogeneous terms consist only of box and triangle integrals without Feynman parameters in the numerator:
\ba 
J_1 &= k_1^2I_{1,1,1,1}^4(1,2,3,4)+I_{1,1,1}^4(3,2,2+3)-I_{1,1,1}^4(3,3+4,4)\,,\nonumber\\
J_2 &= k_{12}^2I_{1,1,1,1}^4(1,2,3,4)+I_{1,1,1}^4(2+3,2,3)-I_{1,1,1}^4(1+4,1,4) \,,\nonumber\\
J_3 &= k_{123}^2 I^4_{1,1,1,1}(1,2,3,4)+I_{1,1,1}^4(2,2+3,2)-I_{1,1,1}^4(2,1,1+2)\,.\label{eq:sources}
\ea 
Using momentum conservation $k_{1234}=0$, the symmetric matrix $G$ can be written as
\ba 
G=\begin{bmatrix}k_1^2 & k_1\!\cdot\! k_{12}& k_1\!\cdot \!k_{123}\\
k_{12}\!\cdot\! k_1 & k_{12}^2 & k_{12}\!\cdot\! k_{123} \\
k_{123}\!\cdot\! k_1 & k_{123}\!\cdot\! k_{12} &  k_{123}^2
\end{bmatrix}\,.
\ea 
It therefore coincides with the Gram matrix $G_{ij}=q_i\!\cdot\! q_j$, with $q_1=k_1$, $q_2=k_{12}=k_1+k_2$, $q_3=k_{123}=k_1+k_2+k_3=-k_4$. The solution is then obtained by inverting the Gram matrix
\ba 
G^{-1}_{ij}=\frac{1}{\det G} \mathrm{adj}(G)_{ij}
\ea 
acting on the source vector $J_1$, $J_2$ and $J_3$. Here 
\ba 
\det G=\frac{1}{3!}\epsilon^{ijk}\epsilon^{lmn}(q_i\cdot q_l)(q_j\!\cdot\! q_m)(q_k\!\cdot\! q_n) = (q_1\wedge q_2 \wedge q_3)^2\,,
\ea 
is the square of the 3-volume of the parallelepiped spanned by $q_1=k_1$, $q_2=k_1+k_2$ and $q_3=k_1+k_2+k_3$ in four-dimensional Euclidean space, and
\ba 
\mathrm{adj}(G)_{ij}=\frac{1}{2}\epsilon_{ikl}\epsilon_{jmn}G_{km}G_{ln} \,.
\ea 
For our case, one obtains
\ba 
\det G= q_1^2q_2^2q_3^2+2(q_1\!\cdot \!q_2) (q_2\!\cdot \!q_3) (q_3\!\cdot\! q_1) -q_1^2 (q_2\!\cdot\! q_3)^2-q_2^2(q_1\!\cdot\! q_3)^2-q_3^2(q_1\!\cdot\! q_2)^2\,.\label{eq:gramdet_sol}
\ea 
and
\ba 
\mathrm{adj}(G)_{11}&= q_2^2q_3^2-(q_2\!\cdot \!q_3)^2\,,\nonumber\\
\mathrm{adj}(G)_{22}&=q_1^2q_3^2-(q_1\!\cdot \!q_3)^2\,,\nonumber\\
\mathrm{adj}(G)_{33}&=q_1^2q_2^2-(q_1\!\cdot \!q_2)^2\,,\nonumber\\
\mathrm{adj}(G)_{12}&=(q_1\!\cdot\! q_3)(q_2\!\cdot\! q_3)-q_3^2(q_1\!\cdot\! q_2)\,\nonumber\\
\mathrm{adj}(G)_{13}&=(q_1\!\cdot \!q_2)(q_2\!\cdot\! q_3)-q_2^2(q_1\!\cdot \!q_3)\,,\nonumber\\
\mathrm{adj}(G)_{23}&=(q_1\!\cdot\! q_2)(q_1\!\cdot\! q_3)-q_1^2(q_2\!\cdot\! q_3)\,.\label{eq:gramadj_sol}
\ea 
The computation of the basis box $I_{1,1,1,1}^4$ and triangle $I_{1,1,1}^4$ integrals in the totally off-shell case is carried out explicitly in Appendix \ref{app:oneloopscalarintegrals}, with the corresponding results given in Eqs.~\eqref{eq:I_4_massless_offshell} and \eqref{eq:I_3_massless_offshell}, respectively. The integrals $I_{2,1,1,1}^4$, $I_{1,2,1,1}^{4}$, $I_{1,1,2,1}^4$ and $I_{1,1,1,2}^4$ are then obtained directly and algebraically from Eqs.~\eqref{eq:gram_solution} and \eqref{eq:sources}, with the inverse Gram matrix determined\footnote{When the Gram determinant vanishes, the integral can be treated separately in the degenerate kinematic configuration, where it reduces or simplifies accordingly.} explicitly through Eqs.~\eqref{eq:gramdet_sol} and \eqref{eq:gramadj_sol}. 

The remaining integrals required to determine completely the six head form factors of the massless rank-4 tensor in fully off-shell kinematics, Eqs.~\eqref{eq:H2111_basisexp}–\eqref{eq:H2341_basisexp}, can be generated recursively in the same way. For instance, the contributions for which the total power of Feynman parameters is 2 consist of the six terms corresponding to index permutations of $I_{2,2,1,1}^4$, together with the four terms corresponding to permutations of $I^4_{3,1,1,1}$. The corresponding relations can be readily derived following the procedure outlined above, together with a judicious use of the constraint $y_{1234}=1$ to obtain additional identities among them, such as
\ba 
I_{2,1,1,2}^4(1,2,3,4)
=
I_{2,1,1,1}^4(1,2,3,4)
-
I_{3,1,1,1}^4(1,2,3,4)
-
I_{2,2,1,1}^4(1,2,3,4)
-
I_{2,1,2,1}^4(1,2,3,4).
\ea
Since the computation of all head form factors involves at most total Feynman-parameter degree four, the recursion terminates at that level.

We note that conventional IBP identities can also be formulated equivalently in Feynman parametric form~\cite{Tarasov:1996br}; see Ref.~\cite{Artico:2023jrc} for a recent discussion. However, in all these standard approaches (based on tensorial reductions and differential equations together with IBP reductions to master integrals), the resulting identities relate integrals with different propagator powers and/or shifted space-time dimensions. By contrast, our construction of the amplitude operates at fixed denominator power and performs the reduction entirely within the space of numerator insertions, yielding a finite closed algebraic system that can be solved explicitly using a recursive relation.

To summarize this section, we  presented here a systematic comparison between the worldline and Feynman diagram approaches in the nontrivial case of light-by-light scattering. We recovered explicitly in the former the expressions in the latter ({\it a la} Karplus and Neuman)  for the six head form factors. We developed a more efficient procedure relative to the latter  to compute all scalar integrals entering their decomposition in fully off-shell kinematics. Our framework thereby extends the original results of Karplus and Neuman to the full off-shell rank-4 vacuum polarization tensor in massless QED.

Specifically, starting from the explicit evaluation of the basis box $I_{1,1,1,1}^4$ and triangle $I_{1,1,1}^4$, we showed that all remaining integrals with Feynman-parameter insertions can be obtained algebraically through a finite linear system derived from differentiation/IBP identities in parameter space. This system is naturally organized in terms of the Gram matrix and its inverse, leading to compact and explicit solutions in terms of lower-point functions. The specialization to other kinematic configurations can similarly be implemented straightforwardly.

\section{Summary and conclusions}
\label{sec:conclusions}
In this work, we showed how the $N$-th rank vacuum polarization tensor in QED acquires a particularly simple and universal structure in a $(0+1)$-dimensional worldline representation. In particular, we demonstrated that this fully off-shell $N$-photon tensor can be written, for arbitrary $N$, as an $\mathcal{N}=1$ supersymmetric combination of bosonic and fermionic worldline Green functions, which is universal for arbitrary $N$. This representation requires neither Wick contractions of the bosonic and fermionic worldlines nor integrations over loop momenta followed by reduction to scalar loop integrals. It also does not rely on an $\mathcal{N}=1$ integration-by-parts (IBP) approach for the reconstruction of the full amplitude, thereby avoiding ambiguities associated with this approach.

We further demonstrated that the fully off-shell worldline representation of the $N$-th rank polarization tensor can be expressed systematically in terms of a remarkably small set of independent head form factors. The reduction in the number of independent tensor structures follows from key ingredients intrinsic to the worldline master formula: these are current conservation, the permutation symmetry of the N-boson insertions, and the explicit $\mathcal{N}=1$ supersymmetry algebra satisfied by the resulting (0+1)-dimensional representation of the theory in terms of worldlines.
Together, these naturally organize the tensor into equivalence classes under the permutation group symmetries of $S_N$ for general $N$, thereby allowing the amplitude to be reconstructed from a minimal basis of independent head form factors. 

We provided explicit expressions for the form factors as well as the rank-4 and rank-6 QED vacuum polarization tensors. A computer program is provided that generalizes this construction to an arbitrary number $N$ of external off-shell photons. We showed that the counting of independent heads for arbitrary $N$ is naturally formulated in terms of their equivalence classes, and we used the Burnside-Cauchy-Frobenius lemma to demonstrate how the multiplicity of independent head form factors grows asymptotically. We find that the number of such irreducible structures scales as $e^{N-1}/\sqrt{N}$, in contrast to the $e^{N-1}N!/\sqrt{N}$  characteristic behavior of conventional perturbation theory employing Feynman diagrams. 

The classification of the $N$-th rank polarization tensors entirely in terms of the small number of head form factors followed from the repeated and systematic use of Ward identities. As a consequence, QED amplitudes computed in background fields can be expressed as gauge invariant combinations of these background fields multiplied by the head form factors. Importantly, the classification of the heads into equivalence classes (or orbits) indicates a similar classification of the relevant gauge invariant field strength operators. In certain cases, as noted, these observations also apply to QCD. Whether the classification has broader applicability to QCD processes is an outstanding question. 

Finally, as a nontrivial check of the formalism, we reproduced the classic Karplus-Neuman light-by-light scattering result directly in the worldline representation. We generalized their results to the fully off-shell case for massless QED outlining a novel systematic integration-by-parts reduction to a minimal set of scalar box and the triangle basis integrals, expressed in terms of an explicitly solvable system of algebraic identities among basis integrals.

Our results confirm that the combination of worldline current permutation symmetries and Ward identities, together with the $\mathcal{N}=1$ worldline supersymmetry, provides an extremely efficient framework for organizing and computing multiphoton and multiloop amplitudes. The method offers a clear path toward higher-point calculations and suggests a viable alternative to the long-standing issue of the factorial growth of tensor structures and Feynman diagrams encountered in conventional perturbative approaches. 

In a follow-up publication, we will discuss the direct computation of the gauge invariant  set of head form factors without ordering the calculation into $N!$ Feynman diagrams. This potentially provides a further $N!$ factorial advantage to the worldline formalism. The framework discussed here therefore provides a systematic and novel way of performing perturbative calculations circumventing $N!$ proliferations of both tensor structures, and of Feynman diagrams, encountered in conventional perturbation theory.

\section*{Acknowledgements}

X.F. is supported by the U.S. Department of Energy Grant Ref. DE-SC0025732, \textit{Novel Holographic Approaches to the Non-perturbative Dynamics
of Proton Spin.} X.F. would like to thank the Isaac Newton Institute for Mathematical Sciences, Cambridge, for support and hospitality during the program \textit{Quantum field theory with boundaries, impurities, and defects} where work on this paper was undertaken. This work was supported by EPSRC grant no EP/R014604/1.

A. T. is supported by the U.S. Department of Energy, Office of Science, Office of Nuclear Physics through Contract No. DE-SC0020081, and within the framework of the Saturated Glue (SURGE) Topical Collaboration in Nuclear Theory. A.T. thanks the Aspen Center for Physics, which is supported by National Science Foundation grant PHY-2210452, where part of this work was performed.

R.V. is supported by the U.S. Department of Energy, Office of Science under contract DE-SC0012704 and within the framework of the Saturated Glue (SURGE) Topical Collaboration in Nuclear Theory. R.V. is also supported at Stony Brook by the Simons Foundation as a co-PI under Award number 994318 (Simons Collaboration on Confinement and QCD Strings). He acknowledges partial support from DOE Grant Ref. DE-SC0025732, \textit{Novel Holographic Approaches to the Non-perturbative Dynamics of Proton Spin.} 
R.V. thanks the UK Royal Society and the Wolfson Foundation for a Visiting Fellowship and the Higgs Center at the University of Edinburgh for their kind hospitality. 

\appendix 

\section{Explicit worldline structure of the rank-4 tensor before symmetry reduction.}
\label{app:rank4_nosymmetryreduction}
We will illustrate here the direct use of  Eqs.~\eqref{eq:pi_mu1_muN_solution} and \eqref{eq:I_mu1_muN} to reconstruct all form factors of the rank-4 QED vacuum polarization tensor solely in terms of bosonic and fermionic worldline Green functions. The construction does not rely on current conservation or current commutation symmetries to reduce the number of form factors from 138 to 6. Instead, we explicitly obtain the 3 tails, 54 shoulders, and 81 heads by direct substitution of the functions defined in Eqs.\eqref{eq:ABCD_def} into Eqs.~\eqref{eq:pi_mu1_muN_solution} and \eqref{eq:I_mu1_muN}.

This representation is equivalent to the streamlined result presented in Appendix~\ref{app:fullresult}, where only six head form factors are required. As emphasized earlier, infinitely many equivalent representations of the tensor exist, differing by terms orthogonal to $k_{\mu_i}$, $i=1,2,3,4$. In the main text we exploit this freedom to avoid computing all contributions derived explicitly here, focusing instead on the evaluation of the head $(N_b=N)$ terms of the master formula in Eq.~\eqref{eq:I_mu1_muN}. The explicit construction in this appendix also illustrates that the reduction to only six head form factors is already a substantial simplification for $N=4$, omitting redundancies, and becomes dramatically more powerful as the number of form factors grows factorially with $N$.

Specializing Eq.~\eqref{eq:I_mu1_muN} to $N=4$ yields six distinct contributions $(N_a,N_b,N_c)$ satisfying $2N_a+N_b+2N_c=4$, namely:
\ba 
(N_a,N_b,N_c)\,:\,\,\,\,\,\,(2,0,0)\,,\,\,\,\,\,\,(0,4,0)\,\,\,\,\,\,\,(0,0,2)\,,\,\,\,\,\,\,(1,2,0)\,,\,\,\,\,\,\,(0,2,1)\,,\,\,\,\,\,\,(1,0,1)\,.
\ea 
For bookkeeping purposes, we therefore write the rank-4 QED vacuum polarization tensor —obtained by specializing Eq.~\eqref{eq:pi_mu1_muN} to N=4— as
\ba
&\Pi_{\mu_1\mu_2\mu_3\mu_4}(1234)\nonumber\\
&= \Pi_{\mu_1\ldots \mu_4}^{(2,0,0)}(1234)+\Pi^{(0,4,0)}_{\mu_1\ldots \mu_4}(1234)+\Pi_{\mu_1\ldots \mu_N}^{(0,0,2)}(1234)+\Pi_{\mu_1\ldots \mu_4}^{(1,2,0)}(1234)+\Pi_{\mu_1\ldots \mu_4}^{(0,2,1)}(1234)+\Pi_{\mu_1\ldots \mu_4}^{(1,0,1)}(1234)\,,\label{eq:pi_Na_Nb_Nc_Nd}
\ea 
with 
\ba
&\Pi_{\mu_1\mu_2\mu_3\mu_4}^{(N_a,N_b,N_c)}(1234) = -(2\pi)^d\delta^d(k_1+\cdots+k_4) \frac{2g^4\mu^{8-2d}}{(4\pi)^{d/2}} \nonumber\\
&\times \int^1_0 d\tau_1 \cdots d\tau_4 \int^\infty_0\frac{d\varepsilon_0}{\varepsilon_0^{1+d/2}}\exp\bigg\{-\varepsilon_0m^2+\frac{\varepsilon_0}{2}\sum_{i,j=1}^4 k_i\cdot k_j G_B^{ij}\bigg\} I_{\mu_1\cdots \mu_4}^{(N_a,N_b,N_c)}\,,\label{eq:pi_mu1_muN_Na_Nb_Nc}
\ea
and
\ba
&I_{\mu_1\mu_2\mu_3\mu_4}^{(N_a,N_b,N_c)}= \frac{(-1)^{N_c}}{N_a!N_b!N_c!N_c!} \epsilon_{i_1j_1\ldots i_{N_a}j_{N_a}m_1\ldots m_{N_b}p_1q_1\ldots p_{N_c}q_{N_c}}\epsilon_{i_1j_1\ldots i_{N_a}j_{N_a}l_1\ldots l_{N_b}r_1s_1\ldots r_{N_c}s_{N_c}}\nonumber\\
&\times \bigg\{\prod_{\alpha=1}^{N_a} A_{i_\alpha j_\alpha}\bigg\}\bigg\{\prod_{\alpha=1}^{N_b} B_{m_\alpha l_\alpha}\bigg\}\bigg\{\prod_{\alpha=1}^{N_c} C_{p_\alpha q_\alpha}\bigg\}\bigg\{\prod_{\alpha=1}^{N_c} D_{r_\alpha s_\alpha}\bigg\}\,.\label{eq:I_Na_Nb_Nc_Nd}
\ea 
We proceed now to the evaluation of the 4-rank tensor in the worldline formalism term-by-term.
\newline 

\underline{\textit{Term $(2,0,0)$.}}
Setting $N_a=2$, $N_b=0$ and $N_c=0$ in Eq.~\eqref{eq:I_Na_Nb_Nc_Nd} gives
\ba
\label{eq:i_2_0_0_def}
&I^{(2,0,0)}_{\mu_1\mu_2\mu_3\mu_4}=\frac{1}{2!}\sum_{i_1,j_1=1}^4\sum_{i_2,j_2=1}^4 \epsilon_{i_1j_1i_2j_2} \epsilon_{i_1j_1i_2j_2}A_{i_1j_1}A_{i_2j_2}\,.
\ea
This sum contains 24 terms. Owing to the symmetry of the coefficients $A_{ij}=A_{ji}$ (see Eqs.~\eqref{eq:ABCD_def} and ~\eqref{eq:g_b_ij_g_f_ij_derivatives}), and the invariance under simultaneous permutations of $i_1\leftrightarrow i_2$ and $j_1\leftrightarrow j_2$, not all terms are distinct. We therefore split the sum in the cases
$i<j$, $i=j$ and $i>j$ for each pair $(i_1,j_i)$ and $(i_2,j_2)$. Noting that $A_{ii}=0$, we obtain
\ba
&I^{(2,0,0)}_{\mu_1\mu_2\mu_3\mu_4}= \frac{1}{2!}\sum_{i_1,j_1=1}^4\sum_{i_2,j_2=1}^4 \epsilon_{i_1j_1i_2j_2} \epsilon_{i_1j_1i_2j_2}A_{i_1j_1}A_{i_2j_2}=2\sum_{i_1<j_1}\sum_{i_2<j_2} \epsilon_{i_1j_1i_2j_2}\epsilon_{i_1j_1i_2j_2} A_{i_1j_1}A_{i_2j_2}\,.
\ea 
Next, we use the symmetry under the simultaneous permutations $i_1\leftrightarrow i_2$ and $j_1\leftrightarrow j_2$ to further rewrite
\ba
I^{(2,0,0)}_{\mu_1\mu_2\mu_3\mu_4}= 4A_{12}A_{34}+4A_{13}A_{24}+4A_{14}A_{23}\,.
\ea
This reduces the sum from $24$ terms down to $3$. Substituting now the form of $A_{ij}$ from Eq.~\eqref{eq:ABCD_def} into this expression one obtains
\ba 
I_{\mu_1\mu_2\mu_3\mu_4}^{(2,0,0)}
=\varepsilon_0^2 \Big\{\delta_{\mu_1\mu_2}\delta_{\mu_3\mu_4}\ddot{G}^B_{12}\ddot{G}^B_{34}+\delta_{\mu_1\mu_3}\delta_{\mu_2\mu_4}\ddot{G}^B_{13}\ddot{G}^B_{24}+\delta_{\mu_1\mu_4}\delta_{\mu_2\mu_3}\ddot{G}^B_{14}\ddot{G}^B_{23}\Big\}\,,
\ea 
These therefore correspond to tail contributions to the expression for the fourth-rank polarization tensor in Eq.~\eqref{eq:Pi_mu1_mu2_mu3_mu4_result}. Plugging this back into Eq.~\eqref{eq:pi_Na_Nb_Nc_Nd}, one obtains 
\ba 
&\Pi^{(2,0,0)}_{\mu_1\mu_2\mu_3\mu_4}(1234)= (2\pi)^d\delta(k_1+k_2+k_3+k_4) \nonumber\\
&\times\bigg\{\eta_{\mu_1\mu_2}\eta_{\mu_3\mu_4}f_{1}(1234)+\eta_{\mu_1\mu_3}\eta_{\mu_2\mu_4}f_2(1234)+\eta_{\mu_1\mu_4}\eta_{\mu_2\mu_3}f_3(1234)\bigg\}\,,
\ea
with 
\ba 
&f_{1}(1234) = -2\frac{g^4\mu^{8-2d}}{(4\pi)^{d/2}}\Gamma\left(2-\frac{d}{2}\right)\int_0^1 d\tau_1d\tau_2d\tau_3d\tau_4 \big[D(1234;\tau_1,\ldots,\tau_4)\big]^{\frac{d}{2}-2}\ddot{G}^B_{12}\ddot{G}^B_{34}\,.
\label{eq:F_1_1234}\\
&f_{2}(1234) = -2\frac{g^4\mu^{8-2d}}{(4\pi)^{d/2}} \Gamma\left(2-\frac{d}{2}\right)\int_0^1 d\tau_1d\tau_2d\tau_3d\tau_4 \big[D(1234;\tau_1,\ldots,\tau_4)\big]^{\frac{d}{2}-2}\ddot{G}^B_{12}\ddot{G}^B_{34}\,.
\label{eq:F_2_1234}\\
&f_{3}(1234) = -2\frac{g^4\mu^{8-2d}}{(4\pi)^{d/2}}\Gamma\left(2-\frac{d}{2}\right)\int_0^1 d\tau_1d\tau_2d\tau_3d\tau_4\big[D(1234;\tau_1,\ldots,\tau_4)\big]^{\frac{d}{2}-2} \ddot{G}^B_{12}\ddot{G}^B_{34}\,.
\label{eq:F_3_1234}
\ea 
Here we have carried out the trivial $\varepsilon_0$-integration, thereby recovering the polynomial denominator in Eq.~\eqref{eq:4_denominator}.
\newline 

\underline{\textit{Term} $(0,4,0)$:}
Setting $N_a=0$, $N_b=4$ and $N_c=0$ in Eq.~\eqref{eq:I_Na_Nb_Nc_Nd} produces
\ba
&I^{(0,4,0)}_{\mu_1\mu_2\mu_3\mu_4}= \frac{1}{4!}\sum_{m_1,m_2,m_3,m_4=1}^4\sum_{l_1,l_2,l_3,l_4=1}^4 \epsilon_{m_1m_2m_3m_4} \epsilon_{l_1l_2l_3l_4}B_{m_1l_1}B_{m_2l_2}B_{m_3l_3}B_{m_4l_4} 
\ea 
The above sum contains $24^2 = 576$ terms although not all of them are distinct. The sum is invariant under simultaneous permutations of $m_i\leftrightarrow m_j$ and $l_i\leftrightarrow l_j$. 
This suggests organizing the sum over the $4!$ possible orderings (for example $m_1 < m_2 < m_3 < m_4$, $m_2<m_1<m_3<m_4$, etc.) which makes it manifest that the resulting terms are equal to each other under the previous symmetry. Accordingly, 
\ba
I^{(0,4,0)}_{\mu_1\mu_2\mu_3\mu_4}=\sum_{l_1,l_2,l_3,l_4=1}^4 \epsilon_{l_1l_2l_3l_4}B_{1l_1}B_{2l_2}B_{3l_3}B_{4l_4}\,.
\ea 
This reduces the sum from $(4!)^2$ contributions to only $4!$. Substituting now the function $B_{ij}$ from Eq.~\eqref{eq:ABCD_def}, we obtain
\ba 
&I^{(0,4,0)}_{\mu_1\mu_2\mu_3\mu_4}=\varepsilon_0^4\sum_{\substack{\alpha_1=2,3,4\\\alpha_2=1,3,4}} \sum_{\substack{\alpha_3=1,2,4\\\alpha_4=1,2,3}} \big(k_{\alpha_1}\big)_{\mu_1}\big(k_{\alpha_2}\big)_{\mu_2}\big(k_{\alpha_3}\big)_{\mu_3}\big(k_{\alpha_4}\big)_{\mu_4} \sum_{l_1,l_2,l_3,l_4=1}^4 \epsilon_{l_1l_2l_3l_4}  \Big\{\delta_{1l_1}\dot{G}^B_{1\alpha_1}+\delta_{l_1\alpha_1}G^F_{\alpha_11}\Big\}\nonumber\\
&\times\Big\{\delta_{2l_2}\dot{G}^B_{2\alpha_2}+\delta_{l_2\alpha_2}G^F_{\alpha_22}\Big\}\Big\{\delta_{3l_3}\dot{G}^B_{3\alpha_3}+\delta_{l_3\alpha_3}G^F_{\alpha_33}\Big\}\Big\{\delta_{4l_4}\dot{G}^B_{4\alpha_4}+\delta_{l_4\alpha_4}G^F_{\alpha_44}\Big\}\,,
\ea 
where we have further restricted the sum over $\alpha_i$ indices, noting that for $\alpha_1=1$, $\alpha_2=2$, $\alpha_3=3$ and $\alpha_4=4$ the corresponding contributions vanish by virtue of $\dot{G}_B^{ii}=G_F^{ii}=0$, see Eqs.~\eqref{eq:g_b_ij_g_f_ij_derivatives}.

Accordingly, one can show that these are the only contributions to the head form factors to the rank-4 tensor as defined in Eq.~\eqref{eq:Pi_mu1_mu2_mu3_mu4_result}, and more generally the class of head contributions  identified for arbitrary $N$ in Eq.~\eqref{eq:H_i1_iN}. Plugging this contribution back into Eq.~\eqref{eq:pi_Na_Nb_Nc_Nd},
\ba 
\Pi^{(0,4,0)}_{\mu_1\mu_2\mu_3\mu_4}(1234) = (2\pi)^d \delta(k_1+k_2+k_3+k_4) \sum_{\substack{\alpha_1=2,3,4\\\alpha_2=1,3,4}} \sum_{\substack{\alpha_3=1,2,4\\\alpha_4=1,2,3}} \big(k_{\alpha_1}\big)_{\mu_1}\big(k_{\alpha_2}\big)_{\mu_2}\big(k_{\alpha_3}\big)_{\mu_3}\big(k_{\alpha_4}\big)_{\mu_4} H_{\alpha_1\alpha_2\alpha_3\alpha_4}(1234) \,,
\ea 
where the head form factor is given by
\ba 
&H_{\alpha_1\alpha_2\alpha_3\alpha_4}(1234) = -2\frac{g^4\mu^{8-2d}}{(4\pi)^{d/2}}\Gamma\left(4-\frac{d}{2}\right)\int^1_0d\tau_1d\tau_2d\tau_3d\tau_4 \big[D(1234;\tau_1,\ldots,\tau_4)\big]^{d/2-4}\nonumber\\
&\times\sum_{l_1,l_2,l_3,l_4=1}^4 \epsilon_{l_1l_2l_3l_4}\Big\{\dot{G}^B_{1\alpha_1}+\delta_{l_1\alpha_1}G^F_{\alpha_11}\Big\}\Big\{\delta_{2l_2}\dot{G}^B_{2\alpha_2}+\delta_{l_2\alpha_2}G^F_{\alpha_22}\Big\}\Big\{\delta_{3l_3}\dot{G}^B_{3\alpha_3}+\delta_{l_3\alpha_3}G^F_{\alpha_33}\Big\}\Big\{\delta_{4l_4}\dot{G}^B_{4\alpha_4}+\delta_{l_4\alpha_4}G^F_{\alpha_44}\Big\}\,,\label{eq:H_alpha1alpha2alpha3alpha4}
\ea 
and the polynomial denominator is given in Eq.~\eqref{eq:4_denominator}.
\newline

\underline{\textit{Term} (0,0,2):}
Setting $N_a=0$, $N_b=0$, and $N_c=2$ in  Eq.~\eqref{eq:I_Na_Nb_Nc_Nd} gives
\ba
&I^{(0,0,2)}_{\mu_1\mu_2\mu_3\mu_4} = \frac{1}{(2!)^2}\sum_{p_1,q_1=1}^4\sum_{p_2,q_2=1}^4\epsilon_{p_1q_1p_2q_2}C_{p_1q_1}C_{p_2q_2}\sum_{r_1,s_1=1}^4\sum_{r_2,s_2=1}^4 \epsilon_{r_1s_1r_2s_2}D_{r_1s_1}D_{r_2s_2}
\ea
The sum contains $(4!)^2 = 576 $ terms. Given that $C_{ij}=-C_{ji}$ and $D_{ij}=-D_{ji}$, as  follows from Eq.~\eqref{eq:ABCD_def}, we can exploit the invariance of the above expression under the exchanges $p_1\leftrightarrow q_1$, $p_2\leftrightarrow q_2$, $r_1\leftrightarrow s_1$ and $r_2\leftrightarrow s_2$ to reduce the number of distinct contributions from $576$ to $36$,
\ba
&I^{(0,0,2)}_{\mu_1\mu_2\mu_3\mu_4}= 4\sum_{p_1<q_1}\sum_{p_2<q_2} \epsilon_{p_1q_1p_2q_2}C_{p_1q_1}C_{p_2q_2}\sum_{r_1<s_1}\sum_{r_2<s_2} \epsilon_{r_1s_1r_2s_2}D_{r_1s_1}D_{r_2s_2}\,,
\ea 
and, using the invariance
under the simultaneous exchanges $p_1\leftrightarrow p_2$ and $q_1\leftrightarrow q_2$, or  $r_1\leftrightarrow r_2$ and $s_1\leftrightarrow s_2$, the remaining $36$ terms can be further reduced to $9$,
\ba
&I^{(0,0,2)}_{\mu_1\mu_2\mu_3\mu_4} = 
16(C_{12}C_{34}-C_{13}C_{24}-+C_{14}C_{23}) (D_{12}D_{34}-D_{13}D_{24}+D_{14}D_{23})\,.
\ea
Hence substituting the values for the functions $C_{ij}$ and $D_{ij}$ from Eq.~\eqref{eq:ABCD_def} one obtains,
\ba 
&I^{(0,0,2)}_{\mu_1\mu_2\mu_3\mu_4}= \varepsilon_0^4 \Big\{\eta_{\mu_1\mu_2}\eta_{\mu_3\mu_4}G^F_{12}G^F_{34}-\eta_{\mu_1\mu_3}\eta_{\mu_2\mu_4}G^F_{13}G^F_{24}+\eta_{\mu_1\mu_4}\eta_{\mu_2\mu_3}G^F_{14}G^F_{23}\Big\}\nonumber\\
&\times \Big\{(k_1\wc k_2)(k_3\wc k_4) G^F_{12}G^F_{34}-(k_1\wc k_3)(k_2\wc k_4)G^F_{13}G^F_{24}+(k_1\wc k_4)(k_2\wc k_3)G^F_{14}G^F_{23}\Big\}\,.
\ea 
These therefore corespond to tail type contributions to the rank-4 tensor. Plugging this back into Eq.~\eqref{eq:pi_Na_Nb_Nc_Nd} and performing the $\varepsilon_0$ integration straightforwardly produces
\ba 
&\Pi^{(0,0,2)}_{\mu_1\mu_2\mu_3\mu_4}(1234)= (2\pi)^d\delta^d(k_1+k_2+k_3+k_4)\nonumber\\
&\times\Big\{\eta_{\mu_1\mu_2}\eta_{\mu_3\mu_4}f_{4}(1234)+\eta_{\mu_1\mu_3}\eta_{\mu_2\mu_4}f_{5}(1234)+\eta_{\mu_1\mu_4}\eta_{\mu_2\mu_3}f_{6}(1234)\Big\}\,,
\ea 
with the corresponding form factors—after performing the $\varepsilon_0$-integration— given by
\ba 
f_4(1234) &= -2\frac{g^4\mu^{8-2d}}{(4\pi)^{d/2}}\Gamma\left(4-\frac{d}{2}\right)\int^1_0d\tau_1d\tau_2d\tau_3d\tau_4 \big[D(1234;\tau_1,\ldots,\tau_4)\big]^{\frac{d}{2}-4}G^F_{12}G^F_{34}\nonumber\\
&\times \Big\{(k_1\wc k_2)(k_3\wc k_4) G^F_{12}G^F_{34}-(k_1\wc k_3)(k_2\wc k_4)G^F_{13}G^F_{24}+(k_1\wc k_4)(k_2\wc k_3)G^F_{14}G^F_{23}\Big\}\,,\\
f_5(1234) &= 2\frac{g^4\mu^{8-2d}}{(4\pi)^{d/2}}\Gamma\left(4-\frac{d}{2}\right)\int^1_0d\tau_1d\tau_2d\tau_3d\tau_4 \big[D(1234;\tau_1,\ldots,\tau_4)\big]^{\frac{d}{2}-4}G^F_{13}G^F_{24}\nonumber\\
&\times \Big\{(k_1\wc k_2)(k_3\wc k_4) G^F_{12}G^F_{34}-(k_1\wc k_3)(k_2\wc k_4)G^F_{13}G^F_{24}+(k_1\wc k_4)(k_2\wc k_3)G^F_{14}G^F_{23}\Big\}\,,\\
f_6(1234) &= -2\frac{g^4\mu^{8-2d}}{(4\pi)^{d/2}}\Gamma\left(4-\frac{d}{2}\right)\int^1_0d\tau_1d\tau_2d\tau_3d\tau_4 \big[D(1234;\tau_1,\ldots,\tau_4)\big]^{\frac{d}{2}-4}G^F_{14}G^F_{23} \nonumber\\
&\times \Big\{(k_1\wc k_2)(k_3\wc k_4) G^F_{12}G^F_{34}-(k_1\wc k_3)(k_2\wc k_4)G^F_{13}G^F_{24}+(k_1\wc k_4)(k_2\wc k_3)G^F_{14}G^F_{23}\Big\}\,,
\ea 
and the polynomial denominator given in Eq.~\eqref{eq:4_denominator}.
\newline 

\underline{\textit{Term} $(1,2,0)$:}
Setting $N_a=1$, $N_b=2$ and $N_c=0$ in  Eq.~\eqref{eq:I_Na_Nb_Nc_Nd} gives
\ba 
&I^{(1,2,0)}_{\mu_1\mu_2\mu_3\mu_4} =\frac{1}{2!} \sum_{i_1,j_1=1}^4\sum_{m_1,m_2=1}^4\sum_{l_1,l_2=1}^4 \epsilon_{i_1j_1m_1m_2}\epsilon_{i_1j_1l_1l_2}A_{i_1j_1}B_{m_1l_1}B_{m_2l_2}\,.
\ea
The sum contains $48$ terms. Using the invariance of the above expression under the exchange $i_1\leftrightarrow j_1$, and under the simultaneous exchange  $m_1\leftrightarrow m_2$ and $l_1\leftrightarrow l_2$, one finds 
\ba
I^{(1,2,0)}_{\mu_1\mu_2\mu_3\mu_4} =2\sum_{i_1<j_1}\sum_{m_1<m_2}\sum_{l_1,l_2=1}^4\epsilon_{i_1j_1m_1m_2}\epsilon_{i_1j_1l_1l_2}A_{i_1j_1}B_{m_1l_1}B_{m_2l_2}\,.
\ea
reducing the $48$ terms in the sum to $6$. After substituting $A_{ij}$ and $B_{ij}$ using Eq.~\eqref{eq:ABCD_def}, we obtain
\ba 
I^{(1,2,0)}_{\mu_1\mu_2\mu_3\mu_4}= \varepsilon_0^3 \sum_{i_1<j_1}\sum_{m_1<m_2}\sum_{\alpha_1,\alpha_2=1}^4 \eta_{\mu_{i_1}\mu_{j_1}}\big(k_{\alpha_1}\big)_{\mu_{m_1}}\big(k_{\alpha_2}\big)_{\mu_{m_2}}  \epsilon_{i_1j_1m_1m_2} \epsilon_{i_1j_1l_1l_2}   \nonumber\\
\times\ddot{G}^B_{i_1j_1}\sum_{l_1,l_2=1}^4 \Big\{\delta_{m_1l_1}\dot{G}^B_{m_1\alpha_1}+\delta_{\alpha_1l_1}G^F_{\alpha_1m_1}\Big\}\Big\{\delta_{m_2l_2}\dot{G}^B_{m_2\alpha_2}+\delta_{\alpha_2l_2}G^F_{\alpha_2m_2}\Big\}\,.
\ea 
As before, the sums can be restricted to $\alpha_1\neq m_1$ and $\alpha_2\neq m_2$, since the corresponding contributions vanish due to $\dot{G}^B_{ii}=G^F_{ii}=0$, see Eqs.~\eqref{eq:g_b_ij_g_f_ij_derivatives}. These correspond to shoulder-type form factor contributions to the rank-4 tensor. Substituting this back into Eq.~\eqref{eq:pi_Na_Nb_Nc_Nd} and performing the $\varepsilon_0$ integration straightforwardly produces
\ba
&\Pi^{(1,2,0)}_{\mu_1\mu_2\mu_3\mu_4}(1234)=(2\pi)^d\delta^d(k_1+k_2+k_3+k_4)\nonumber\\
\times \Bigg\{ &\sum_{\substack{\alpha_3=1,2,4\\\alpha_4=1,2,3}}\delta_{\mu_1\mu_2} \big(k_{\alpha_3}\big)_{\mu_3} \big(k_{\alpha_4}\big)_{\mu_4} g_{1,\alpha_3\alpha_4}(1234)+\sum_{\substack{\alpha_2=1,3,4\\\alpha_4=1,2,3}}\delta_{\mu_1\mu_3} \big(k_{\alpha_2}\big)_{\mu_2} \big(k_{\alpha_4}\big)_{\mu_4} g_{2,\alpha_2\alpha_4}(1234)\nonumber\\
+& \sum_{\substack{\alpha_2=1,3,4\\\alpha_3=1,2,4}}\delta_{\mu_1\mu_4}\big(k_{\alpha_2}\big)_{\mu_2}\big(k_{\alpha_3}\big)_{\mu_3}g_{3,\alpha_2\alpha_3}(1234)+\sum_{\substack{\alpha_1=2,3,4\\\alpha_4=1,2,3}}\delta_{\mu_2\mu_3}\big(k_{\alpha_1}\big)_{\mu_1}\big(k_{\alpha_4}\big)_{\mu_4}g_{4,\alpha_1\alpha_4}(1234)\nonumber\\
+&\sum_{\substack{\alpha_1=2,3,4\\\alpha_3=1,2,4}}\delta_{\mu_2\mu_4}\big(k_{\alpha_1}\big)_{\mu_1}\big(k_{\alpha_3}\big)_{\mu_3} g_{5,\alpha_1\alpha_3}(1234)+\sum_{\substack{\alpha_1=2,3,4\\\alpha_2=1,3,4}}\delta_{\mu_3\mu_4}\big(k_{\alpha_1}\big)_{\mu_1}\big(k_{\alpha_2}\big)_{\mu_2}g_{6,\alpha_1\alpha_2}(1234) \Bigg\}\,,
\ea 
where
\ba
&g_{1,\alpha_3\alpha_4}(1234)=-2\frac{g^4\mu^{8-2d}}{(4\pi)^{d/2}}\Gamma\left(3-\frac{d}{2}\right)\int^1_0d\tau_1d\tau_2d\tau_3d\tau_4 \big[D(1234;\tau_1,\ldots,\tau_4)\big]^{\frac{d}{2}-3}\nonumber\\
&\times \ddot{G}^B_{12}\sum_{l_3,l_4=3,4}  \epsilon_{12l_3l_4} \Big\{\delta_{3l_3}\dot{G}^B_{3\alpha_3}+\delta_{\alpha_3l_3}G^F_{\alpha_33}\Big\}\Big\{\delta_{4l_4}\dot{G}^B_{4\alpha_4}+\delta_{\alpha_4l_4}G^F_{\alpha_4l_4}\Big\}\,,\nonumber\\
&g_{2,\alpha_2\alpha_4}(1234)=-2\frac{g^4\mu^{8-2d}}{(4\pi)^{d/2}}\Gamma\left(3-\frac{d}{2}\right)\int^1_0d\tau_1d\tau_2d\tau_3d\tau_4 \big[D(1234;\tau_1,\ldots,\tau_4)\big]^{\frac{d}{2}-3}\nonumber\\
&\times\ddot{G}^B_{13} \sum_{l_2,l_4=2,4}  \epsilon_{1l_23l_4} \Big\{\delta_{2l_2}\dot{G}^B_{2\alpha_2}+\delta_{\alpha_2l_2}G^F_{\alpha_22}\Big\}\Big\{\delta_{4l_4}\dot{G}^B_{4\alpha_4}+\delta_{\alpha_4l_4}G^F_{\alpha_4l_4}\Big\}\,,\\
&g_{3,\alpha_2\alpha_3}(1234)=-2\frac{g^4\mu^{8-2d}}{(4\pi)^{d/2}}\Gamma\left(3-\frac{d}{2}\right)\int^1_0d\tau_1d\tau_2d\tau_3d\tau_4 \big[D(1234;\tau_1,\ldots,\tau_4)\big]^{\frac{d}{2}-3}\nonumber\\
&\times\ddot{G}^B_{14} \sum_{l_2,l_3=2,3}  \epsilon_{1l_2l_34} \Big\{\delta_{2l_2}\dot{G}^B_{2\alpha_2}+\delta_{\alpha_2l_2}G^F_{\alpha_22}\Big\}\Big\{\delta_{3l_3}\dot{G}^B_{3\alpha_3}+\delta_{\alpha_3l_3}G^F_{\alpha_3l_3}\Big\}\,,\\
&g_{4,\alpha_1\alpha_4}(1234)=-2\frac{g^4\mu^{8-2d}}{(4\pi)^{d/2}}\Gamma\left(3-\frac{d}{2}\right)\int^1_0d\tau_1d\tau_2d\tau_3d\tau_4 \big[D(1234;\tau_1,\ldots,\tau_4)\big]^{\frac{d}{2}-3}\nonumber\\
&\times\ddot{G}^B_{23} \sum_{l_1,l_4=1,4}  \epsilon_{l_1 23l_4} \Big\{\delta_{1l_1}\dot{G}^B_{1\alpha_1}+\delta_{\alpha_1l_1}G^F_{\alpha_11}\Big\}\Big\{\delta_{4l_4}\dot{G}^B_{4\alpha_4}+\delta_{\alpha_4l_4}G^F_{\alpha_4l_4}\Big\}\,,\\
&g_{5,\alpha_1\alpha_3}(1234)=-2\frac{g^4\mu^{8-2d}}{(4\pi)^{d/2}}\Gamma\left(3-\frac{d}{2}\right)\int^1_0d\tau_1d\tau_2d\tau_3d\tau_4 \big[D(1234;\tau_1,\ldots,\tau_4)\big]^{\frac{d}{2}-3}\nonumber\\
&\times\ddot{G}^B_{24} \sum_{l_1,l_3=1,3}  \epsilon_{l_1 2l_34} \Big\{\delta_{1l_1}\dot{G}^B_{1\alpha_1}+\delta_{\alpha_1l_1}G^F_{\alpha_11}\Big\}\Big\{\delta_{3l_3}\dot{G}^B_{3\alpha_3}+\delta_{\alpha_3l_3}G^F_{\alpha_3l_3}\Big\}\,,\\
&g_{6,\alpha_1\alpha_2}(1234)=-2\frac{g^4\mu^{8-2d}}{(4\pi)^{d/2}}\Gamma\left(3-\frac{d}{2}\right)\int^1_0d\tau_1d\tau_2d\tau_3d\tau_4 \big[D(1234;\tau_1,\ldots,\tau_4)\big]^{\frac{d}{2}-3}\nonumber\\
&\times\ddot{G}^B_{34} \sum_{l_1,l_2=1,2}  \epsilon_{l_1 l_234} \Big\{\delta_{1l_1}\dot{G}^B_{1\alpha_1}+\delta_{\alpha_1l_1}G^F_{\alpha_11}\Big\}\Big\{\delta_{2l_2}\dot{G}^B_{2\alpha_2}+\delta_{\alpha_2l_2}G^F_{\alpha_2l_2}\Big\}\,,
\ea 
with the polynomial denominator the one in Eq.~\eqref{eq:4_denominator}.
\newline 

\underline{\textit{Term} $(0,2,1)$:}
Setting $N_a=0$, $N_b=2$ and $N_c=1$ in  Eq.~\eqref{eq:I_Na_Nb_Nc_Nd} produces 
\ba 
&I^{(0,2,1)}_{\mu_1\mu_2\mu_3\mu_4}  = -\frac{1}{2}\sum_{m_1,m_2=1}^4 \sum_{p_1,q_1=1}^4\sum_{r_1,s_1=1}^4 \epsilon_{k_1k_2p_1q_1}\epsilon_{l_1l_2r_1s_1} B_{m_1m_1}B_{k_2l_2}C_{p_1q_1}D_{r_1s_1}\,.
&
\ea 
The sum contains $576$ terms, and can be reduced to $144$ by applying the same symmetry considerations as in the previous case,
\ba 
I^{(0,2,1)}_{\mu_1\mu_2\mu_3\mu_4}= -4\sum_{m_1<m_2}\sum_{p_1<q_1}\sum_{r_1<s_1} \epsilon_{m_1m_2p_1q_1}\epsilon_{l_1l_2r_1s_1} B_{m_1l_1}B_{m_2l_2}C_{p_1q_1}D_{r_1s_1} \,.
\ea 
Replacing now the coefficient functions $B_{ij}$, $C_{ij}$ and $D_{ij}$ in Eq.~\eqref{eq:ABCD_def} yields,
\ba 
&I^{(0,2,1)}_{\mu_1\mu_2\mu_3\mu_4} = -\varepsilon_0^4\sum_{m_1<m_2}\sum_{p_1<q_1}\sum_{r_1<s_1}  \eta_{\mu_{p_1}\mu_{q_1}}\big(k_{\alpha_1}\big)_{\mu_{m_1}}\big(k_{\alpha_2}\big)_{\mu_{m_2}}  \epsilon_{m_1m_2p_1q_1}G^F_{p_1q_1}\nonumber\\
&\times \sum_{r_1<s_1}\sum_{l_1,l_2=1}^4 \epsilon_{l_1l_2r_1s_1}\Big\{\delta_{m_1l_1}\dot{G}^B_{m_1\alpha_1}+\delta_{l_1\alpha_1}G^F_{\alpha_1m_1}\Big\}\Big\{\delta_{m_2l_2}\dot{G}^B_{m_2\alpha_2}+\delta_{l_2\alpha_2}G^F_{\alpha_2m_2}\Big\} (k_{r_1}\wc k_{s_1})G^F_{r_1s_1}
\ea
These correspond, as in the previous case, to shoulder-type form factor contributions to the rank-4 tensor. Substituting this back into Eq.~\eqref{eq:pi_Na_Nb_Nc_Nd} and performing the $\varepsilon_0$ integration straightforwardly produces
\ba 
&\Pi^{(0,2,1)}_{\mu_1\mu_2\mu_3\mu_4}(1234)=(2\pi)^d\delta^d(k_1+k_2+k_3+k_4)\nonumber\\
\times \bigg\{ &\sum_{\substack{\alpha_3=1,2,4\\\alpha_4=1,2,3}}\delta_{\mu_1\mu_2} \big(k_{\alpha_3}\big)_{\mu_3} \big(k_{\alpha_4}\big)_{\mu_4} g_{7,\alpha_3\alpha_4}(1234)+\sum_{\substack{\alpha_2=1,3,4\\\alpha_4=1,2,3}}\delta_{\mu_1\mu_3} \big(k_{\alpha_2}\big)_{\mu_2} \big(k_{\alpha_4}\big)_{\mu_4} g_{8,\alpha_2\alpha_4}(1234)\nonumber\\
+& \sum_{\substack{\alpha_2=1,3,4\\\alpha_3=1,2,4}}\delta_{\mu_1\mu_4}\big(k_{\alpha_2}\big)_{\mu_2}\big(k_{\alpha_3}\big)_{\mu_3}g_{9,\alpha_2\alpha_3}(1234)+\sum_{\substack{\alpha_1=2,3,4\\\alpha_4=1,2,3}}\delta_{\mu_2\mu_3}\big(k_{\alpha_1}\big)_{\mu_1}\big(k_{\alpha_4}\big)_{\mu_4} g_{10,\alpha_1\alpha_4}(1234)\nonumber\\
+&\sum_{\substack{\alpha_1=2,3,4\\\alpha_3=1,2,4}}\delta_{\mu_2\mu_4}\big(k_{\alpha_1}\big)_{\mu_1}\big(k_{\alpha_3}\big)_{\mu_3}g_{11,\alpha_1\alpha_3}(1234)+\sum_{\substack{\alpha_1=2,3,4\\\alpha_2=1,3,4}}\delta_{\mu_3\mu_4}\big(k_{\alpha_1}\big)_{\mu_1}\big(k_{\alpha_2}\big)_{\mu_2}g_{12,\alpha_1\alpha_2}(1234) \bigg\}\,,
\ea 
where
\ba
&g_{7,\alpha_3\alpha_4}(1234)=-2\frac{g^4\mu^{8-2d}}{(4\pi)^{d/2}}\Gamma\left(4-\frac{d}{2}\right)\int^1_0d\tau_1d\tau_2d\tau_3d\tau_4 \big[D(1234;\tau_1,\ldots,\tau_4)\big]^{\frac{d}{2}-4}\nonumber\\
&\times G^F_{12}\sum_{l_3,l_4=1}^4\sum_{r_1<s_1} \epsilon_{l_3l_4r_1s_1} (k_{r_1}\wc k_{s_1}) \Big\{\delta_{3l_3}\dot{G}^B_{3\alpha_3}+\delta_{\alpha_3l_3}G^F_{\alpha_33}\Big\}\Big\{\delta_{4l_4}\dot{G}^B_{4\alpha_4}+\delta_{\alpha_4l_4}G^F_{\alpha_4l_4}\Big\}\,,\nonumber\\
&g_{8,\alpha_2\alpha_4}(1234)=+2\frac{g^4\mu^{8-2d}}{(4\pi)^{d/2}}\Gamma\left(4-\frac{d}{2}\right)\int^1_0d\tau_1d\tau_2d\tau_3d\tau_4 \big[D(1234;\tau_1,\ldots,\tau_4)\big]^{\frac{d}{2}-4}\nonumber\\
&\times G^F_{13} \sum_{l_2,l_4=1}^4\sum_{r_1<s_1} \epsilon_{l_2l_4r_1s_1} (k_{r_1}\wc k_{s_1})   \epsilon_{1l_23l_4} \Big\{\delta_{2l_2}\dot{G}^B_{2\alpha_2}+\delta_{\alpha_2l_2}G^F_{\alpha_22}\Big\}\Big\{\delta_{4l_4}\dot{G}^B_{4\alpha_4}+\delta_{\alpha_4l_4}G^F_{\alpha_4l_4}\Big\}\,,\\
&g_{9,\alpha_2\alpha_3}(1234)=-2\frac{g^4\mu^{8-2d}}{(4\pi)^{d/2}}\Gamma\left(4-\frac{d}{2}\right)\int^1_0d\tau_1d\tau_2d\tau_3d\tau_4 \big[D(1234;\tau_1,\ldots,\tau_4)\big]^{\frac{d}{2}-4}\nonumber\\
&\times G^F_{14} \sum_{l_2,l_3=1}^4 \sum_{r_1<s_1} \epsilon_{l_2l_3r_1s_1} (k_{r_1}\wc k_{s_1})  \Big\{\delta_{2l_2}\dot{G}^B_{2\alpha_2}+\delta_{\alpha_2l_2}G^F_{\alpha_22}\Big\}\Big\{\delta_{3l_3}\dot{G}^B_{3\alpha_3}+\delta_{\alpha_3l_3}G^F_{\alpha_3l_3}\Big\}\,,\\
&g_{10,\alpha_1\alpha_4}(1234)=+2\frac{g^4\mu^{8-2d}}{(4\pi)^{d/2}}\Gamma\left(4-\frac{d}{2}\right)\int^1_0d\tau_1d\tau_2d\tau_3d\tau_4 \big[D(1234;\tau_1,\ldots,\tau_4)\big]^{\frac{d}{2}-4}\nonumber\\
&\times G^F_{23}\sum_{l_1,l_4=1}^4 \sum_{r_1<s_1} \epsilon_{l_1l_4r_1s_1} (k_{r_1}\wc k_{s_1})  \Big\{\delta_{1l_1}\dot{G}^B_{1\alpha_1}+\delta_{\alpha_1l_1}G^F_{\alpha_11}\Big\}\Big\{\delta_{4l_4}\dot{G}^B_{4\alpha_4}+\delta_{\alpha_4l_4}G^F_{\alpha_4l_4}\Big\}\,,\\
&g_{11,\alpha_1\alpha_3}(1234)=-2\frac{g^4\mu^{8-2d}}{(4\pi)^{d/2}}\Gamma\left(4-\frac{d}{2}\right)\int^1_0d\tau_1d\tau_2d\tau_3d\tau_4 \big[D(1234;\tau_1,\ldots,\tau_4)\big]^{\frac{d}{2}-4}\nonumber\\
&\times G^F_{24} \sum_{l_1,l_3=1}^4  \sum_{r_1<s_1} \epsilon_{l_1l_3r_1s_1} (k_{r_1}\wc k_{s_1})  \Big\{\delta_{1l_1}\dot{G}^B_{1\alpha_1}+\delta_{\alpha_1l_1}G^F_{\alpha_11}\Big\}\Big\{\delta_{3l_3}\dot{G}^B_{3\alpha_3}+\delta_{\alpha_3l_3}G^F_{\alpha_3l_3}\Big\}\,,\\
&g_{12,\alpha_1\alpha_2}(1234)=+2\frac{g^4\mu^{8-2d}}{(4\pi)^{d/2}}\Gamma\left(4-\frac{d}{2}\right)\int^1_0d\tau_1d\tau_2d\tau_3d\tau_4 \big[D(1234;\tau_1,\ldots,\tau_4)\big]^{\frac{d}{2}-4}\nonumber\\
&\times G^F_{34} \sum_{l_1,l_2=1}^4  \sum_{r_1<s_1} \epsilon_{l_1l_2r_1s_1} (k_{r_1}\wc k_{s_1})  \Big\{\delta_{1l_1}\dot{G}^B_{1\alpha_1}+\delta_{\alpha_1l_1}G^F_{\alpha_11}\Big\}\Big\{\delta_{2l_2}\dot{G}^B_{2\alpha_2}+\delta_{\alpha_2l_2}G^F_{\alpha_2l_2}\Big\}\,,
\ea 
and the denominator given by Eq.~\eqref{eq:4_denominator}.
\newline 

\underline{\textit{Term} $(1,0,1)$:}
Setting $N_a=1$, $N_b=0$, and $N_c=1$ in Eq.~\eqref{eq:I_Na_Nb_Nc_Nd} and using the same symmetry considerations to reduce the number of terms produces
\ba 
&I^{(1,0,1)}_{\mu_1\mu_2\mu_3\mu_4}= -\sum_{i_1,j_1=1}^4 \epsilon_{i_1j_1p_1q_1}\epsilon_{i_1j_1r_1s_1}A_{i_1j_1}C_{p_1q_1}D_{r_1s_1}= -8 \sum_{i_1<j_1}\sum_{p_1<q_1}\sum_{r_1<s_1} \epsilon_{i_1j_1p_1q_1}\epsilon_{i_1j_1r_1s_1}A_{i_1j_1}C_{p_1q_1}D_{r_1s_1}
\ea 
Substituting now the coefficient functions $A_{ij}$, $C_{ij}$ and $D_{ij}$ using Eq.~\eqref{eq:ABCD_def}, and noticing $\big(G^F_{ij}\big)^2=1$, this leads to
\ba 
&I^{(1,0,1)}_{\mu_1\mu_2\mu_3\mu_4}= -\varepsilon_0^3 \bigg\{\eta_{\mu_1\mu_2}\eta_{\mu_3 \mu_4}\Big( k_3\wc k_4 \ddot{G}^B_{12} +k_1\wc k_2 \ddot{G}^B_{34} \Big)\nonumber\\
&+\eta_{\mu_1\mu_3}\eta_{\mu_2\mu_4}\Big(k_2\wc k_4\ \ddot{G}^B_{13}+k_1\wc k_3\ddot{G}^B_{24}\Big)+\eta_{\mu_1\mu_4}\eta_{\mu_2\mu_3}\Big(k_2\wc k_3 \ddot{G}^B_{14}+k_2\wc k_3 \ddot{G}^B_{23}\Big)\bigg\}
\ea 
These are tail-type form factor contributions to the tensor. 
After substituting back this expression into Eq.~\eqref{eq:pi_Na_Nb_Nc_Nd} and performing the $\varepsilon_0$ integration one obtains
\ba 
&\Pi_{\mu_1\mu_2\mu_3\mu_4}^{(1,0,1)}(1234) = (2\pi)^d\delta^d(k_1+k_2+k_3+k_4)\nonumber\\
&\times\Big\{ \eta_{\mu_1\mu_2}\eta_{\mu_3\mu_4} f_7(1234)+\eta_{\mu_1\mu_3}\eta_{\mu_2\mu_4}f_8(1234)+ \eta_{\mu_1\mu_4}\eta_{\mu_2\mu_3}f_9(1234)\Big\}\,,
\ea 
with
\ba 
&f_7(1234)= 2\frac{g^4\mu^{8-2d}}{(4\pi)^{d/2}}\Gamma\left(3-\frac{d}{2}\right)\int d\tau_1d\tau_2d\tau_3d\tau_4 \big[D(1234;\tau_1,\ldots,\tau_4)\big]^{\frac{d}{2}-3} \Big(\ddot{G}^B_{12} k_{3}\wc k_{4}+\ddot{G}_{34}^B k_{1}\wc k_{2}\Big)\,,\\
&f_8(1234)= 2\frac{g^4\mu^{8-2d}}{(4\pi)^{d/2}}\Gamma\left(3-\frac{d}{2}\right)\int d\tau_1d\tau_2d\tau_3d\tau_4 \big[D(1234;\tau_1,\ldots,\tau_4)\big]^{\frac{d}{2}-3} \Big(\ddot{G}^B_{13} k_{2}\wc k_{4}+\ddot{G}_{24}^B k_{1}\wc k_{3}\Big)\,,\\
&f_9(1234)= 2\frac{g^4\mu^{8-2d}}{(4\pi)^{d/2}}\Gamma\left(3-\frac{d}{2}\right)\int d\tau_1d\tau_2d\tau_3d\tau_4 \big[D(1234;\tau_1,\ldots,\tau_4)\big]^{\frac{d}{2}-3} \Big(\ddot{G}^B_{14} k_{2}\wc k_{3}+\ddot{G}_{23}^B k_{1}\wc k_{4}\Big)\,.
\ea 
\newline 

\underline{\textit{Final result:}} We can now collect the resulting form factors according to their tail, shoulder, and head classification, as in Eq.~\eqref{eq:Pi_mu1_mu2_mu3_mu4_result}, as follows:
\ba
&F_1(1234)=f_1(1234)+f_4(1234)+f_7(1234)\,,\nonumber\\ &F_2(1234)=f_2(1234)+f_5(1234)+f_8(1234)\,,\\
&F_3(1234)=f_3(1234)+f_6(1234)+f_9(1234)\,,
\ea 
and
\ba
&G_{1,\alpha_3\alpha_4}(1234)=g_{1,\alpha_3\alpha_4}(1234)+g_{7,\alpha_3\alpha_4}(1234)\,,\,\,\,G_{2,\alpha_2\alpha_4}(1234)=g_{2,\alpha_2\alpha_4}(1234)+g_{8,\alpha_2\alpha_4}(1234)\,,\nonumber\\
&G_{3,\alpha_2\alpha_3}(1234)=g_{3,\alpha_2\alpha_3}(1234)+g_{9,\alpha_2\alpha_3}(1234)\,,\,\,\, G_{4,\alpha_1\alpha_4}(1234)=g_{4,\alpha_1\alpha_4}(1234)+g_{10,\alpha_1\alpha_4}(1234)\,,\\
&G_{5,\alpha_1\alpha_3}(1234)=g_{5,\alpha_1\alpha_3}(1234)+g_{11,\alpha_1\alpha_3}(1234)\,,\,\,\, G_{6,\alpha_1\alpha_2}(1234)=g_{6,\alpha_1\alpha_2}(1234)+g_{12,\alpha_1\alpha_2}(1234)\,.
\ea 
Note that the head form factor  $H_{\alpha_1\alpha_2\alpha_3\alpha_4}(1234)$ arises uniquely from the $N_b=4$ contribution and is given therefore directly by Eq.~\eqref{eq:H_alpha1alpha2alpha3alpha4}, as stated in Section \ref{sec:nthrankpoltensors}.

\section{Transverse tensor form decomposition into heads using current conservation.}
\label{app:transversedecomposition}
In this Appendix, we will streamline the procedure discussed in the main text for the transverse decomposition of the tensor with respect to the Ward identities in Eqs.~\eqref{eq:ward_1}-\eqref{eq:ward_4}. We will also provide the complete list of solutions expressing the tails and shoulders in terms of the heads, following the systematic approach oultlined in Sec. \ref{subsec:N4}. From Eq.\eqref{eq:ward_1}, collecting the terms that do not involve metric-tensor structures, one finds the additional relations,
\ba 
(3)_2(1)_3(1)_4\times \bigg[G_{2,31}(1234)+G_{3,31}(1234)+\sum_{i_1\neq 1} (1\wc i_1) H_{i_1311}(1234)\bigg]&=0\,,\label{eq:G_ward1_5}\\
(3)_2(1)_3(i_4)_4 \times\bigg[G_{2,3i_4}(1234)+\sum_{i_1\neq 1} (1\wc i_1)H_{i_131i_4}(1234)\bigg]&=0\,,\,\,\,i_4=2,3\,,\label{eq:G_ward1_6}\\
(3)_2(i_3)_3(1)_4\times \bigg[G_{3,3i_3}(1234)+\sum_{i_1\neq 1} (1\wc i_1) H_{i_13i_31}(1234)\bigg]&=0\,,\,\,\,i_3=2,4\,,\label{eq:G_ward1_7}\\
(3)_2(i_3)_3(i_4)_4\times \bigg[\sum_{i_1\neq 1}(1\wc i_1)H_{i_13i_3i_4}(1234)\bigg]&=0\,,\,\,\,i_3=2,4\,,\,\,\,i_4=2,3\,.\label{eq:G_ward1_8}
\ea 
and
\ba 
(4)_2(1)_3(1)_4\times\bigg[G_{2,41}(1234)+G_{3,41}(1234)+\sum_{i_1\neq 1} (1\wc i_1) H_{i_1411}(1234)\bigg]&=0\,,\label{eq:G_ward1_9}\\
(4)_2(i_3)_3(1)_4\times\bigg[G_{3,4i_3}(1234)+\sum_{i_1\neq 1} (1\wc i_1) H_{i_14i_31}(1234)\bigg]&=0\,,\,\,\, i_3=2,4\,,\label{eq:G_ward1_10}\\
(4)_2(1)_3(i_4)_4\times\bigg[G_{2,4i_4}(1234)+\sum_{i_1\neq 1} (1\wc i_1)H_{i_141i_4}(1234)\bigg]&=0\,,\,\,\, i_4=2,3\,.\label{eq:G_ward1_11}\\
(4)_2(i_3)_3(i_4)_4\times\bigg[\sum_{i_1\neq 1}(1\wc i_1) H_{i_14i_3i_4}(1234)\bigg]&=0\,,\,\,\, i_3=2,4\,,\,\,\, i_4=2,3\,.\label{eq:G_ward1_12}
\ea 

The same procedure can be straightforwardly extended to obtain analogous relations from the three other Ward identities in Eqs.\eqref{eq:ward_2}–\eqref{eq:ward_4}. By successive elimination, the tails can then be expressed in terms of the shoulders, and the shoulders in terms of the heads. Not all identities among heads, shoulders, and tails obtained in this way are independent. Moreover, the reduction is not unique: different choices lead to alternative but equivalent representations of the fourth-rank tensor, related by identities among the head form factors implied by the Ward identities. These later relations can be further used to cast the tensor in a more compact form, as done by Karplus and Neuman \cite{Karplus:1950zza}. However, this complicates the analysis of the pole structures when the tensor appears within a large amplitude, in kinematical regimes of relevance to discuss non-perturbative physics, such as the Regge \cite{Dolen:1965gfx} or Bjorken limits \cite{Tarasov:2019rfp}. 

For example, the fourth Ward identity in Eq.\eqref{eq:ward_4} yields relations such as
\ba 
(2)_1(1)_2(2)_3\Big[(4\wc 1) H_{2121}(1234)+(4\cdot 2)H_{2122}(1234)+(4\wc 3)H_{2123}(1234)\Big]=0\,
\ea 
which, after substituting $H_{2122}(1234)=H_{2111}(2134)$ using the permutation group symmetries of the equivalence class in Eq.\eqref{eq:class_1_representative}, proves—as conjectured—the vanishing of Eq.(44a) in Karplus and Neuman \cite{Karplus:1950zza}. Analogously, 
\ba 
(2)_1(3)_2(1)_3\Big[(4\wc 1)H_{2311}(1234)+(4\wc 2)H_{2312}(1234)+(4\wc 3)H_{2313}(1234)\Big]=0\,.
\ea 
which, after substituting $H_{2313}(1234)=H_{2311}(3124)$ and $H_{2312}(1234)=H_{2311}(2314)$ from Eq.~\eqref{eq:class_4_representative}, proves the vanishing of Eq.~(44b) in their paper.

As observed in \cite{Dolen:1965gfx}, the use of these additional relations, although formally valid, introduces collinear singularities which are spurious; the corresponding scalar products of photon momenta $k_i\!\cdot\! k_j$, $(i,j=1,2,3,4)$ give singular expressions in the amplitude, but cancel in the complete expression for the polarization tensor. This quantity is of course infrared safe in QED (for arbitrary $N$) precisely due to the current conservation implied by Eqs.~\eqref{eq:ward_1}-\eqref{eq:ward_4}.

For completeness, we provide below the full list of solutions. The shoulder form factors $G_{1, i_3 i_4}(1234)$ with $i_3\neq 3$ and $i_4\neq 4$ can be expressed in terms of the head form factors as
\ba 
G_{1,11}(1234)&= -\sum_{ i_2\neq 2} (2\wc i_2)H_{2 i_211}(1234)\,,\label{eq:sol_g1_11}\\
G_{1,12}(1234)&=-\sum_{ i_1\neq 1}(1\wc i_1)H_{ i_1112}(1234)+\sum_{ i_3\neq 3}(3\wc i_3) H_{31 i_32}(1234)\,,\label{eq:sol_g1_12}\\
G_{1,13}(1234)&=-\sum_{ i_2\neq 2} (2\wc i_2)H_{2 i_213}(1234)\,,\label{eq:sol_g1_13}\\
G_{1,21}(1234)&=-\sum_{ i_1\neq 1} (1\wc i_1)H_{ i_1121}(1234)+\sum_{ i_4\neq 4}(4\wc i_4)H_{412 i_4}(1234)\,,\label{eq:sol_g1_21}\\
G_{1,22}(1234)&=-\sum_{ i_1\neq 1}(1\wc i_1)H_{ i_1122}(1234)\,,\label{eq:sol_g1_22}\\
G_{1,23}(1234)&=-\sum_{ i_1\neq 1} (1\wc i_1) H_{ i_1123}(1234)\,,\label{eq:sol_g1_23}\\
G_{1,41}(1234)&=-\sum_{ i_2\neq 2} (2\wc i_2)H_{2 i_241}(1234)\,,\label{eq:sol_g1_41}\\
G_{1,42}(1234)&=-\sum_{ i_1\neq 1}(1\wc i_1) H_{ i_1142}(1234)\,,\label{eq:sol_g1_42}\\
G_{1,43}(1234)&=-\sum_{ i_2\neq 2} (2\wc i_2)H_{2 i_243}(1234)\,,\label{eq:sol_g1_43}
\ea 
Analogously, the solutions for the shoulder form factors $G_{2, i_2 i_4}(1234)$ with $i_2\neq 2$ and $i_4\neq 4$ are:
\ba 
G_{2,11}(1234)&=-\sum_{ i_3\neq 3} (3\wc  i_3) H_{31 i_31}(1234)\,,\label{eq:sol_g2_11}\\
G_{2,12}(1234)&=-\sum_{ i_3\neq 3}(3\wc i_3) H_{31 i_32}(1234)\,,\label{eq:sol_g2_12}\\
G_{2,13}(1234)&=-\sum_{ i_3\neq 3}(3\wc i_3) H_{31 i_33}(1234)+\sum_{ i_4\neq 4} (4\wc i_4)H_{314 i_4}(1234)\,,\label{eq:sol_g2_13}\\
G_{2,31}(1234)&=-\sum_{ i_1\neq 1} (1\wc i_1) H_{ i_1311}(1234)+\sum_{ i_4\neq 4} (4\wc i_4) H_{431 i_4}(1234)\,,\label{eq:sol_g2_31}\\
G_{2,32}(1234)&=-\sum_{ i_1\neq 1}(1\wc i_1)H_{ i_1312}(1234)\,,\label{eq:sol_g2_32}\\
G_{2,33}(1234)&=-\sum_{ i_1\neq 1}(1\wc i_1) H_{ i_1313}(1234)\,,\label{eq:sol_g2_33}\\
G_{2,41}(1234)&=-\sum_{ i_3\neq 3} (3\wc i_3) H_{34 i_31}(1234)\,,\label{eq:sol_g2_41}\\
G_{2,42}(1234)&=-\sum_{ i_1\neq 1} (1\wc i_1) H_{ i_1412}(1234)\,,\label{eq:sol_g2_42}\\
G_{2,43}(1234)&=-\sum_{ i_1\neq 1}(1\wc i_1) H_{ i_1413}(1234)\,.\label{eq:sol_g2_43}
\ea 
Solutions for $G_{3, i_2 i_3}(1234)$ with $i_2\neq 2$ and $i_3\neq 3$:
\ba 
G_{3,11}(1234)&=-\sum_{ i_4\neq 4} (4\wc i_4) H_{411 i_4}(1234)\,,\label{eq:sol_g3_11}\\
G_{3,12}(1234)&=-\sum_{ i_4\neq 4} (4\wc  i_4) H_{412 i_4}(1234)\,,\label{eq:sol_g3_12}\\
G_{3,14}(1234)&=-\sum_{ i_4\neq 4} (4\wc  i_4) H_{414 i_4}(1234)+\sum_{ i_3\neq 3}(3\wc i_3) H_{41 i_33}(1234)\,,\label{eq:sol_g3_14}\\
G_{3,31}(1234)&=-\sum_{ i_4\neq 4} (4\wc i_4) H_{431 i_4}(1234)\,,\label{eq:sol_g3_31}\\
G_{3,32}(1234)&=-\sum_{ i_4\neq 4} (4\wc  i_4) H_{432 i_4}(1234)\,,\label{eq:sol_g3_32}\\
G_{3,34}(1234)&=-\sum_{ i_1\neq 1} (1\wc  i_1) H_{ i_1341}(1234)\,,\label{eq:sol_g3_34}\\
G_{3,41}(1234)&=-\sum_{ i_4\neq 4} (4\wc i_4)H_{441 i_4}(1234)+\sum_{ i_2\neq 2} (2\wc i_2) H_{4 i_212}(1234)\,,\label{eq:sol_g3_41}\\
G_{3,42}(1234)&=-\sum_{ i_1\neq 1}(1\wc i_1) H_{ i_1421}(1234)\,,\label{eq:sol_g3_42}\\
G_{3,44}(1234)&=-\sum_{ i_1\neq 1} (1\wc i_1) H_{ i_1441}(1234)\,.\label{eq:sol_g3_44}
\ea 
Solutions for $G_{4, i_1 i_4}(1234)$ with $i_1\neq 1$ and $i_4\neq 4$:
\ba 
G_{4,21}(1234)&=-\sum_{ i_3\neq 3} (3\wc i_3) H_{23 i_31}(1234)\,,\label{eq:sol_g4_21}\\
G_{4,22}(1234)&=-\sum_{ i_3\neq 3} (3\wc i_3) H_{23 i_32}(1234)\,,\label{eq:sol_g4_22}\\
G_{4,23}(1234)&=-\sum_{ i_3\neq 3} (3\wc i_3) H_{23 i_33}(1234)+\sum_{ i_4\neq 4} (4\wc i_4) H_{234 i_4}(1234)\,,\label{eq:sol_g4_23}\\
G_{4,31}(1234)&=-\sum_{ i_2\neq 2} (2\wc i_2) H_{3 i_221}(1234)\,,\label{eq:sol_g4_31}\\
G_{4,32}(1234)&=-\sum_{ i_2\neq 2} (2\wc i_2) H_{3 i_222}(1234)+\sum_{ i_4\neq 4} (4\wc i_4)H_{342 i_4}(1234)\,,\label{eq:sol_g4_32}\\
G_{4,33}(1234)&=-\sum_{ i_2\neq 2} (2\wc i_2) H_{3 i_223}(1234)\,,\label{eq:sol_g4_33}\\
G_{4,41}(1234)&=-\sum_{ i_2\neq_2} (2\wc i_2) H_{4 i_221}(1234)\,,\label{eq:sol_g4_41}\\
G_{4,42}(1234)&=-\sum_{ i_3\neq 3} (3\wc i_3) H_{43 i_32}(1234)\,,\label{eq:sol_g4_42}\\
G_{4,43}(1234)&=-\sum_{ i_2\neq 2} (2\wc  i_2) H_{4 i_223}(1234)\,.\label{eq:sol_g4_43}
\ea 
Solutions for $G_{5, i_1 i_3}(1234)$ with $i_1\neq 1$ and $i_3\neq 3$:
\ba
G_{5,21}(1234)&=-\sum_{ i_4\neq 4} (4\wc i_4) H_{241 i_4}(1234)\,,\label{eq:sol_g5_21}\\
G_{5,22}(1234)&=-\sum_{ i_4\neq 4} (4\wc i_4) H_{242 i_4}(1234)\,,\label{eq:sol_g5_22}\\
G_{5,24}(1234)&=-\sum_{ i_4\neq 4} (4\wc i_4) H_{244 i_4}(1234)+\sum_{ i_3\neq 3} (3\wc i_3) H_{24 i_33}(1234)\,,\label{eq:sol_g5_24}\\
G_{5,31}(1234)&=-\sum_{ i_2\neq 2} (2\wc i_2) H_{3 i_212}(1234)\label{eq:sol_g5_31}\,,\\
G_{5,32}(1234)&=-\sum_{ i_4\neq 4} (4\wc  i_4)H_{342 i_4}(1234)\,,\label{eq:sol_g5_32}\\
G_{5,34}(1234)&=-\sum_{ i_2\neq 2} (2\wc i_2)H_{3 i_242}(1234)\,,\label{eq:sol_g5_34}\\
G_{5,41}(1234)&=-\sum_{ i_2\neq 2}(2\wc i_2)H_{4 i_212}(1234)\,,\label{eq:sol_g5_41}\\
G_{5,42}(1234)&=-\sum_{ i_2\neq 2} (2\wc i_2) H_{4 i_222}(1234)+\sum_{ i_3\neq 3}(3\wc i_3) H_{43 i_32}(1234)\,,\label{eq:sol_g5_42}\\
G_{5,44}(1234)&=-\sum_{ i_2\neq 2} (2\wc i_2) H_{4 i_242}(1234)\,\label{eq:sol_g5_44}.
\ea 
Solutions for $G_{6, i_1 i_2}(1234)$ with $i_1\neq 1$ and $i_2\neq 2$:
\ba 
G_{6,21}(1234)&=-\sum_{ i_3\neq 3} (3\wc i_3) H_{21 i_33}(1234)\,,\label{eq:sol_g6_21}\\
G_{6,23}(1234)&=-\sum_{ i_4\neq 4} (4\wc i_4) H_{234 i_4}(1234)\,,\label{eq:sol_g6_23}\\
G_{6,24}(1234)&=-\sum_{ i_3\neq 3} (3\wc i_3) H_{24 i_33}(1234)\,,\label{eq:sol_g6_24}\\
G_{6,31}(1234)&=-\sum_{ i_4\neq 4} (4\wc i_4) H_{314 i_4}(1234)\,,\label{eq:sol_g6_31}\\
G_{6,33}(1234)&=-\sum_{ i_4\neq 4} (4\wc i_4)H_{334 i_4}(1234)\,,\label{eq:sol_g6_33}\\
G_{6,34}(1234)&=-\sum_{ i_4\neq 4}(4\wc i_4)H_{344 i_4}(1234)+\sum_{ i_2\neq 2} (2\wc i_2) H_{3 i_242}(1234)\,,\label{eq:sol_g6_34}\\
G_{6,41}(1234)&=-\sum_{ i_3\neq 3} (3\wc i_3) H_{41 i_33}(1234)\,,\label{eq:sol_g6_41}\\
G_{6,43}(1234)&=-\sum_{ i_3\neq 3} (3\wc i_3) H_{43 i_33}(1234)+\sum_{ i_2\neq 2} (2\wc i_2) H_{4 i_223}(1234)\,,\label{eq:sol_g6_43}\\
G_{6,44}(1234)&=-\sum_{ i_3\neq 3} (3\wc i_3) H_{44 i_33}(1234)\,.\label{eq:sol_g6_44}
\ea 
Solution for $F_1(1234)$:
\ba 
F_1(1234)=\sum_{ i_3\neq 3} (1\wc 2)(3\wc  i_3) H_{21 i_33}(1234)+\sum_{ i_4\neq 4} (1\wc 3)(4\wc i_4) H_{314 i_4}(1234)+\sum_{ i_3\neq 3} (1\wc 4)(3\wc  i_3) H_{41 i_33}(1234)\,.\label{eq:sol_f1}
\ea 
Solution for $F_2(1234)$:
\ba 
F_2(1234)=\sum_{ i_4\neq 4} (1\wc 2)(4\wc  i_4)H_{241 i_4}(1234)+\sum_{ i_2\neq 2} (1\wc 3)(2\wc  i_2) H_{3 i_212}(1234)+\sum_{ i_2\neq 2} (1\wc 4)(2\wc i_2) H_{4 i_212}(1234)\,.\label{eq:sol_f2}
\ea 
Solution for $F_3(1234)$:
\ba 
F_3(1234)=\sum_{ i_3\neq 3} (1\wc 2)(3\wc  i_3) H_{23 i_31}(1234)+\sum_{ i_2\neq 2} (1\wc 3)(2\wc i_2)H_{3 i_221}(1234)+\sum_{ i_2\neq 2} (1\wc 4)(2\wc i_2)H_{4 i_221}(1234)\,.\label{eq:sol_f3}
\ea 
This completes the determination of all tail and shoulder form factors of the $N=4$ tensor in terms of the head form factors.
\section[Full list of equivalence classes under permutation group symmetries of the 4-th rank tensor.]{Full list of equivalence classes under permutation group symmetries of the $4$-th rank tensor.}
\label{app:notational_invariance}
In Section~\ref{subsec:N4}, we observed that the compact master formula for the worldline representation of the head form factors of Eq.~\eqref{eq:H_i1_iN} satisfies the permutation group symmetry in Eq.~\eqref{eq:transposition_def}, in agreement with the fact that the $\mathcal{N}=1$ supersymmetric charged worldline currents trivially commute in the rank$-N$ tensor of Eq.~\eqref{eq:pi_mu1_muN}. Using Eq.~\eqref{eq:transposition_def}, we generated the equivalence class corresponding to the head form factor representative $H_{2111}(1234)$. In this appendix we list the five remaining equivalence classes defined by the action of this symmetry group. The elements of these classes are later used to express the original tensor in terms of the reduced set of six head form-factor representatives, as done in Appendix~\ref{app:fullresult}.
 
Starting with $H_{2121}(1234)$, a sequential application of the symmetry-group action in Eq.~\eqref{eq:transposition_def} generates the second equivalence class,
\ba
&H_{2121}(1234)=H_{2112}(1243)=H_{3311}(1324)=H_{4411}(1342)=H_{3113}(1423)=H_{4141}(1432)\nonumber\\
=&H_{2112}(2134)=H_{2121}(2143)=H_{3113}(2314)=H_{4141}(2341)=H_{4411}(2431)=H_{3311}(2413)\nonumber\\
=&H_{3322}(3124)=H_{4422}(3142)=H_{2323}(3214)=H_{2442}(3241)=H_{4343}(3412)=H_{3443}(3421)\nonumber\\
=&H_{2323}(4123)=H_{2442}(4132)=H_{3322}(4213)=H_{4422}(4231)=H_{4343}(4321)=H_{3443}(4312)\,.\label{eq:class_2_representative}
\ea
Likewise starting from another element not contained in Eq.~\eqref{eq:class_1_representative} or \eqref{eq:class_2_representative}, for instance $H_{2123}(1234)$,  Eq.~\eqref{eq:transposition_def} generates 
\ba
&H_{2123}(1234)=H_{2142}(1243)=H_{3312}(1324)=H_{4421}(1342)=H_{3413}(1423)=H_{4341}(1432)\nonumber\\
=&H_{2113}(2134)=H_{2141}(2143)=H_{3112}(2314)=H_{4121}(2341)=H_{3411}(2413)=H_{4311}(2431)\nonumber\\
=&H_{3321}(3124)=H_{4412}(3142)=H_{2321}(3214)=H_{2412}(3241)=H_{4143}(3412)=H_{3143}(3421)\nonumber\\
=&H_{4323}(4123)=H_{3442}(4132)=H_{4322}(4213)=H_{3422}(4231)=H_{2343}(4321)=H_{2443}(4312)\,.\label{eq:class_3_representative}
\ea 
We repeat this procedure starting with $H_{2311}(1234)$, $H_{2143}(1234)$ and $H_{2341}(1234)$ to obtain, 
\ba
&H_{2311}(1234)=H_{2411}(1243)=H_{3121}(1324)=H_{4112}(1342)=H_{3141}(1423)=H_{4113}(1432)\nonumber\\
=&H_{3122}(2134)=H_{4122}(2143)=H_{2313}(2314)=H_{2441}(2341)=H_{4313}(2413)=H_{3441}(2431)\nonumber\\
=&H_{2312}(3124)=H_{2421}(3142)=H_{3123}(3214)=H_{4142}(3241)=H_{3341}(3412)=H_{4413}(3421)\nonumber\\
=&H_{2342}(4123)=H_{2423}(4132)=H_{3423}(4213)=H_{4342}(4231)=H_{4423}(4321)=H_{3342}(4312)\,,\label{eq:class_4_representative}
\ea
as well as 
\ba
&H_{2143}(1234)=H_{2143}(1243)=H_{3412}(1324)=H_{4321}(1342)=H_{3412}(1423)=H_{4321}(1432)\nonumber\\
=&H_{2143}(2134)=H_{2143}(2143)=H_{3412}(2314)=H_{4321}(2341)=H_{4321}(2431)=H_{3412}(2413)\nonumber\\
=&H_{4321}(3124)=H_{3412}(3142)=H_{4321}(3214)=H_{3412}(3241)=H_{2143}(3421)=H_{2143}(3412)\nonumber\\
=&H_{4321}(4123)=H_{3412}(4132)=H_{4321}(4213)=H_{3412}(4231)=H_{2143}(4312)=H_{2143}(4321)\,,\label{eq:class_5_representative}
\ea 
and
\ba
&H_{2341}(1234)=H_{2413}(1243)=H_{3421}(1324)=H_{4312}(1342)=H_{3142}(1423)=H_{4123}(1432)\nonumber\\
=&H_{3142}(2134)=H_{4123}(2143)=H_{2413}(2314)=H_{2341}(2341)=H_{4312}(2413)=H_{3421}(2431)\nonumber\\
=&H_{4312}(3124)=H_{3421}(3142)=H_{4123}(3214)=H_{3142}(3241)=H_{2341}(3412)=H_{2413}(3421)\nonumber\\
=&H_{2341}(4123)=H_{2413}(4132)=H_{3421}(4213)=H_{4312}(4231)=H_{3142}(4312)=H_{4123}(4321)\,,\label{eq:class_6_representative}
\ea 
This completes the list, allowing any of the 81 head form factors to be written now as a number of inverse permutation in the momenta arguments of the six class representatives above. We choose as representatives the first element of each class: $H_{2111}(1234)$, $H_{2121}(1234)$, $H_{2123}(1234)$, $H_{2311}(1234)$, $H_{2143}(1234)$, and $H_{2341}(1234)$. The choice is arbitrary, and any alternative set would lead to an equivalent result. The list provided here incorporates corrections to typographical errors in the original manuscript of Karplus and Neuman~\cite{Karplus:1950zza}, previously identified in~\cite{Dolen:1965gfx}. 

Note that repeated heads appear within each equivalence class, sharing the same indices but differing by permutations of the photon momentum arguments, as already observed in Section~\ref{subsec:N4}. This leads to equivalence classes of different cardinalities, so counting the number of independent heads is not a trivial task. 
We provide in this work a computational algorithm that automates this procedure and gives the results for general $N$. We emphasize that the central result of this work, the worldline master formula for the tensor form factors in Eq.~\eqref{eq:H_i1_iN}, automatically satisfies the aforementioned permutation group symmetries, and thus a systematic way of finding the minimal set of independent form factors.

\section{Complete expression for the rank-4 QED vacuum polarization tensor.}
\label{app:fullresult}
We present the result in tabular form, with the terms organized according to the same six equivalence classes of head form factors listed above. 
First, the expressions for the tails (in Eqs.~\eqref{eq:sol_f1}-\eqref{eq:sol_f3}) and shoulders (in Eqs.~\eqref{eq:sol_g1_11}-\eqref{eq:sol_g6_44}), each given in terms of the heads, are substituted into Eq.~\eqref{eq:Pi_mu1_mu2_mu3_mu4_result}.
For each head form factor, we then indicate its replacement using Eq.~\eqref{eq:class_1_representative} and Eqs.~\eqref{eq:class_2_representative}–\eqref{eq:class_6_representative}, together with the coefficient multiplying it in the polarization tensor. The rank-$4$ tensor $\Pi_{1234}(1234)$ is then obtained by summing all the terms listed below. As emphasized earlier, since the symmetries and identities employed are completely general, the resulting expression holds for the rank-4 tensor at any order in perturbation theory in QED, with modifications affecting only the explicit definition of the form factors $H_{ijkl}(1234)$.
\newline 

\underline{\textit{Class 1}}
\ba 
&\Big[H_{2111}(1234)\to H_{2111}(1234)\Big] \times \Big[(2)_{1}(1)_{2}(1)_{3}(1)_{4}-(1\wc 2) \delta_{12}(1)_{3}(1)_{4}\Big]\,,\nonumber\\
&\Big[H_{3111}(1234)\to H_{2111}(1324)\Big]\times \Big[(3)_{1}(1)_{2}(1)_{3}(1)_{4}-(1\wc 3) \delta_{13} (1)_{2} (1)_{4}\Big]\,,\nonumber\\
&\Big[H_{4111}(1234)\to H_{2111}(1423)\Big]\times\Big[ (4)_{1}(1)_{2}(1)_{3}(
1)_{4}-(1\wc 4) \delta_{14} (1)_{2}(1)_{3}\Big]\,,\nonumber\\
&\Big[H_{2122}(1234)\to H_{2111}(2143)\Big]\times\Big[ (2)_{1}(1)_{2}(2)_{3}(2)_{4}-(1\wc 2) \delta_{12} (2)_{3}(2)_{4}\Big]\,,\nonumber\\
&\Big[H_{3313}(1234)\to H_{2111}(3124)\Big]\times\Big[ (3)_{1}(3)_{2}(1)_{3}(3)_{4}-(1\wc 3) \delta_{13} (3)_{2}(3)_{4}\Big]\,,\nonumber\\
&\Big[H_{4441}(1234)\to H_{2111}(4123)\Big]\times\Big[ (4)_{1}(4)_{2}(4)_{3}(1)_{4}-(1\wc 4) \delta_{14} (4)_{2}(4)_{3}\Big]\,,\nonumber\\
&\Big[H_{2322}(1234)\to H_{2111}(2314)\Big]\times\Big[ (2)_{1}(3)_{2}(2)_{3}(2)_{4}-(2\wc 3)\delta_{23} (2)_{1}(2)_{4}\Big]\,,\nonumber\\
&\Big[H_{2422}(1234)\to H_{2111}(2413)\Big]\times\Big[ (2)_{1}(4)_{2}(2)_{3}(2)_{4}-(2\wc 4) \delta_{24} (2)_{1}(2)_{3}\Big]\,,\nonumber\\
&\Big[H_{3323}(1234)\to H_{2111}(3214)\Big]\times\Big[ (3)_{1}(3)_{2}(2)_{3}(3)_{4}-(2\wc 3) \delta_{23} (3)_{1}(3)_{4}\Big]\,,\nonumber\\
&\Big[H_{4442}(1234)\to H_{2111}(4231)\Big]\times\Big[ (4)_{1}(4)_{2}(4)_{3}(2)_{4}-(2\wc 4) \delta_{24} (4)_{1}(4)_{3}\Big]\,,\nonumber\\
&\Big[H_{4443}(1234)\to H_{2111}(4321)\Big]\times\Big[ (4)_{1}(4)_{2}(4)_{3}(3)_{4}-(3\wc 4) \delta_{34} (4)_{1}(4)_{2}\Big]\,,\nonumber\\
&\Big[H_{3343}(1234)\to H_{2111}(3412)\Big]\times\Big[ (3)_{1}(3)_{2}(4)_{3}(3)_{4} - (3\wc 4) \delta_{34} (3)_{1}(3)_{2 }\Big]\,.\label{eq:class_1_poltensorterms}
\ea 
\newline

\underline{\textit{Class 2}}
\ba 
&\Big[H_{2121}(1234)\to H_{2121}(1234)\Big]\times\Big[ (2)_{1}(1)_{2}(2)_{3}(1)_{4}-(1\wc 2) \delta_{12} (2)_{3}(4)_{4}\Big]\,,\nonumber\\
&\Big[H_{2112}(1234)\to H_{2121}(1243)\Big]\times\Big[ (2)_{1}(1)_{2}(1)_{3}(2)_{4}-(1\wc 2) \delta_{12} (1)_{3}(2)_{4}\Big]\,,\nonumber\\
&\Big[H_{3311}(1234)\to H_{2121}(3142)\Big]\times\Big[ (3)_{1}(3)_{2}(1)_{3}(1)_{4}-(1\wc 3) \delta_{13} (3)_{2}(1)_{4}\Big]\,,\nonumber\\
&\Big[H_{4411}(1234)\to H_{2121}(4132)\Big]\times\Big[ (4)_{1}(4)_{2}(1)_{3}(1)_{4}-(1\wc 4) \delta_{14} (1)_{3} (4)_{2}\Big]\,,\nonumber\\
&\Big[H_{3113}(1234)\to H_{2121}(3124)\Big]\times\Big[ (3)_{1}(1)_{2}(1)_{3}(3)_{4}-(1\wc 3) \delta_{13} (1)_{2}(3)_{4}\Big]\,,\nonumber\\
&\Big[H_{4141}(1234)\to H_{2121}(4123)\Big]\times\Big[ (4)_{1}(1)_{2}(4)_{3}(1)_{4}-(1\wc 4)  \delta_{14}(1)_{2}(4)_{3}\Big]\,,\nonumber\\
&\Big[H_{3322}(1234)\to H_{2121}(3241)\Big]\times\Big[ (3)_{1}(3)_{2}(2)_{3}(2)_{4}-(2\wc 3) \delta_{23} (3)_{1}(2)_{4}\Big]\,,\nonumber\\
&\Big[H_{4422}(1234)\to H_{2121}(2413)\Big]\times\Big[ (4)_{1}(4)_{2}(2)_{3}(2)_{4}-(2\wc 4) \delta_{24} (4)_{1}(2)_{3}\Big]\,,\nonumber\\
&\Big[H_{2323}(1234)\to H_{2121}(2341)\Big]\times\Big[ (2)_{1}(3)_{2}(2)_{3}(3)_{4}-(2\wc 3) \delta_{23}(2)_{1}(3)_{4} \Big]\,,\nonumber\\
&\Big[H_{2442}(1234)\to H_{2121}(2431)\Big]\times\Big[ (2)_{1}(4)_{2}(4)_{3}(2)_{4}-(2\wc 4) \delta_{24} (2)_{1}(4)_{3}\Big]\,,\nonumber\\
&\Big[H_{4343}(1234)\to H_{2121}(3412)\Big]\times\Big[ (4)_{1}(3)_{2}(4)_{3}(3)_{4}-(4\wc 3)\delta_{34} (4)_{1}(3)_{2}  \Big]\,,\nonumber\\
&\Big[H_{3443}(1234)\to H_{2121}(4312)\Big]\times\Big[ (3)_{1}(4)_{2}(4)_{3}(3)_{4}-(3\wc 4) \delta_{34} (3)_{1}(4)_{2}\Big]\,.\label{eq:class_2_poltensorterms}
\ea 
\newline

\underline{\textit{Class 3}}
\ba 
\Big[H_{2123}(1234)\to H_{2123}(1234)\Big]\times\Big[ &(2)_{1}(1)_{2}(2)_{3}(3)_{4}-(1\wc 2)\delta_{12}(2)_{3}(3)_{4}-(2\wc 3) \delta_{34}(2)_{1}(1)_{2}\nonumber\\
&+(1\wc 2)(2\wc 3) \delta_{12}\delta_{34}\Big]\,,\nonumber\\
\Big[H_{2141}(1234)\to H_{2123}(2143)\Big]\times \Big[&(2)_{1}(1)_{2}(4)_{3}(1)_{4}-(1\wc 2)\delta_{12}(4)_{3}(1)_{4}\Big]\,,\nonumber\\
\Big[H_{2142}(1234)\to H_{2123}(1243)\Big]\times\Big[ &(2)_{1}(1)_{2}(4)_{3}(2)_{4}-(1\wc 2) \delta_{12}(4)_{3}(2)_{4}\Big]\,,\nonumber\\
\Big[H_{3312}(1234)\to H_{2123}(1324)\Big]\times\Big[ &(3)_{1}(3)_{2}(1)_{3}(2)_{4}-(1\wc 3) \delta_{13}(3)_{2}(2)_{4} -(2\wc 3) \delta_{24} (3)_{1}(3)_{3}\nonumber\\
&+(2\wc 3) (1\wc 3) \delta_{13}\delta_{24}\Big]\,,\nonumber\\
\Big[H_{4421}(1234)\to H_{2123}(1423)\Big]\times\Big[ &(4)_{1}(4)_{2}(2)_{3}(1)_{4}-(1\wc 4) \delta_{14} (4)_{2}(4)_{3}-(2\wc 4) \delta_{23}(4)_{1}(1)_{4} \nonumber\\
&+(1\wc 4)(2\wc 4)\delta_{24}\delta_{23}\Big]\,,\nonumber\\
\Big[H_{3413}(1234)\to H_{2123}(1342)\Big]\times\Big[ &(4)_{1}(4)_{2}(2)_{3}(1)_{4}-(1\wc 3) \delta_{13} (4)_{2}(3)_{4}\Big]\nonumber\\
\Big[H_{4341}(1234)\to H_{2123}(1432)\Big]\times\Big[ &(4)_{1}(3)_{2}(4)_{3}(1)_{4}-(1\wc 4) \delta_{14}(3)_{2}(4)_{3}\Big]\nonumber\\
\Big[H_{2113}(1234)\to H_{2123}(2134)\Big]\times\Big[ &(2)_{1}(1)_{2}(1)_{3}(3)_{4}-(1\wc 2) (1)_{3}(3)_{4} \delta_{12}-(1\wc 3) \delta_{34}(2)_{1}(1)_{2}\nonumber\\
&+(1\wc 2)(1\wc 3) \delta_{12}\delta_{34}\nonumber\\
\Big[H_{2141}(1234)\to H_{2123}(2143)\Big]\times\Big[ &(2)_{1}(1)_{2}(4)_{3}(1)_{4}-(1\wc 2)\delta_{12} (4)_{3}(1)_{4}\Big]\nonumber\\
\Big[H_{3112}(1234)\to H_{2123}(3124)\Big]\times\Big[ &(3)_{1}(1)_{2}(1)_{3}(2)_{4}-(1\wc 2) \delta_{24}(3)_{1}(1)_{3}-(1\wc 3) \delta_{13} (1)_{2}(2)_{4} \nonumber\\
&+ (1\wc 2)(1\wc 3) \delta_{13}\delta_{24}\Big]\,,\nonumber\\
\Big[H_{4121}(1234)\to H_{2123}(4123)\Big]\times\Big[ &(4)_{1}(1)_{2}(2)_{3}(1)_{4}-(1\wc 2) \delta_{23}(4)_{1}(1)_{4}-(1\wc 4) (1)_{2}(2)_{3} \delta_{14}\nonumber\\
&+(1\wc 2) (1\wc 4) \delta_{14} \delta_{23}\Big]\nonumber\\
\Big[H_{3411}(1234)\to H_{2123}(3142)\Big]\times\Big[ &(1)_{1}(2)_{2}(3)_{3}(4)_{4}-(1\wc 3) \delta_{13}(4)_{2}(1)_{4}\Big]\nonumber\\
\Big[H_{4311}(1234)\to H_{2123}(4132)\Big]\times\Big[ &(4)_{1}(3)_{2}(1)_{3}(2)_{4}-(1\wc 4) \delta_{14} (3)_{2}(1)_{3}\Big]\nonumber\\
\Big[H_{3321}(1234)\to H_{2123}(2314)\Big]\times\Big[ &(3)_{1}(3)_{2}(2)_{3}(1)_{4}-(2\wc 3)\delta_{23} (3)_{1}(1)_{4} +(1\wc 3)(2\wc 3) \delta_{14}\delta_{23}\Big]\nonumber\\
\Big[H_{4412}(1234)\to H_{2321}(2413)\Big]\times\Big[ &(4)_{1}(4)_{2}(1)_{3}(2)_{4}-(1\wc 4)\delta_{13} (4)_{2}(2)_{4} -(2\wc 4) \delta_{24} (4)_{1}(1)_{3}\nonumber\\
&+(1\wc 4) (2\wc 4) \delta_{13}\delta_{24}\Big]\,.\nonumber\\
\Big[H_{2321}(1234)\to H_{2123}(3214)\Big]\times\Big[ &(2)_{1}(3)_{2}(2)_{3}(1)_{4}-(2\wc 3) \delta_{23}(2)_{1}(1)_{4} +(1\wc 2)(2\wc 3) \delta_{14}\delta_{23}\Big]\nonumber\\
\Big[H_{2412}(1234)\to H_{2123}(4213)\Big]\times\Big[ &(2)_{1}(4)_{2}(1)_{3}(2)_{4}-(1\wc 2) \delta_{13}(4)_{2}(2)_{4}-(2\wc 4) \delta_{24} (2)_{1}(1)_{3}\nonumber\\
&+\delta_{13}(1\wc 2) (2\wc 4)\Big]\,,\nonumber\\
\Big[H_{4143}(1234)\to H_{2123}(3412)\Big]\times\Big[ &(4)_{1}(1)_{2}(4)_{3}(3)_{4}-(3\wc 4) \delta_{34} (4)_{1}(1)_{2} +(1\wc 4)(3\wc 4) \delta_{12}\delta_{34}\Big]\,,\nonumber\\
\Big[H_{3143}(1234)\to H_{2123}(4312)\Big]\times\Big[ &(3)_{1}(1)_{2}(4)_{3}(3)_{4}-(3\wc 4) (3)_{1}(1)_{2} \delta_{34}+(1\wc 3)(3\wc 4) \delta_{12}\delta_{34}\Big]\nonumber\\
\Big[H_{4323}(1234)\to H_{2123}(2341)\Big]\times\Big[ &(4)_{1}(3)_{2}(2)_{3}(3)_{4}-(2\wc 3)\delta_{23} (4)_{1}(3)_{4}-(3\wc 4) \delta_{14} (3)_{2}(2)_{3}\Big]\nonumber\\
\Big[H_{3442}(1234)\to H_{2123}(2431)\Big]\times\Big[ &(3)_{1}(4)_{2}(4)_{3}(2)_{4}-(2\wc 3) \delta_{23}(4)_{1}(3)_{4} -(3\wc 4) \delta_{14} (3)_{2}(2)_{3}\Big]\nonumber\\
\Big[H_{4322}(1234)\to H_{2123}(3241)\Big]\times\Big[ &(4)_{1}(3)_{2}(2)_{3}(2)_{4}-(2\wc 3) \delta_{23}(4)_{1}(2)_{4}-(2\wc 4) \delta_{14} (3)_{2}(2)_{3}\Big]\nonumber\\
\Big[H_{3422}(1234)\to H_{2123}(4231)\Big]\times\Big[ &(3)_{1}(4)_{2}(2)_{3}(2)_{4}-(2\wc 4) \delta_{24}(3)_{1}(2)_{3} \Big]\nonumber\\
\Big[H_{2343}(1234)\to H_{2123}(4321)\Big]\times\Big[ &(2)_{1}(3)_{2}(4)_{3}(3)_{4}-(2\wc 3) \delta_{12} (4)_{3}(3)_{4}-(3\wc 4) \delta_{34} (2)_{1}(3)_{2}\Big]\nonumber\\
\Big[H_{2443}(1234)\to H_{2123}(3421)\Big]\times\Big[ &(2)_{1}(4)_{2}(4)_{3}(3)_{4}-(2\wc 4) \delta_{12}(4)_{3}(3)_{4}-(3\wc 4) \delta_{34}(2)_{1}(4)_{2}\Big]\,.\label{eq:class_3_poltensorterms}
\ea 
\newline

\underline{\textit{Class 4}}
\ba 
\Big[H_{2311}(1234)\to H_{2311}(1234)\Big]\times\Big[ &(2)_{1}(3)_{2}(1)_{3}(1)_{4}-(1\wc 2) \delta_{13}(3)_{2}(1)_{4}-(1\wc 3)\delta_{23}(2)_{1}(1)_{4}\nonumber\\
&-(2\wc 3)\delta_{12} (1)_{3}(1)_{4}+(1\wc 2)(1\wc 3) \delta_{14}\delta_{23}\Big]\,,\nonumber\\
\Big[H_{2411}(1234)\to H_{2311}(1243)\Big]\times\Big[ &(2)_{1}(3)_{2}(1)_{3}(1)_{4}-(1\wc 4)\delta_{24}(2)_{1}(1)_{3}\nonumber\\
&-(2\wc 4) \delta_{12} (1)_{3}(1)_{4}+(1\wc 2)(1\wc 4) \delta_{13}\delta_{24}\Big]\,,\nonumber\\
\Big[H_{3121}(1324)\to H_{2311}(1324)\Big]\times\Big[ &(3)_{1}(1)_{2}(2)_{3}(1)_{4}-(1\wc 2) \delta_{23} (3)_{1} (1)_{4} -(1\wc 3) \delta_{12} (2)_{3}(1)_{4}\nonumber\\
&+(2\wc 3) \delta_{13}(1)_{2}(1)_{4}+(1\wc 2)(1\wc 3) \delta_{14}\delta_{23}\Big]\,,\nonumber\\
\Big[H_{4112}(1234)\to H_{2311}(1423)\Big]\times\Big[ &(4)_{1}(1)_{2}(1)_{3}(2)_{4}-(1\wc 2) (4)_{1}(1)_{3} \delta_{24} +(1\wc 2) \delta_{14} (4)_{2}(1)_{3}\nonumber\\
& -(1\wc 4)\delta_{12} (1)_{3}(2)_{4}-(2\wc 4) \delta_{14} (1)_{2} (1)_{3} \delta_{14} +(1\wc 2) (1\wc 4) \delta_{13}\delta_{24}\Big]\,,\nonumber\\
\Big[H_{3141}(1234)\to H_{2311}(1342)\Big]\times\Big[ &(3)_{1}(1)_{2}(4)_{3}(1)_{4}+(1\wc 4) \delta_{13} (1)_{2}(3)_{4}-(1\wc 4) \delta_{34} (3)_{1} (1)_{2} \nonumber\\
&-(3\wc 4) (1)_{2}(1)_{4} (3\wc 4) \delta_{13}+(1\wc 3) (1\wc 4) \delta_{12}\delta_{34}\Big]\,,\nonumber\\
\Big[H_{4113}(1234)\to H_{2311}(1432)\Big]\times\Big[ &(4)_{1}(1)_{2}(1)_{3}(3)_{4}-(1\wc 3) \delta_{34} (4)_{1}(1)_{2}+(1\wc 3) \delta_{14} (1)_{2}(4)_{3} \nonumber\\
&-(3\wc 4) \delta_{14} (1)_{2}(1)_{3} +(1 \wc 3) (1\wc 4) \delta_{12}\delta_{34}\Big]\,,\nonumber\\
\Big[H_{3122}(1234)\to H_{2311}(2134) \Big]\times\Big[ &(3)_{1}(1)_{2}(2)_{3}(2)_{4}-(1\wc 2) \delta_{23} (3)_{1}(2)_{4}-(1\wc 3) \delta_{12}(2)_{3}(2)_{4} \nonumber\\
&+(2\wc 3) \delta_{12} (1)_{3}(2)_{4} -(2\wc 3) \delta_{13}(1)_{2}(2)_{4} \Big]\,,\nonumber\\
\Big[H_{4122}(1234)\to H_{2311}(2143)\Big]\times\Big[ &(4)_{1}(1)_{2}(2)_{3}(2)_{4}-(1\wc 2) \delta_{24} (4)_{1}(2)_{3} -(1\wc 4) \delta_{12} (2)_{3}(2)_{4}\nonumber\\
&+(2\wc 4) \delta_{12} (2)_{3}(1)_{4} -(2\wc 4) \delta_{14} \delta_{14}(1)_{2}(2)_{3} \Big]\,,\nonumber\\
\Big[H_{2313}(1234)\to H_{2311}(3124)\Big]\times\Big[ &(2)_{1}(3)_{2}(1)_{3}(3)_{4}-(1\wc 2) \delta_{13}(3)_{2}(3)_{4} -(1\wc 3) \delta_{23}(2)_{1}(3)_{4}\nonumber\\
&-(2\wc 3) \delta_{12}(1)_{3}(3)_{4}\Big]\,,\nonumber\\
\Big[H_{2441}(1234)\to H_{2311}(4123)\Big]\times\Big[ &(2)_{1}(4)_{2}(1)_{3}(1)_{4}-(1\wc 4)\delta_{24}(2)_{1}(1)_{3} -(2\wc 4)\delta_{12} (1)_{3}(1)_{4} \Big]\,,\nonumber\\
\Big[H_{4313}(1234)\to H_{2311}(3142)\Big]\times\Big[ &(4)_{1}(3)_{2}(1)_{3}(3)_{4}-(1\wc 3) \delta_{34}(4)_{1}(3)_{2}-(1\wc 4) \delta_{13}(3)_{2}(3)_{4}\nonumber\\
&+(3\wc 4) \delta_{13} (3)_{2}(1)_{4}\Big]\,,\nonumber\\
\Big[H_{3441}(1234)\to H_{2311}(4132)\Big]\times\Big[ &(3)_{1}(4)_{2}(4)_{3}(1)_{4}-(1\wc 3) \delta_{14} (4)_{2}(4)_{3}-(1\wc 4) \delta_{34} (3)_{1}(4)_{2}\nonumber\\
&-(3\wc 4) (4)_{2}(1)_{4}\delta_{13}\Big]\,,\nonumber\\
\Big[H_{2312}(1234)\to H_{2311}(2314)\Big]\times\Big[ &(2)_{1}(3)_{2}(1)_{3}(2)_{4}-(1\wc 2) \delta_{13} (3)_{2}(2)_{4}-(1\wc 3) \delta_{23} (2)_{1}(2)_{4}\Big]\,,\nonumber\\
\Big[H_{2421}(1234)\to H_{2311}(2413)\Big]\times\Big[ &(2)_{1}(4)_{2}(2)_{3}(1)_{4}-(1\wc 2) \delta_{14} (4)_{2}(2)_{3} -(1\wc 4) \delta_{24}(2)_{1}(2)_{3}\Big]\,,\nonumber\\
\Big[H_{3123}(1234)\to H_{2311}(3214)\Big]\times\Big[ &(3)_{1}(1)_{2}(1)_{3}(1)_{4}-(1\wc 2) \delta_{23} (3)_{1}(3)_{4}-(1\wc 3) \delta_{12} (2)_{3}(3)_{2} \nonumber\\
&-(2\wc 3) \delta_{13}(1)_{2}(3)_{4} \Big]\,,\nonumber\\
\Big[H_{4142}(1234)\to H_{2311}(4213)\Big]\times\Big[ &(4)_{1}(1)_{2}(4)_{3}(2)_{4}-(1\wc 2) \delta_{24} (4)_{1}(4)_{3} -\delta_{12} (1\wc 4) (4)_{3}(2)_{4}\nonumber\\
&-(2\wc 4) \delta_{14}(1)_{2}(4)_{3}\Big]\,,\nonumber\\
\Big[H_{3341}(1234)\to H_{2311}(3412)\Big]\times\Big[ &(3)_{1}(3)_{2}(4)_{3}(1)_{4}-(1\wc 3) \delta_{14} (3)_{2}(4)_{3}-(1\wc 4)\delta_{34} (3)_{1}(3)_{2}\Big]\,,\nonumber\\
\Big[H_{4413}(1234)\to H_{2311}(4312)\Big]\times\Big[ &(4)_{1}(4)_{2}(1)_{3}(3)_{4}-(1\wc 3) \delta_{34} (4)_{1}(4)_{2}-(1\wc 4)\delta_{13} (4)_{2}(3)_{4}\nonumber\\
&-(3\wc 4)\delta_{14}(4)_{2}(1)_{3}\Big]\,,\nonumber\\
\Big[H_{2342}(1234)\to H_{2311}(2341)\Big]\times\Big[ & (2)_{1}(3)_{2}(4)_{3}(2)_{4}-(2\wc 4) \delta_{23} (2)_{1}(3)_{4} -(2\wc 4) \delta_{34} (2)_{1}(3)_{2}\nonumber\\
&-(3\wc 4)\delta_{23} (2)_{1}(2)_{4}\Big]\,,\nonumber\\
\Big[H_{2423}(1234)\to H_{2311}(2431)\Big]\times\Big[ & (2)_{1}(4)_{2}(2)_{3}(3)_{4}-(2\wc 3) \delta_{34}(2)_{1}(4)_{2}+(2\wc 3) \delta_{24}(2)_{1}(4)_{3}\nonumber\\
&-(3\wc 4) \delta_{24}(2)_{1}(2)_{3}\Big]\,,\nonumber\\
\Big[H_{3423}(1234)\to H_{2311}(3241)\Big]\times\Big[ &(3)_{1}(4)_{2}(2)_{3}(3)_{4}-(2\wc 3) \delta_{34} (2)_{1}(4)_{2}-(2\wc 3) \delta_{24}(2)_{1}(4)_{3}\nonumber\\
&-(3\wc 4) \delta_{24} (2)_{1}(2)_{3} \Big]\,,\nonumber\\
\Big[H_{4342}(1234)\to H_{2311}(4231)\Big]\times\Big[ &(4)_{1}(3)_{2}(4)_{3}(2)_{4}-(2\wc 3) \delta_{24} (4)_{1}(4)_{3}-(3\wc 4) \delta_{23} (4)_{1}(2)_{4}\nonumber\\
&+(3\wc 4) \delta_{24}(4)_{1}(2)_{3}\Big]\,,\nonumber\\
\Big[H_{4423}(1234)\to H_{2311}(4321)\Big]\times\Big[ &(4)_{1}(4)_{2}(2)_{3}(3)_{4}-(2\wc 3) \delta_{34} (4)_{1}(4)_{2}-(2\wc 4) \delta_{23} (4)_{1}(3)_{4}\nonumber\\
&+(2\wc 4) \delta_{34}(4)_{1}(3)_{2}\Big]\,,\nonumber\\
\Big[H_{3342}(1234)\to H_{2311}(3421)\Big]\times\Big[ &(2)_{1}(3)_{2}(1)_{3}(1)_{4}-(2\wc 3) \delta_{24}(3)_{1}(4)_{3}\nonumber\\
&+(2\wc 3) \delta_{34} (3)_{1}(4)_{2}\Big]\,.\label{eq:class_4_poltensorterms}
\ea 
\newline

\underline{\textit{Class 5}}
\ba 
\Big[H_{2143}(1234)\to H_{2143}(1234)\Big]\times\Big[ &(2)_{1}(1)_{2}(4)_{3}(3)_{4}-(1\wc 2) \delta_{12}(4)_{3} (3)_{4}-(3\wc 4)\delta_{34} (2)_{1} (1)_{2}\nonumber\\
&+ (1\wc 2) (3\wc 4) \delta_{12}\delta_{34}\Big]\,,\nonumber\\
\Big[H_{3412}(1234)\to H_{2143}(1324)\Big]\times\Big[ &(3)_{1}(4)_{2}(1)_{3}(2)_{4}-(1\wc 3) \delta_{13}(4)_{2}(2)_{4}-(2\wc 4)\delta_{24} (3)_{1}(1)_{3}\nonumber\\
&+(1\wc 3) (2\wc 4) \delta_{13}\delta_{24}\Big]\,,\nonumber\\
\Big[H_{4321}(1234)\to H_{2143}(4321)\Big]\times\Big[ &(4)_{1}(3)_{2}(2)_{3}(1)_{4}-(2\wc 3) \delta_{23}(4)_{1}(1)_{4}-(1\wc 4) \delta_{14}(3)_{2}(2)_{3}\nonumber\\
&+(1\wc 4) (2\wc 3) \delta_{14}\delta_{23}\Big]\,.\label{eq:class_5_poltensorterms}
\ea 
\newline

\underline{\textit{Class 6}}
\ba 
\Big[H_{2341}(1234)\to H_{2341}(1234)\Big]\times\Big[ &(2)_{1}(3)_{2}(4)_{3}(1)_{4}-(1\wc 2) \delta_{14} (3)_{2}(4)_{3}-(2\wc 3) \delta_{12} (4)_{3}(1)_{4} \nonumber\\
&+ (1\wc 4) \delta_{23} (2)_{1}(3)_{4} -(1\wc 4) \delta_{34}(2)_{1}(3)_{2}-(3\wc 4) \delta_{23} (2)_{1}(1)_{4}\nonumber\\
&+(1\wc 2) (3\wc 4) \delta_{14}\delta_{23}\Big]\,,\nonumber\\
\Big[H_{2413}(1234)\to H_{2341}(1243)\Big]\times\Big[ &(2)_{1}(4)_{2}(1)_{3}(3)_{4}-(1\wc 2) \delta_{13} (4)_{2}(3)_{4} -(1\wc 3) \delta_{34} (2)_{1}(4)_{2} \nonumber\\
&+(1\wc 3) \delta_{24} (2)_{1}(4)_{3}-(2\wc 4)\delta_{12} (1)_{3}(3)_{4}- (3\wc 4) \delta_{24}(2)_{1}(1)_{3} \nonumber\\
&+(1\wc 2)(3\wc 4) \delta_{14}\delta_{24}\Big]\,,\nonumber\\
\Big[H_{3421}(1234)\to H_{2341}(1324)\Big]\times\Big[ &(3)_{1}(4)_{2}(2)_{3}(1)_{4}-(1\wc 3) \delta_{14} (4)_{2}(2)_{3} -(2\wc 3) \delta_{13} (2\wc 3) (4)_{2}(1)_{4}\nonumber\\
&+(1\wc 4) \delta_{23} (3)_{1}(2)_{4}-(1\wc 4) \delta_{24}(3)_{1}(2)_{3}-(2\wc 4) \delta_{23} (3)_{1}(1)_{4} \nonumber\\
&+(1\wc 3) (2\wc 4) \delta_{14}\delta_{23}\Big]\,\nonumber\\
\Big[H_{4312}(1234)\to H_{2341}(1423)\Big]\times\Big[ &(4)_{1}(3)_{2}(1)_{3}(2)_{4}-(1\wc 3) \delta_{23} (4)_{1}(2)_{4}+(1\wc 3) \delta_{24} (4)_{1}(2)_{3} \nonumber\\
&-(2\wc 3)(4)_{1}(1)_{3}-(1\wc 4) \delta_{13} (3)_{2}(2)_{4} + (2\wc 3) \delta_{14} (4)_{2}(1)_{3}\nonumber\\
&+(2\wc 4) \delta_{13} (3)_{2}(1)_{4}-(2\wc 4) \delta_{14}(3)_{2}(1)_{3} +(1\wc 4)(2\wc 3) \delta_{13}\delta_{24}\Big]\,,\nonumber\\
\Big[H_{3142}(1234)\to H_{2341}(1342)\Big]\times\Big[ &(3)_{1}(1)_{2}(4)_{3}(2)_{4}-(1\wc 2) \delta_{24} (3)_{1}(4)_{3}-(1\wc 3) \delta_{12} (4)_{3}(2)_{4} \nonumber\\
&+(1\wc 2) \delta_{34} (3)_{1}(4)_{2}+ \delta_{13} (2\wc 4) (1)_{2} (3)_{4} -(2\wc 4) \delta_{34} (3)_{1}(1)_{2} \nonumber\\
&+(3\wc 4) \delta_{12} (1)_{3} (2)_{4}-(3\wc 4) \delta_{13} (1)_{2}(2)_{4}+(1\wc 3) (2\wc 4) \delta_{12} \delta_{34}\Big]\,,\nonumber\\
\Big[H_{4123}(1234)\to H_{2341}(1432)\Big]\times\Big[ &(1)_{1}(2)_{2}(3)_{3}(4)_{4}-(1\wc 2) \delta_{23} (4)_{1}(3)_{4}+(1\wc 2) \delta_{34} (4)_{1}(3)_{2}\nonumber\\
&-(2\wc 3) (4)_{1}(1)_{2} \delta_{34} -(1\wc 4) \delta_{12} (2)_{3}(3)_{4}+(2\wc 3) \delta_{14} (1)_{2} (4)_{3} \nonumber\\
&+(3\wc 4) \delta_{12} (2)_{3}(1)_{4} -(3\wc 4) \delta_{14} (1)_{2}(2)_{3} +(1\wc 4) (2\wc 3) \delta_{12} \delta_{34}\Big]\,.\label{eq:class_6_poltensorterms}
\ea 

\section[The form factors of the N=6-th rank tensor.]{The form factors of the $N=6$-th rank tensor.}
\label{app:N6list}
We list here the results for the numerators of the 40 independent 6-th rank heads:
\ba 
P_{2 1 1 1 1 1} &= \dot{G}^B_{1 2}\dot{G}^B_{2 1}\dot{G}^B_{3 1}\dot{G}^B_{4 1}\dot{G}^B_{5 1}\dot{G}^B_{6 1} -\dot{G}^B_{3 1}\dot{G}^B_{4 1}\dot{G}^B_{5 1}\dot{G}^B_{6 1}G^F_{1 2}G^F_{2 1}
\\
P_{2 1 1 1 1 2} &= \dot{G}^B_{1 2}\dot{G}^B_{2 1}\dot{G}^B_{3 1}\dot{G}^B_{4 1}\dot{G}^B_{5 1}\dot{G}^B_{6 2} -\dot{G}^B_{3 1}\dot{G}^B_{4 1}\dot{G}^B_{5 1}\dot{G}^B_{6 2}G^F_{1 2}G^F_{2 1}
\\
P_{2 1 1 1 1 3} &= \dot{G}^B_{1 2}\dot{G}^B_{2 1}\dot{G}^B_{3 1}\dot{G}^B_{4 1}\dot{G}^B_{5 1}\dot{G}^B_{6 3} -\dot{G}^B_{3 1}\dot{G}^B_{4 1}\dot{G}^B_{5 1}\dot{G}^B_{6 3}G^F_{1 2}G^F_{2 1}
\\
P_{2 1 1 1 2 2} &= \dot{G}^B_{1 2}\dot{G}^B_{2 1}\dot{G}^B_{3 1}\dot{G}^B_{4 1}\dot{G}^B_{5 2}\dot{G}^B_{6 2} -\dot{G}^B_{3 1}\dot{G}^B_{4 1}\dot{G}^B_{5 2}\dot{G}^B_{6 2}G^F_{1 2}G^F_{2 1}
\\
P_{2 1 1 1 2 3} &= \dot{G}^B_{1 2}\dot{G}^B_{2 1}\dot{G}^B_{3 1}\dot{G}^B_{4 1}\dot{G}^B_{5 2}\dot{G}^B_{6 3} -\dot{G}^B_{3 1}\dot{G}^B_{4 1}\dot{G}^B_{5 2}\dot{G}^B_{6 3}G^F_{1 2}G^F_{2 1}
\\
P_{2 1 1 1 2 5} &= \dot{G}^B_{1 2}\dot{G}^B_{2 1}\dot{G}^B_{3 1}\dot{G}^B_{4 1}\dot{G}^B_{5 2}\dot{G}^B_{6 5} -\dot{G}^B_{3 1}\dot{G}^B_{4 1}\dot{G}^B_{5 2}\dot{G}^B_{6 5}G^F_{1 2}G^F_{2 1}
\\
P_{2 1 1 1 3 3} &= \dot{G}^B_{1 2}\dot{G}^B_{2 1}\dot{G}^B_{3 1}\dot{G}^B_{4 1}\dot{G}^B_{5 3}\dot{G}^B_{6 3} -\dot{G}^B_{3 1}\dot{G}^B_{4 1}\dot{G}^B_{5 3}\dot{G}^B_{6 3}G^F_{1 2}G^F_{2 1}
\\
P_{2 1 1 1 3 4} &= \dot{G}^B_{1 2}\dot{G}^B_{2 1}\dot{G}^B_{3 1}\dot{G}^B_{4 1}\dot{G}^B_{5 3}\dot{G}^B_{6 4} -\dot{G}^B_{3 1}\dot{G}^B_{4 1}\dot{G}^B_{5 3}\dot{G}^B_{6 4}G^F_{1 2}G^F_{2 1}
\\
P_{2 1 1 1 3 5} &= \dot{G}^B_{1 2}\dot{G}^B_{2 1}\dot{G}^B_{3 1}\dot{G}^B_{4 1}\dot{G}^B_{5 3}\dot{G}^B_{6 5} -\dot{G}^B_{3 1}\dot{G}^B_{4 1}\dot{G}^B_{5 3}\dot{G}^B_{6 5}G^F_{1 2}G^F_{2 1}
\\
P_{2 1 1 1 6 5} &= \dot{G}^B_{1 2}\dot{G}^B_{2 1}\dot{G}^B_{3 1}\dot{G}^B_{4 1}\dot{G}^B_{5 6}\dot{G}^B_{6 5} -\dot{G}^B_{1 2}\dot{G}^B_{2 1}\dot{G}^B_{3 1}\dot{G}^B_{4 1}G^F_{5 6}G^F_{6 5} \nonumber \\&-\dot{G}^B_{3 1}\dot{G}^B_{4 1}\dot{G}^B_{5 6}\dot{G}^B_{6 5}G^F_{1 2}G^F_{2 1} + \dot{G}^B_{3 1}\dot{G}^B_{4 1}G^F_{1 2}G^F_{2 1}G^F_{5 6}G^F_{6 5}
\\
P_{2 1 1 2 3 3} &= \dot{G}^B_{1 2}\dot{G}^B_{2 1}\dot{G}^B_{3 1}\dot{G}^B_{4 2}\dot{G}^B_{5 3}\dot{G}^B_{6 3} -\dot{G}^B_{3 1}\dot{G}^B_{4 2}\dot{G}^B_{5 3}\dot{G}^B_{6 3}G^F_{1 2}G^F_{2 1}
\\
P_{2 1 1 2 3 4} &= \dot{G}^B_{1 2}\dot{G}^B_{2 1}\dot{G}^B_{3 1}\dot{G}^B_{4 2}\dot{G}^B_{5 3}\dot{G}^B_{6 4} -\dot{G}^B_{3 1}\dot{G}^B_{4 2}\dot{G}^B_{5 3}\dot{G}^B_{6 4}G^F_{1 2}G^F_{2 1}
\\
P_{2 1 1 2 3 5} &= \dot{G}^B_{1 2}\dot{G}^B_{2 1}\dot{G}^B_{3 1}\dot{G}^B_{4 2}\dot{G}^B_{5 3}\dot{G}^B_{6 5} -\dot{G}^B_{3 1}\dot{G}^B_{4 2}\dot{G}^B_{5 3}\dot{G}^B_{6 5}G^F_{1 2}G^F_{2 1}
\\
P_{2 1 1 2 6 5} &= \dot{G}^B_{1 2}\dot{G}^B_{2 1}\dot{G}^B_{3 1}\dot{G}^B_{4 2}\dot{G}^B_{5 6}\dot{G}^B_{6 5} -\dot{G}^B_{1 2}\dot{G}^B_{2 1}\dot{G}^B_{3 1}\dot{G}^B_{4 2}G^F_{5 6}G^F_{6 5} \nonumber \\&-\dot{G}^B_{3 1}\dot{G}^B_{4 2}\dot{G}^B_{5 6}\dot{G}^B_{6 5}G^F_{1 2}G^F_{2 1} + \dot{G}^B_{3 1}\dot{G}^B_{4 2}G^F_{1 2}G^F_{2 1}G^F_{5 6}G^F_{6 5}
\\
P_{2 1 1 3 3 3} &= \dot{G}^B_{1 2}\dot{G}^B_{2 1}\dot{G}^B_{3 1}\dot{G}^B_{4 3}\dot{G}^B_{5 3}\dot{G}^B_{6 3} -\dot{G}^B_{3 1}\dot{G}^B_{4 3}\dot{G}^B_{5 3}\dot{G}^B_{6 3}G^F_{1 2}G^F_{2 1}
\\
P_{2 1 1 3 3 4} &= \dot{G}^B_{1 2}\dot{G}^B_{2 1}\dot{G}^B_{3 1}\dot{G}^B_{4 3}\dot{G}^B_{5 3}\dot{G}^B_{6 4} -\dot{G}^B_{3 1}\dot{G}^B_{4 3}\dot{G}^B_{5 3}\dot{G}^B_{6 4}G^F_{1 2}G^F_{2 1}
\\
P_{2 1 1 3 4 4} &= \dot{G}^B_{1 2}\dot{G}^B_{2 1}\dot{G}^B_{3 1}\dot{G}^B_{4 3}\dot{G}^B_{5 4}\dot{G}^B_{6 4} -\dot{G}^B_{3 1}\dot{G}^B_{4 3}\dot{G}^B_{5 4}\dot{G}^B_{6 4}G^F_{1 2}G^F_{2 1}
\\
P_{2 1 1 3 4 5} &= \dot{G}^B_{1 2}\dot{G}^B_{2 1}\dot{G}^B_{3 1}\dot{G}^B_{4 3}\dot{G}^B_{5 4}\dot{G}^B_{6 5} -\dot{G}^B_{3 1}\dot{G}^B_{4 3}\dot{G}^B_{5 4}\dot{G}^B_{6 5}G^F_{1 2}G^F_{2 1}
\\
P_{2 1 1 3 6 5} &= \dot{G}^B_{1 2}\dot{G}^B_{2 1}\dot{G}^B_{3 1}\dot{G}^B_{4 3}\dot{G}^B_{5 6}\dot{G}^B_{6 5} -\dot{G}^B_{1 2}\dot{G}^B_{2 1}\dot{G}^B_{3 1}\dot{G}^B_{4 3}G^F_{5 6}G^F_{6 5} \nonumber \\&-\dot{G}^B_{3 1}\dot{G}^B_{4 3}\dot{G}^B_{5 6}\dot{G}^B_{6 5}G^F_{1 2}G^F_{2 1} + \dot{G}^B_{3 1}\dot{G}^B_{4 3}G^F_{1 2}G^F_{2 1}G^F_{5 6}G^F_{6 5}
\\
P_{2 1 1 5 4 4} &= \dot{G}^B_{1 2}\dot{G}^B_{2 1}\dot{G}^B_{3 1}\dot{G}^B_{4 5}\dot{G}^B_{5 4}\dot{G}^B_{6 4} -\dot{G}^B_{1 2}\dot{G}^B_{2 1}\dot{G}^B_{3 1}\dot{G}^B_{6 4}G^F_{4 5}G^F_{5 4}\nonumber \\&-\dot{G}^B_{3 1}\dot{G}^B_{4 5}\dot{G}^B_{5 4}\dot{G}^B_{6 4}G^F_{1 2}G^F_{2 1} + \dot{G}^B_{3 1}\dot{G}^B_{6 4}G^F_{1 2}G^F_{2 1}G^F_{4 5}G^F_{5 4}
\\
P_{2 1 1 5 6 4} &= \dot{G}^B_{1 2}\dot{G}^B_{2 1}\dot{G}^B_{3 1}\dot{G}^B_{4 5}\dot{G}^B_{5 6}\dot{G}^B_{6 4} + \dot{G}^B_{1 2}\dot{G}^B_{2 1}\dot{G}^B_{3 1}G^F_{4 6}G^F_{5 4}G^F_{6 5} \nonumber\\&-\dot{G}^B_{3 1}\dot{G}^B_{4 5}\dot{G}^B_{5 6}\dot{G}^B_{6 4}G^F_{1 2}G^F_{2 1} -\dot{G}^B_{3 1}G^F_{1 2}G^F_{2 1}G^F_{4 6}G^F_{5 4}G^F_{6 5}
\\
P_{2 1 4 3 6 5} &= \dot{G}^B_{1 2}\dot{G}^B_{2 1}\dot{G}^B_{3 4}\dot{G}^B_{4 3}\dot{G}^B_{5 6}\dot{G}^B_{6 5} -\dot{G}^B_{1 2}\dot{G}^B_{2 1}\dot{G}^B_{3 4}\dot{G}^B_{4 3}G^F_{5 6}G^F_{6 5} \nonumber\\
&-\dot{G}^B_{1 2}\dot{G}^B_{2 1}\dot{G}^B_{5 6}\dot{G}^B_{6 5}G^F_{3 4}G^F_{4 3} + \dot{G}^B_{1 2}\dot{G}^B_{2 1}G^F_{3 4}G^F_{4 3}G^F_{5 6}G^F_{6 5} \nonumber\\&-\dot{G}^B_{3 4}\dot{G}^B_{4 3}\dot{G}^B_{5 6}\dot{G}^B_{6 5}G^F_{1 2}G^F_{2 1} + \dot{G}^B_{3 4}\dot{G}^B_{4 3}G^F_{1 2}G^F_{2 1}G^F_{5 6}G^F_{6 5}\nonumber \\&+ \dot{G}^B_{5 6}\dot{G}^B_{6 5}G^F_{1 2}G^F_{2 1}G^F_{3 4}G^F_{4 3} -G^F_{1 2}G^F_{2 1}G^F_{3 4}G^F_{4 3}G^F_{5 6}G^F_{6 5}
\\
P_{2 1 4 5 3 3} &= \dot{G}^B_{1 2}\dot{G}^B_{2 1}\dot{G}^B_{3 4}\dot{G}^B_{4 5}\dot{G}^B_{5 3}\dot{G}^B_{6 3} + \dot{G}^B_{1 2}\dot{G}^B_{2 1}\dot{G}^B_{6 3}G^F_{3 5}G^F_{4 3}G^F_{5 4} \nonumber\\&-\dot{G}^B_{3 4}\dot{G}^B_{4 5}\dot{G}^B_{5 3}\dot{G}^B_{6 3}G^F_{1 2}G^F_{2 1} -\dot{G}^B_{6 3}G^F_{1 2}G^F_{2 1}G^F_{3 5}G^F_{4 3}G^F_{5 4}
\\
P_{2 1 4 5 6 3} &= \dot{G}^B_{1 2}\dot{G}^B_{2 1}\dot{G}^B_{3 4}\dot{G}^B_{4 5}\dot{G}^B_{5 6}\dot{G}^B_{6 3} -\dot{G}^B_{1 2}\dot{G}^B_{2 1}G^F_{3 6}G^F_{4 3}G^F_{5 4}G^F_{6 5} \nonumber\\&-\dot{G}^B_{3 4}\dot{G}^B_{4 5}\dot{G}^B_{5 6}\dot{G}^B_{6 3}G^F_{1 2}G^F_{2 1} + G^F_{1 2}G^F_{2 1}G^F_{3 6}G^F_{4 3}G^F_{5 4}G^F_{6 5}
\\
P_{2 3 1 1 1 1} &= \dot{G}^B_{1 2}\dot{G}^B_{2 3}\dot{G}^B_{3 1}\dot{G}^B_{4 1}\dot{G}^B_{5 1}\dot{G}^B_{6 1} + \dot{G}^B_{4 1}\dot{G}^B_{5 1}\dot{G}^B_{6 1}G^F_{1 3}G^F_{2 1}G^F_{3 2}
\\
P_{2 3 1 1 1 2} &= \dot{G}^B_{1 2}\dot{G}^B_{2 3}\dot{G}^B_{3 1}\dot{G}^B_{4 1}\dot{G}^B_{5 1}\dot{G}^B_{6 2} + \dot{G}^B_{4 1}\dot{G}^B_{5 1}\dot{G}^B_{6 2}G^F_{1 3}G^F_{2 1}G^F_{3 2}
\\
P_{2 3 1 1 1 3} &= \dot{G}^B_{1 2}\dot{G}^B_{2 3}\dot{G}^B_{3 1}\dot{G}^B_{4 1}\dot{G}^B_{5 1}\dot{G}^B_{6 3} + \dot{G}^B_{4 1}\dot{G}^B_{5 1}\dot{G}^B_{6 3}G^F_{1 3}G^F_{2 1}G^F_{3 2}
\\
P_{2 3 1 1 1 4} &= \dot{G}^B_{1 2}\dot{G}^B_{2 3}\dot{G}^B_{3 1}\dot{G}^B_{4 1}\dot{G}^B_{5 1}\dot{G}^B_{6 4} + \dot{G}^B_{4 1}\dot{G}^B_{5 1}\dot{G}^B_{6 4}G^F_{1 3}G^F_{2 1}G^F_{3 2}
\\
P_{2 3 1 1 2 3} &= \dot{G}^B_{1 2}\dot{G}^B_{2 3}\dot{G}^B_{3 1}\dot{G}^B_{4 1}\dot{G}^B_{5 2}\dot{G}^B_{6 3} + \dot{G}^B_{4 1}\dot{G}^B_{5 2}\dot{G}^B_{6 3}G^F_{1 3}G^F_{2 1}G^F_{3 2}
\\
P_{2 3 1 1 2 4} &= \dot{G}^B_{1 2}\dot{G}^B_{2 3}\dot{G}^B_{3 1}\dot{G}^B_{4 1}\dot{G}^B_{5 2}\dot{G}^B_{6 4} + \dot{G}^B_{4 1}\dot{G}^B_{5 2}\dot{G}^B_{6 4}G^F_{1 3}G^F_{2 1}G^F_{3 2}
\\
P_{2 3 1 1 2 5} &= \dot{G}^B_{1 2}\dot{G}^B_{2 3}\dot{G}^B_{3 1}\dot{G}^B_{4 1}\dot{G}^B_{5 2}\dot{G}^B_{6 5} + \dot{G}^B_{4 1}\dot{G}^B_{5 2}\dot{G}^B_{6 5}G^F_{1 3}G^F_{2 1}G^F_{3 2}
\\
P_{2 3 1 1 4 4} &= \dot{G}^B_{1 2}\dot{G}^B_{2 3}\dot{G}^B_{3 1}\dot{G}^B_{4 1}\dot{G}^B_{5 4}\dot{G}^B_{6 4} + \dot{G}^B_{4 1}\dot{G}^B_{5 4}\dot{G}^B_{6 4}G^F_{1 3}G^F_{2 1}G^F_{3 2}
\\
P_{2 3 1 1 4 5} &= \dot{G}^B_{1 2}\dot{G}^B_{2 3}\dot{G}^B_{3 1}\dot{G}^B_{4 1}\dot{G}^B_{5 4}\dot{G}^B_{6 5} + \dot{G}^B_{4 1}\dot{G}^B_{5 4}\dot{G}^B_{6 5}G^F_{1 3}G^F_{2 1}G^F_{3 2}
\\
P_{2 3 1 5 6 4} &= \dot{G}^B_{1 2}\dot{G}^B_{2 3}\dot{G}^B_{3 1}\dot{G}^B_{4 5}\dot{G}^B_{5 6}\dot{G}^B_{6 4} + \dot{G}^B_{1 2}\dot{G}^B_{2 3}\dot{G}^B_{3 1}G^F_{4 6}G^F_{5 4}G^F_{6 5} \nonumber\\&+ \dot{G}^B_{4 5}\dot{G}^B_{5 6}\dot{G}^B_{6 4}G^F_{1 3}G^F_{2 1}G^F_{3 2} + G^F_{1 3}G^F_{2 1}G^F_{3 2}G^F_{4 6}G^F_{5 4}G^F_{6 5}
\\
P_{2 3 4 1 1 1} &= \dot{G}^B_{1 2}\dot{G}^B_{2 3}\dot{G}^B_{3 4}\dot{G}^B_{4 1}\dot{G}^B_{5 1}\dot{G}^B_{6 1} -\dot{G}^B_{5 1}\dot{G}^B_{6 1}G^F_{1 4}G^F_{2 1}G^F_{3 2}G^F_{4 3}
\\
P_{2 3 4 1 1 2} &= \dot{G}^B_{1 2}\dot{G}^B_{2 3}\dot{G}^B_{3 4}\dot{G}^B_{4 1}\dot{G}^B_{5 1}\dot{G}^B_{6 2} -\dot{G}^B_{5 1}\dot{G}^B_{6 2}G^F_{1 4}G^F_{2 1}G^F_{3 2}G^F_{4 3}
\\
P_{2 3 4 1 1 3} &= \dot{G}^B_{1 2}\dot{G}^B_{2 3}\dot{G}^B_{3 4}\dot{G}^B_{4 1}\dot{G}^B_{5 1}\dot{G}^B_{6 3} -\dot{G}^B_{5 1}\dot{G}^B_{6 3}G^F_{1 4}G^F_{2 1}G^F_{3 2}G^F_{4 3}
\\
P_{2 3 4 1 1 5} &= \dot{G}^B_{1 2}\dot{G}^B_{2 3}\dot{G}^B_{3 4}\dot{G}^B_{4 1}\dot{G}^B_{5 1}\dot{G}^B_{6 5} -\dot{G}^B_{5 1}\dot{G}^B_{6 5}G^F_{1 4}G^F_{2 1}G^F_{3 2}G^F_{4 3}
\\
P_{2 3 4 5 1 1} &= \dot{G}^B_{1 2}\dot{G}^B_{2 3}\dot{G}^B_{3 4}\dot{G}^B_{4 5}\dot{G}^B_{5 1}\dot{G}^B_{6 1} + \dot{G}^B_{6 1}G^F_{1 5}G^F_{2 1}G^F_{3 2}G^F_{4 3}G^F_{5 4}
\\
P_{2 3 4 5 6 1} &= \dot{G}^B_{1 2}\dot{G}^B_{2 3}\dot{G}^B_{3 4}\dot{G}^B_{4 5}\dot{G}^B_{5 6}\dot{G}^B_{6 1} -G^F_{1 6}G^F_{2 1}G^F_{3 2}G^F_{4 3}G^F_{5 4}G^F_{6 5}
\ea

\section[One-loop scalar integrals for N-point Feynman diagrams]{One-loop scalar integrals for $N$-point Feynman diagrams.}
\label{app:oneloopscalarintegrals}
In this appendix, we discuss the calculation of the one-loop scalar Euclidean N-point functions used throughout the paper. Closed-form results for these integrals in Minkowski spacetime have long been available, starting with the seminal work of ’tHooft and Veltman\cite{tHooft:1978jhc}, and subsequently improved and implemented in publicly available software such as LoopTools~\cite{Hahn:1998yk}, QCDLoop~\cite{Ellis:2007qk,Carrazza:2016gav}, and Collier~\cite{Denner:2016kdg}, among others. These implementations provide efficient and numerically stable evaluations across a wide range of kinematic configurations. Comprehensive overviews of these methods and results can be found in~\cite{vanOldenborgh:1989wn,Denner:1991kt,Ellis:2007qk}. 

Although some of the results discussed here are well known, we were unable to find a closed form expression in the literature in at least one of the cases discussed in the paper. The reference~\cite{Denner:2010tr} provides the analytic result for the scalar four-point integral for arbitrary internal masses, so the massive on-shell box, relevant to light-by-light scattering is, in principle, obtained as a limiting expression of that result. In practice, however, extracting from their formulas a closed form expression in terms of dilogarithms for this specific kinematics is nontrivial. Ref.~\cite{Davydychev:1990jt} also provides explicit expressions for massive $N$-point functions in terms of generalized hypergeometric functions. Again, the limiting case of interest is nontrivial, and the results require analytic continuation to specific kinematic regions. 

For completeness then, we revisit here these calculations and present a detailed derivation of the three cases of phenomenological relevance for our goal. We will illustrate the technical structure of such computations, which will be important to compare with results from our forthcoming worldline calculation in \cite{PaperII}. As mentioned in the main text, direct computation of the worldline integrals avoid the $N!$ decomposition into Feynman diagrams.

We start by introducing the one-loop scalar $N$-point function:
\ba 
\mathcal{I}_{n_1,\ldots,n_N}^d(m_1,\ldots,m_N;k_1,\ldots,k_N)= \int\frac{d^dp}{(2\pi)^d} \frac{1}{\big[(p+k_1)^2+m_1^2]^{n_1}}\cdots\frac{1}{\big[(p+k_1+\cdots+k_N)^2+m_N^2]^{n_N}}\label{eq:Npoint_scalar_function}
\ea 
We denote the loop momentum by $p$ and the external momenta by $k_i$, with $i=1,\ldots,N$, all taken as incoming and satisfying $k_1+\cdots+k_N=0$. 
We present our results in Euclidean spacetime, where all Lorentz invariants are positive definite. The corresponding expressions in Minkowski spacetime~\cite{Bern:1993kr} can be obtained by a Wick rotation of the final result, with branch cuts restored through the usual $i\epsilon$ prescription in logarithms and related functions. 

A standard method for evaluating Eq.~\eqref{eq:Npoint_scalar_function} is to introduce Schwinger parameters:
\ba 
&\mathcal{I}_{n_1,\ldots,n_N}^d(m_1,\ldots,m_N;k_1,\ldots,k_N)=\frac{1}{\Gamma(n_1)}\cdots\frac{1}{\Gamma(n_N)} \int^\infty_{-\infty} ds_1\cdots ds_N \theta(s_1)\cdots \theta(s_N) s_1^{n_1-1}\cdots s_N^{n_{N}-1} \nonumber\\
&\times\int\frac{d^dp}{(2\pi)^d} e^{-s_1(p+k_1)^2-s_1m_1^2-\cdots-s_N(p+k_1+\cdots+k_N)^2-s_Nm_N^2}\label{eq:Npoint_scalar_function_schwinger}
\ea 
The introduction of Heaviside step functions is convenient for passing to the Feynman parameter representation, where the integration domain is constrained by Dirac delta functions that are consistently imposed only when the original integrals extend over $\mathbb{R}^n$. The integration over $p$ then becomes Gaussian and can be carried out straightforwardly. As is well known, the Feynman parameter representation is subsequently obtained by inserting one (or several) factors of the form
\ba 
\label{eq:unity_factor}
1=\int^{\infty}_{-\infty} \frac{ds}{s}\theta(s)\delta(1-s'/s)\,,
\ea 
with the parameter $s'>0$ chosen to be a suitable combination of the Schwinger parameters. Different choices lead to different, but equivalent, Feynman parameter representations of the same integral. A standard choice is to treat all scalar propagators in Eq.~\eqref{eq:Npoint_scalar_function} on equal footing by combining all of them through Feynman parameters at once. This corresponds to setting $s' = s_{123\ldots N}$ in Eq.~\eqref{eq:unity_factor}, where
\ba 
s_{123\ldots N} = s_1 + s_2 + s_3 + \cdots + s_N\,.
\ea 
The Feynman parameter representation then follows immediately by rescaling each $s_i \to y_i=s_i/s$ and performing the trivial integration over $s$.

Other choices consist of combining propagators sequentially. In this approach, one inserts a Dirac delta via Eq.~\eqref{eq:unity_factor} for each subset of propagators to be treated on equal footing, choosing s' as the sum of the corresponding Schwinger parameters. Additional Dirac delta constraints are then introduced to combine the different subsets. Depending on the specific problem, one or the other method may be more convenient; both correspond to different implementations of the Cheng–Wu theorem. (See for example, \cite{Smirnov:2012}.) We quote the result for general $N$ in the standard form:
\ba 
&\mathcal{I}_{n_1,\ldots,n_N}^d(m_1,\ldots,m_N;k_1,\ldots,k_N)=\frac{\Gamma(n_{1\ldots N}-d/2)}{\Gamma(n_1)\cdots \Gamma(n_N)}\frac{1}{(4\pi)^{d/2}}\int^{+\infty}_{-\infty} dy_1\cdots dy_{N} \theta(y_1)\cdots\theta(y_N)\delta(1-y_{12\ldots N})\nonumber\\
&\times y_1^{n_1-1}\cdots y_N^{n_N-1}\bigg[\sum_{i=1}^N m_i^2 y_i
+
\sum_{1\le i<j\le N} y_i y_j\,(k_{i+1}+\cdots+k_j)^2\bigg]^{d/2-n_{1\ldots N}}\,.
\label{eq:I_N_feynmanpar}
\ea 
Note that only for $n_i=1$, $i=1,2,3,4$, does the basis of scalar integrals naturally appearing in the worldline construction of the rank-$N$ tensor in our work (introduced in Eq. \eqref{eq:J_definition} for $N=4$), coincide exactly with the integral basis in Eq. \eqref{eq:I_N_feynmanpar} used in conventional differential equation and IBP reduction approaches:
\ba
I_{1,\ldots,1}^d(123\ldots N)= \mathcal{I}_{1,\ldots,1}^d(m,\ldots,m;k_1,\ldots,k_N)\,.
\ea 
In this specific case, both representations describe the same scalar $N$-point integral with $n_1+\ldots+n_N=N$ propagators. For any other choice of $n_i$,  Eq.~\eqref{eq:Npoint_scalar_function} describes integrals with $N$ propagators raised to powers $n_i$, while $I_{n_1,\ldots,n_N}^d(123\ldots N)$ has the power of the combined denominator fixed to the value $N-d/2$ in all cases.

We will now focus on the scalar box (the one-loop scalar four-point function), which enters the evaluation of the rank-4 tensor. We present the derivation for the cases relevant to our computation: the on-shell configuration for the light-by-light scattering amplitude and its massless limit, as well as the massless four-point and three-point amplitudes with all four off-shell photon momenta.
\subsection{Massive one-loop scalar four-point function: on-shell kinematics.}
\label{subsec:massive_onshell_4point}
In Minkowski spacetime, this corresponds to the kinematic configuration $(k_i^M)^2=0$, $i=1,2,3,4$. In the center-of-mass frame one then has $\bm{k}_1+\bm{k}_2=0$ and therefore $\omega_2=\sqrt{\bm{k}_2^2}=\omega_1$. Momentum conservation implies $k_3^M+k_4^M=-k_1^M-k_2^M=-(2\omega_1,\bm{0})$, hence $\bm{k}_4^M=-\bm{k}_3^M$. Consequently $\omega_4=\sqrt{\bm{k}_4^2}=\omega_3=-\omega_1$, and therefore
\ba 
k_1^M=(\omega_1,\bm{k}_1)\,,\,\,\, k_2^M=(\omega_1,-\bm{k}_1)\,,\,\,\, k_3^M=(-\omega_1,\bm{k}_3)\,,\,\,\, k_4^M=(-\omega_1,-\bm{k}_3)
\ea 
The only independent invariants are therefore the Mandelstam variables, which in Minkowski spacetime read:
\ba 
s=2k_1^M\wc k_2^M = 4\omega_1^2\,,\,\,\,\, t=2k_2^M\wc k_3^M=-2\omega_1^2(1-\cos\theta_{32})=-4\omega_1^2\sin^2\theta_{32}/2\,.
\label{eq:invariants}
\ea 
where $\theta_{32}$ is the photon scattering angle. 
In the Euclidean region, we therefore evaluate $\mathcal{I}_{1,1,1,1}^d$ as a function only of  $k_{12}^2=(k_1+k_2)^2$ and $k_{23}^2=(k_2+k_3)^2$. Upon Wick rotation to Minkowski spacetime, these quantities correspond to the Mandelstam invariants $s$ and $t$, respectively. 

Rather than employing the standard representation of Eq.~\eqref{eq:I_N_feynmanpar}, it is more convenient to combine the propagators sequentially in Eq.~\eqref{eq:Npoint_scalar_function_schwinger}. We first pair propagators $1$ and $2$, and $3$ and $4$, by inserting two factors of unity with $s'=s_{12}$ and $s''=s_{34}$, respectively, and then a third one with $s'''=s'+s'$ to combine both pairs. Setting $d=4$ (since for $m\neq 0$, the integral is finite and free of IR poles), one readily obtains
\ba 
&I_{1,1,1,1}^4\left(m;1234\right) = \frac{1}{(4\pi)^2} \frac{1}{(4m^2)^2} \int_0^1 dy\, y(1-y)\nonumber\\
&\times \int^1_0 dz dx \frac{1}{\Big[1/4+xzy(1-y)k_{23}^2/4m^2+(1-x)(1-z)y(1-y)k_{12}^2/4m^2\Big]^2}\,.
\ea 
Introducing the shorthand $u=k_{23}^2y(1-y)/4m^2\ge0$ and $v=k_{12}^2y(1-y)/4m^2\ge 0$, the integrals over $x$ and $z$ can be performed straightforwardly,
\ba 
\int_0^1 dx \int_0^1 dz \frac{1}{\big[1/4+xzu+(1-x)(1-z)v\big]^2}=\frac{4}{4uv+u+v}\Big\{\log(1+4u)+\log(1+4v)\Big\}\,.
\ea 
Then
\ba 
&I_{1,1,1,1}^4\left(m;1234\right)= \nonumber\\
&= \frac{1}{(4\pi)^2}\frac{1}{k_{12}^2k_{23}^2}\int^1_0 dy \frac{1}{m^2/k_{12}^2+m^2/k_{23}^2+y(1-y)}\Big\{\log\Big[1+y(1-y)k_{23}^2/m^2\Big]+\log\Big[1+y(1-y)k_{12}^2/m^2\Big]\Big\}\,.
\ea 
We adopt the standard branch cut convention for the logarithm, with the cut placed along the negative real axis and the argument defined in $(-\pi,\pi]$. Accordingly, for real and positive $x$, one has $\log(-x)=\log x + i\pi$. The arguments of the logarithms above vanish at
\ba 
y_{t}^{\pm}=\frac{1}{2}\pm \frac{1}{2}\bigg[1+\frac{4m^2}{k_{23}^2}\bigg]^{1/2}\,,\,\, y_{s}^{\pm} =\frac{1}{2}\pm \frac{1}{2}\bigg[1+\frac{4m^2}{k_{12}^2}\bigg]^{1/2}\,,
\label{eq:y_t_y_s_roots}
\ea 
with $y_{t,s}^+\in[1,\infty)$ and $y_{t,s}^-\in(-\infty,0]$. Therefore, for $y\in[0,1]$ they can be written as 
\ba 
\log\bigg[1+\frac{k_{23}^2}{m^2}y(1-y)\bigg]=\log\bigg[\frac{k_{23}^2}{m^2}(y_{t}^+-y)(y-y_t^-)\bigg] = \log \frac{k_{23}^2}{m^2}+\log (y_{t}^+-y)+\log (y-y_t^-)\,,
\ea 
without crossing the branch cut, and all quantities remain real and positive. The same expansion applies to the terms depending on $k_{12}^2$. 

In contrast, the denominator has zeros at
\ba 
y_{ts}^\pm=\frac{1}{2}\pm \frac{1}{2}\bigg[1+4m^2\bigg(\frac{1}{k_{23}^2}+\frac{1}{k_{12}^2}\bigg)\bigg]^{1/2}\,,
\label{eq:y_ts_roots}
\ea 
with $y_{ts}^+\in [1,\infty)$  and $y_{ts}^-\in (-\infty,0]$. Therefore 
\ba 
\frac{1}{m^2/k_{12}^2+m^2/k_{23}^2+y(1-y)}=\frac{1}{(y-y_{ts}^-)(y_{ts}^+-y)}=\frac{1}{y_{ts}^+-y_{ts}^-}\bigg\{\frac{1}{y_{ts}^+-y}+\frac{1}{y-y_{ts}^-}\bigg\}\,.
\ea 
Then the $y$-integral can be now re-expressed as
\ba 
&I_{1,1,1,1}^4\left(m;1234\right) = \frac{1}{(4\pi)^2}\frac{1}{k_{12}^2k_{23}^2}\frac{1}{\big[1+4m^2(k_{23}^2+k_{12}^2)/k_{12}^2k_{23}^2\big]^{1/2}}\int^1_0 dy\bigg\{\frac{1}{y_{ts}^+-y}+\frac{1}{y-y_{ts}^-}\bigg\}\nonumber\\
&\times\bigg\{\log\frac{k_{23}^2}{m^2}+\log(y_t^+-y)+\log(y-y_t^-)+\log\frac{k_{12}^2}{m^2}+\log (y_s^+-y)+\log(y-y_s^-)\bigg\}\,.
\ea 
The final integration over $y$ is straightforward, but leads to a rather lengthy expression. For clarity and bookkeeping purposes, we present the result as
\ba 
I_{1,1,1,1}^4\left(m;1234\right) = \frac{1}{(4\pi)^2}\frac{1}{k_{12}^2k_{23}^2}\frac{1}{\big[1+4m^2(k_{12}^2+k_{23}^2)/k_{12}^2k_{23}^2\big]^{1/2}}\times \sum_{i=1}^{12}\mathscr{I}_i\,,\label{eq:I4_massiveonshell_result}
\ea 
with
\ba 
\mathscr{I}_1&= \log \frac{k_{23}^2}{m^2}\int^1_0 \frac{dy}{y_{ts}^+-y}=\log \frac{k_{23}^2}{m^2}\log\bigg[-\frac{y_{ts}^+}{y_{ts}^-}\bigg]\,,\nonumber\\
\mathscr{I}_2&=\int^1_0 \frac{dy}{y_{ts}^+-y} \log\left[y_{t}^+-y\right]\nonumber\\
&=\text{Li}_2\bigg[\frac{y_{t}^+}{y_{t}^+-y_{ts}^+}\bigg]-\text{Li}_2\bigg[\frac{y_{t}^-}{y_{ts}^+-y_{t}^+}\bigg]+\log y_{t}^+ \log \bigg[ \frac{y_{ts}^+}{y_{ts}^+-y_{t}^+}\bigg]-\log \left[-y_{t}^-\right] \log \bigg[\frac{y_{ts}^-}{y_{t}^+-y_{ts}^+}\bigg]\,,\nonumber\\
\mathscr{I}_3&=\int^1_0 \frac{dy}{y_{ts}^+-y}\log\left[y-y_t^-\right]\nonumber\\
&=\mathrm{Li}_2\left[\frac{y_{t}^-}{y_t^--y_{ts}^+}\right]-\mathrm{Li}_2\left[\frac{y_{t}^+}{y_{ts}^+-y_{t}^-}\right]+\log[-y_t^-]\log\bigg[\frac{y_{ts}^+}{y_{ts}^+-y_t^-}\bigg]-\log[y_t^+]\log\bigg[\frac{y_{ts}^-}{y_{ts}^+-y_t^-}\bigg]\,,\nonumber\\
\mathscr{I}_4&= \log \frac{k_{12}^2}{m^2}\int^1_0 \frac{dy}{y_{ts}^+-y}=\log \frac{k_{12}^2}{m^2}\log\bigg[-\frac{y_{ts}^+}{y_{ts}^-}\bigg]\,,\nonumber\\
\mathscr{I}_5&=\int^1_0 \frac{dy}{y_{ts}^+-y} \log\big[y_{s}^+-y\big]\nonumber\\
&=\text{Li}_2\bigg[\frac{y_{s}^+}{y_{s}^+-y_{ts}^+}\bigg]-\text{Li}_2\bigg[\frac{y_{s}^-}{y_{ts}^+-y_{s}^+}\bigg]+\log y_{s}^+ \log \bigg[ \frac{y_{ts}^+}{y_{ts}^+-y_{s}^+}\bigg]-\log \left[-y_{s}^-\right] \log \bigg[\frac{y_{ts}^-}{y_{s}^+-y_{ts}^+}\bigg]\,,\nonumber\\
\mathscr{I}_6&=\int^1_0 \frac{dy}{y_{ts}^+-y}\log \big[y-y_{s}^-\big]\nonumber\\
&=\mathrm{Li}_2\left[\frac{y_s^-}{y_s^--y_{ts}^+}\right]-\mathrm{Li}_2\left[\frac{y_s^+}{y_{ts}^+-y_s^-}\right]+\log\left[-y_s^-\right]\log\left[\frac{y_{ts}^+}{y_{ts}^+-y_{s}^-}\right]-\log\left[y_s^+\right]\log\left[\frac{y_{ts}^-}{y_{s}^--y_{ts}^+}\right]\,,\nonumber\\
\mathscr{I}_7&=\log \frac{k_{23}^2}{m^2}\int^1_0 \frac{dy}{y-y_{ts}^-} =\log \frac{k_{23}^2}{m^2}\log \bigg[-\frac{y_{ts}^+}{y_{ts}^-}\bigg]\,,\nonumber\\
\mathscr{I}_8&=\int_0^1 \frac{dy}{y-y_{ts}^-}\log \big[y_{t}^+-y\big]\nonumber\\
&=\text{Li}_2\bigg[\frac{y_{t}^-}{y_{ts}^--y_{t}^+}\bigg]-\text{Li}_2\bigg[\frac{y_{t}^+}{y_t^+-y_{ts}^-}\bigg]+\log \left[-y_{t}^-\right] \log \bigg[\frac{y_{ts}^+}{y_{t}^+-y_{ts}^-}\bigg]-\log y_{t}^+ \log \bigg[ \frac{y_{ts}^-}{y_{ts}^--y_{t}^+}\bigg]\,,\nonumber\\
\mathscr{I}_9&=\int^1_0 \frac{dy}{y-y_{ts}^-}\log\big[y-y_{t}^-\big]\nonumber\\
&=\text{Li}_2 \bigg[\frac{y_{t}^+}{y_{ts}^--y_{t}^-}\bigg]-\text{Li}_2\bigg[\frac{y_{t}^-}{y_{t}^--y_{ts}^-}\bigg]+\log[y_t^+]\log\bigg[\frac{y_{ts}^+}{y_{t}^--y_{ts}^-}\bigg]-\log[-y_t^-]\log\bigg[\frac{y_{ts}^-}{y_{ts}^--y_t^-}\bigg]\,,\nonumber\\
\mathscr{I}_{10}&=\log \frac{k_{12}^2}{m^2}\int^1_0 \frac{dy}{y-y_{ts}^-} =\log \frac{k_{12}^2}{m^2}\log \bigg[-\frac{y_{ts}^+}{y_{ts}^-}\bigg]\,,\nonumber\\
\mathscr{I}_{11}&=\int_0^1 \frac{dy}{y-y_{ts}^-}\log \big[y_{s}^+-y\big]\nonumber\\
&=\text{Li}_2\bigg[\frac{y_{s}^-}{y_{ts}^--y_{s}^+}\bigg]-\text{Li}_2\bigg[\frac{y_{s}^+}{y_s^+-y_{ts}^-}\bigg]+\log \left[-y_{s}^-\right] \log \bigg[\frac{y_{ts}^+}{y_{s}^+-y_{ts}^-}\bigg]-\log [y_{s}^+] \log \left[ \frac{y_{ts}^-}{y_{ts}^--y_{s}^+}\right]\,,\nonumber\\
\mathscr{I}_{12}&=\int^1_0 \frac{dy}{y-y_{ts}^-}\log\left[y-y_{s}^-\right]\nonumber\\
&=\text{Li}_2\left[\frac{y_{s}^+}{y_{ts}^--y_{s}^-}\right]-\text{Li}_2 \left[\frac{y_{s}^-}{y_{s}^--y_{ts}^-}\right]+\log y_s^+ \log\left[\frac{y_{ts}^+}{y_{s}^--y_{ts}^-}\right]-\log \left[-y_{s}^-\right]\log \left[\frac{y_{ts}^-}{y_{ts}^--y_{s}^-}\right]\,,
\ea 
with $y_{t,s}^{\pm}$ and $y_{ts}^\pm$ given by the roots in Eqs.~\eqref{eq:y_t_y_s_roots} and \eqref{eq:y_ts_roots}. This completes the evaluation of the basis one-loop scalar four-point integral in massive on-shell case entering the computation of the head form factors. 
\subsection{Massless one-loop scalar four-point function: on-shell kinematics.}
The (infrared-divergent) massless limit can be obtained directly from Eq.~\eqref{eq:I4_massiveonshell_result} by taking $m \to 0$. In this limit, the fermion mass acts as an infrared regulator. Expanding the roots in Eqs.~\eqref{eq:y_t_y_s_roots} and \eqref{eq:y_ts_roots} for small ratios $m^2/k_{23}^2$ and $m^2/k_{12}^2$,
\ba 
y_t^+= 1+\frac{m^2}{k_{23}^2}+\cdots\,,\,\,\,\,\,\, y_{t}^-=-\frac{m^2}{k_{23}^2}+\cdots\,,\,\,\,\,\,\,
y_s^+= 1+\frac{m^2}{k_{12}^2}+\cdots\,,\,\,\,\,\,\, y_{s}^-=-\frac{m^2}{k_{12}^2}+\cdots\,,
\ea 
and
\ba 
y_{ts}^+= 1+\frac{m^2}{k_{23}^2}+\frac{m^2}{k_{12}^2}\cdots\,,\,\,\,\,\,\, y_{ts}^-=-\frac{m^2}{k_{23}^2}-\frac{m^2}{k_{12}^2}+\cdots\,.
\ea 
When $m^2\to 0$ one finds
\ba
\lim_{m^2\to 0} I_{1,1,1,1}^4\left(m;1234\right)  = \frac{1}{(4\pi)^2}\frac{1}{k_{12}^2k_{23}^2}\times \sum_{i=1}^{12}\mathscr{I}_i\,,
\ea 
now with
\ba 
\mathscr{I}_1&\simeq  \log\bigg[\frac{m^2}{k_{23}^2}\bigg]\log\left[\frac{m^2}{k_{23}^2}+\frac{m^2}{k_{12}^2}\right]\,,\nonumber\\
\mathscr{I}_2&\simeq \frac{1}{2}\log^2\bigg[\frac{m^2}{k_{12}^2}\bigg]-\log\bigg[\frac{m^2}{k_{12}^2}\bigg]\log\bigg[\frac{m^2}{k_{23}^2}+\frac{m^2}{k_{12}^2}\bigg]-\mathrm{Li}_2\bigg[-\frac{k_{12}^2}{k_{23}^2}\bigg]-\log\frac{k_{12}^2}{k_{23}^2}\log\bigg[1+\frac{k_{12}^2}{k_{23}^2}\bigg]-\frac{\pi^2}{6},\nonumber\\
\mathscr{I}_3&\simeq-\frac{\pi^2}{6}\,,\nonumber\\
\mathscr{I}_4&\simeq\log\bigg[\frac{m^2}{k_{12}^2}\bigg]\log\left[\frac{m^2}{k_{23}^2}+\frac{m^2}{k_{12}^2}\right]\,,\nonumber\\
\mathscr{I}_5&\simeq -\frac{1}{2}\log^2\bigg[\frac{m^2}{k_{23}^2}\bigg]-\log\bigg[\frac{m^2}{k_{12}^2}\bigg]\log\bigg[1+\frac{k_{23}^2}{k_{12}^2}\bigg]-\mathrm{Li}_2\bigg[-\frac{k_{23}^2}{k_{12}^2}\bigg]-\frac{\pi^2}{6}\,,\nonumber\\
\mathscr{I}_6&\simeq-\frac{\pi^2}{6}\,,\nonumber\\
\mathscr{I}_7&\simeq\log\bigg[\frac{m^2}{k_{23}^2}\bigg]\log\left[\frac{m^2}{k_{23}^2}+\frac{m^2}{k_{12}^2}\right]\,,\nonumber\\
\mathscr{I}_8&\simeq-\frac{\pi^2}{6}\,,\nonumber\\
\mathscr{I}_9&\simeq -\frac{1}{2}\log^2\bigg[\frac{m^2}{k_{12}^2}\bigg]-\log\frac{m^2}{k_{23}^2}\log\bigg[1+\frac{k_{12}^2}{k_{23}^2}\bigg]-\mathrm{Li}_2\bigg[-\frac{k_{12}^2}{k_{23}^2}\bigg]-\frac{\pi^2}{6}\,,\nonumber\\
\mathscr{I}_{10}&\simeq\log\bigg[\frac{m^2}{k_{12}^2}\bigg]\log\left[\frac{m^2}{k_{23}^2}+\frac{m^2}{k_{12}^2}\right]\,,\nonumber\\
\mathscr{I}_{11}&\simeq-\frac{\pi^2}{6}\,,\nonumber\\
\mathscr{I}_{12}&\simeq-\frac{1}{2}\log^2\bigg[\frac{m^2}{k_{23}^2}\bigg]-\log\bigg[\frac{m^2}{k_{12}^2}\bigg]\log\bigg[1+\frac{k_{23}^2}{k_{12}^2}\bigg]-\mathrm{Li}_2\left[-\frac{k_{23}^2}{k_{12}^2}\right]-\frac{\pi^2}{6}\,.
\ea 
Summing all contributions leads to 
\ba 
\label{eq:I_4_massless_onshell1}
I_{1,1,1,1}^4(0;1234)
\sim
\frac{1}{8\pi^2}\frac{1}{k_{12}^2k_{23}^2}
\left[
\log\frac{k_{12}^2}{m^2}\log\frac{k_{23}^2}{m^2}
-\frac{\pi^2}{2}
\right]
\qquad (m^2\to0).
\ea 
with the expected double-logarithmic infrared divergences.

Alternatively, one may set $m^2=0$ and evaluate the integral in $d=4-2\epsilon$ dimensions to regulate the infrared singularity. Introducing a renormalization scale $\mu^2$ with the appropriate power ensures that the logarithms are dimensionless at the end of the calculation for this diagram,
\ba 
&I_{1,1,1,1}^d\left(0;1234\right)=\frac{(\mu^2)^{2-d/2}}{(4\pi)^{d/2}}\Gamma(4-d/2)\int^\infty_{-\infty} d^4y \theta(y_1)\cdots \theta(y_4) \delta(1-y_1-y_2-y_3-y_4) \frac{1}{\big[k_{12}^2y_2y_4+k_{23}^2y_1y_3\big]^{4-d/2}}\,.
\ea
As in the massive case, it is convenient to employ the alternative parametric representation obtained by sequentially combining propagators in pairs (see the discussion in Section~\ref{subsec:massive_onshell_4point}),
\ba 
&I_{1,1,1,1}^d\left(0;1234\right)=\frac{(\mu^2)^{2-d/2}}{(4\pi)^{d/2}}\Gamma(4-d/2) \int_0^1 dy \big[y(1-y)\big]^{-3+d/2} \int_0^1 dx dz \frac{1}{\big[xzk_{23}^2+(1-x)(1-z)k_{12}^2]^{4-d/2}}\,.
\ea 
Under this change of variables, the integrand factorizes completely, allowing the integration over the Feynman parameters to be carried out more easily,
\ba
\label{eq:I_4_massless_onshell2}
&I_{1,1,1,1}^{4-2\epsilon}\left(0;1234\right)= \frac{1}{(4\pi)^2} \frac{\Gamma(1+\epsilon)\Gamma^2(1-\epsilon)}{\Gamma(1-2\epsilon)} \bigg[\frac{2}{\epsilon^2}\bigg(\frac{(4\pi\mu^2)^\epsilon}{s^\epsilon}+\frac{(4\pi\mu^2)^\epsilon}{t^\epsilon}\bigg)-\log^2\frac{k_{23}^2}{k_{12}^2}-\pi^2\bigg]+\mathcal{O}(\epsilon)\,.
\ea 
thereby recovering the well-known result quoted in \cite{Bern:1993kr}. Fully expanding in $\epsilon$ one therefore obtains
\ba 
I^d_{1,1,1,1}(0;1234)
=
\frac{1}{(4\pi)^2}
\left[
\frac{4}{\epsilon^2}
+\frac{2}{\epsilon}\left(\log\frac{4\pi\mu^2}{s}+\log\frac{4\pi\mu^2}{t}\right)
-\log^2\frac{s}{t}-\pi^2
\right].
\ea 
The equivalence between the two expressions (Eqs.~\eqref{eq:I_4_massless_onshell1} and \eqref{eq:I_4_massless_onshell2}) follows from an appropriate identification of the infrared poles  $1/(d-4)$ and $\log m^2$; see \cite{Marciano:1974tv} for a similar discussion. In particular
\ba 
\log m^2
\equiv 
-\frac{2}{\epsilon}+\log(4\pi\mu^2)-\gamma_E\,,\quad \text{or}\quad
m^2
\equiv 
(4\pi e^{-\gamma_E}\mu^2)\,e^{-2/\epsilon}.
\ea 
The branch cuts are recovered by identifying $k_{12}^2\to s$ and $k_{23}^2\to t$ and using
\ba 
\log(-s) \to \log |s|-i\pi\theta(s)\,,\,\,\,\, \log(-t)\to \log|t|-i\pi\theta(t)\,.
\ea
We refer to \cite{Bern:1993kr} for the final result of the Wick rotation.
\subsection{Massless scalar three-point and four-point functions: fully off-shell kinematics.}
We now consider the case in which all external legs are off-shell and the fermion mass in the loop can be neglected. Setting $N=4$, $m=0$, and $d=4$ (since the integral is convergent in this configuration), one obtains from Eq.~\eqref{eq:Npoint_scalar_function_schwinger},
\ba 
&I_{1,1,1,1}^4(0;1234)=\frac{1}{(4\pi)^2}\int^\infty_{-\infty }d^4y \theta(y_1)\cdots \theta(y_4) \delta(1-y_1-y_2-y_3-y_4)\nonumber\\
&\times\Big[y_1y_2k_2^2+y_1y_3k_{23}^2+y_1y_4k_1^2+y_2y_3k_3^2+y_2y_4k_{12}^2+y_3y_4k_4^2\Big]^{-2}\,.\label{eq:I_4_offshell_massless_def}
\ea 
As in the previous cases, the evaluation simplifies under a different parametrization, here obtained by combining three propagators (see the discussion in Section~\ref{subsec:massive_onshell_4point}) and integrating over the remaining parameter associated with the fourth propagator. This corresponds to the change of integration variables~\cite{Usyukina:1992jd} $y_3=u$, $y_i=(1-u)u_i$, $i=1,2,4$, in Eq.~\eqref{eq:I_4_offshell_massless_def}. The Jacobian of the transformation yields $\det J=(1-u)^3$, and the integral becomes
\ba 
&I_{1,1,1,1}^4(0;1234)=\frac{1}{(4\pi)^2}\int^\infty_{-\infty }du du_1du_2du_4 \theta((1-u)u_1)\theta((1-u)u_2)\theta(u)\theta((1-u)u_4) \delta(-1+u_1+u_2+u_4)\nonumber\\
&\times\Big[u_1u_2k_2^2+u_1u_4k_1^2+u_2u_4k_{12}^2+u\big(u_1k_{23}^2+u_2k_3^2+u_4k_4^2-u_1u_2k_2^2-u_1u_4k_1^2-u_2u_4k_{12}^2\big)\Big]^{-2}\,.
\ea 
The $u$-integration can then be carried out; it is restricted to $u\in [0,1]$ by the Heaviside functions and yields a fully factorized form of the integral
\ba 
&I_{1,1,1,1}^4(0;1234)=\frac{1}{(4\pi)^2}\int^\infty_{-\infty }du_1du_2du_4\theta(u_1)\theta(u_2)\theta(u_4) \delta(-1+u_1+u_2+u_4)\nonumber\\
&\times\frac{1}{u_1u_2k_2^2+u_1u_4k_1^2+u_2u_4k_{12}^2}\frac{1}{u_1k_{23}^2+u_2k_3^2+u_4k_4^2}\,.
\ea 
As previously, it is convenient to insert an additional factor of unity of the form 
\ba 
1=\int^\infty_{-\infty}\frac{ds}{s}\theta(s)\delta(1-\alpha/s)\,.
\ea 
This fixes one of the Feynman parameters to unity, while the remaining two extend over $[0,\infty)$ and play the role of Schwinger parameters. Choosing $\alpha=u_4$ and rescaling $u_i \to s\,u_i$, the integrals over $u_4$ and $s$ can be performed trivially using the two Dirac delta functions. This yields (see \cite{Hodges:2010kq} for an alternative derivation in the context of momentum-twistor geometry),
\ba  
&I_{1,1,1,1}^4(0;1234)=\frac{1}{(4\pi)^2}\int^\infty_{0}du_1du_2 \frac{1}{u_1u_2k_2^2+u_1k_1^2+u_2k_{12}^2}\,\,\frac{1}{u_1k_{23}^2+u_2k_3^2+k_4^2}\,.
\ea 
This can be recast as
\ba 
I_{1,1,1,1}^4(0;1234)=\frac{1}{(4\pi)^2}\int^\infty_{0}du_1du_2 \frac{1}{1+u_1+u_2}\,\,\frac{1}{\alpha u_1u_2+\beta u_1+\gamma u_2}\,,
\ea 
with 
\ba 
\alpha=k_2^2k_4^2\,,\,\,\, \beta=k_1^2k_3^2\,,\,\,\, \gamma=k_{12}^2k_{23}^2\,.\label{eq:alphabetagamma}
\ea 
The integral over $u_2$ becomes straightforward once the two denominators are decomposed into partial fractions,
\ba 
I_{1,1,1,1}^4(0;1234)=\frac{1}{(4\pi)^2}\int^\infty_{0}du_1\frac{1}{\alpha u_1^2+(\alpha-\beta+\gamma)u_1+\gamma}\log \frac{(1+u_1)(\gamma+\alpha u_1)}{\beta u_1}\,.
\ea 
The integral over $u_1$ can be evaluated using the same method. However, we first perform the change of variables $u_1 \to c/(a\,u_1)$ to simplify the expresion to
\ba 
I_{1,1,1,1}^4(0;1234)=\frac{1}{(4\pi)^2}\int^\infty_{0}du_1\frac{2\log(1+u_1)+\log \alpha/\beta}{\alpha u_1^2+(\alpha-\beta+\gamma)u_1+\gamma}=\frac{1}{(4\pi)^2}\int^\infty_{0}du_1\frac{2\log(1+u_1)+\log \alpha/\beta}{\alpha(u_1+u_1^+)(u_1+u_1^-)}\,,
\ea 
where $-u_1^{\pm}$ denote the roots of the quadratic polynomial appearing in the denominator of the first expression,
\ba 
\label{eq:u_1_pm}
u_1^{\pm} = \frac{\alpha-\beta+\gamma \mp \sqrt{\alpha^2+\beta^2+\gamma^2-2\alpha\beta-2\beta\gamma-2\alpha\gamma}}{2\alpha}= \frac{\alpha-\beta+\gamma \mp \lambda^{1/2}}{2\alpha}\,,
\ea 
and we introduced the Käll\'en function $\lambda$, which can be expressed in terms of the Lorentz invariants using Eq.~\eqref{eq:alphabetagamma} as 
\ba 
\lambda
=
\big(k_{12}^2k_{23}^2 - k_1^2k_3^2 - k_2^2k_4^2\big)^2
-4k_1^2k_2^2k_3^2k_4^2\,.
\ea 
Hence,
\ba 
u_1^\pm
=
\frac{k_2^2k_4^2-k_1^2k_3^2+k_{12}^2k_{23}^2
\mp \sqrt{\big(k_{12}^2k_{23}^2 - k_1^2k_3^2 - k_2^2k_4^2\big)^2
-4k_1^2k_2^2k_3^2k_4^2}}
{2\,k_2^2k_4^2}.
\ea 
The $u_1$-integration can now be carried out  straightforwardly in the region of Euclidean spacetime (where $\lambda<0$) either by partial fractions or by residues, yielding
\ba 
&I_{1,1,1,1}^4(0;1234)\label{eq:I_4_massless_offshell}\\
&=\frac{1}{(4\pi)^2} \frac{1}{\sqrt{\lambda}} \bigg\{\log\frac{\alpha}{\beta}\log\frac{u_1^+}{u_1^-}-\log^2(-1+u_1^-)+\log^2(-1+u_1^+)-2\mathrm{Li}_2\bigg[\frac{1}{1-u_1^-}\bigg]+2\mathrm{Li}_2\bigg[\frac{1}{1-u_1^+}\bigg]\bigg\}\nonumber\\
&=\frac{1}{(4\pi)^2}
\frac{1}{\sqrt{\lambda}}
\left[
2\operatorname{Li}_2(1-u_1^-)-2\operatorname{Li}_2(1-u_1^+)
+
\log\frac{\alpha}{\beta}\log\frac{u_1^+}{u_1^-}
\right].
\ea 
This completes the calculation, where $u_1^{\pm}$ are the roots given in Eqs.~\eqref{eq:u_1_pm} expressed in terms of the kinematic invariants of Eq.~\eqref{eq:alphabetagamma}. The result is valid for $\lambda<0$, and can be analytically continued to other kinematic regions and to $\lambda=0$ following the procedure outlined in \cite{Chavez:2012kn}. A more detailed discussion of the singularity structure of the different kinematic regions may be found in \cite{Hodges:2010kq}; we do not discuss these issues further here. The above calculation gives in closed form the fully off-shell massless limit of $I_{1,1,1,1}^4$, with $u_1^{\pm}$ the roots given by Eqs.~\eqref{eq:u_1_pm} expressed in terms of the kinematic invariants of Eq.~\eqref{eq:alphabetagamma}. 

We next consider the off-shell three-point (triangle) scalar integral appearing in the recursion relation in Eq.~\eqref{eq:gram_solution}, which enters the calculation of the basis integrals $I_{n_1,n_2,n_3,n_4}^4$ with shifted powers $n_i\neq 1$. It is given by specializing  Eq.~\eqref{eq:I_N_feynmanpar} to $N=3$
\ba 
I^4_{1,1,1}(0;123)=\frac{1}{(4\pi)^2}\int^\infty_{-\infty}dy_1dy_2dy_3\theta(y_1)\theta(y_2)\theta(y_3)\delta(-1+y_1+y_2+y_3)\frac{1}{y_1y_2k_1^2+y_1y_3k_2^2+y_2y_3k_3^2}\,.
\ea 
Inserting a factor one of the form in Eq.~\eqref{eq:unity_factor} with $s'=y_3$ and re-scaling Feynman parameters according to $y_i\to \lambda y_i$ leads to the factorized form
\ba 
I^4_{1,1,1}(0;123)=\frac{1}{(4\pi)^2}\int^\infty_0 dy_1 dy_2 \frac{1}{1+y_1+y_2}\frac{1}{y_1y_2\rho+y_1\sigma+y_2\kappa}\,.
\ea 
now with parameters
\ba 
\rho=k_1^2\,,\quad \sigma=k_2^2\,,\quad \kappa=k_3^2\,.
\ea 
Therefore the integral is of the same type as $I^4_{1,1,1,1}(0;1234)$ producing
\ba
I_{1,1,1}^4(0;1234)=\frac{1}{(4\pi)^2}
\frac{1}{\sqrt{\chi}}
\left[
2\operatorname{Li}_2(1-y_1^-)-2\operatorname{Li}_2(1-y_1^+)
+
\log\frac{\rho}{\sigma}\log\frac{y_1^+}{y_1^-}
\right],\label{eq:I_3_massless_offshell}
\ea 
where 
\ba 
\chi=4\Big[(k_1\!\cdot\!k_2)-k_1^2k_2^2\Big]\,,\quad 
u_1^\pm
=
\frac{k_1^2+k_1\cdot k_2 \mp \sqrt{(k_1\cdot k_2)^2-k_1^2k_2^2}}{k_1^2}.
\ea 
\bibliography{main}
\bibliographystyle{utphys.bst}
\end{document}